\documentclass[preprintnumbers,amsmath,amssymb,pra,showkeys]{revtex4-1}
\usepackage{color}
\usepackage{graphicx}
\usepackage{algorithmic}

\def\ie{i.e.,}

\def\bravert{\egroup\,\vrule\,\bgroup}

{\catcode`\|=\active
  \gdef\Twoint#1{\left(\mathcode`\|"8000\let|\bravert {#1}\right)}}
{\catcode`\|=\active
  \gdef\Braket#1{\left<\mathcode`\|"8000\let|\bravert {#1}\right>}}

\newcommand{\beq}{\begin{equation}}
\newcommand{\eeq}{\end{equation}}
\newcommand{\beqa}{\begin{eqnarray}}
\newcommand{\eeqa}{\end{eqnarray}}
\newcommand{\bea}{\begin{array}}
\newcommand{\eea}{\end{array}}

\newcommand{\bef}{\begin{figure}}
\newcommand{\ef}{\end{figure}}
\newcommand{\bc}{\begin{center}}
\newcommand{\ec}{\end{center}}
\newcommand{\bt}{\begin{table}}
\newcommand{\et}{\end{table}}
\newcommand{\btb}{\begin{tabular}}
\newcommand{\etb}{\end{tabular}}
\newcommand{\dir}{{\tt {DIRAC }}}

\newcommand{\map}{{\emph{a priori }}}

\begin{document}
\title {\itshape{The Integral Screened Configuration Interaction Method}}

\author{Lasse Kragh S{\o}rensen$^{a}$$^{\ast}$,\vspace{6pt}
        Sebastian Bauch$^{b}$\vspace{6pt} and
        Lars Bojer Madsen$^{c}$\vspace{6pt} \\
$^{a}${\em{Department of Chemistry - {\AA}ngstr{\"o}m Laboratory,
             Uppsala University,
             S-75105 Uppsala,
             Sweden}};\\
$^{b}${\em{Institut f\"ur Theoretische Physik und Astrophysik, 
             Christian-Albrechts-Universit\"at zu Kiel, 
             Leibnizstrasse 15, 
             D-24098 Kiel, 
             Germany}};\\
$^{c}${\em{Department of Physics and Astronomy,
             Aarhus Universitet, 
             Ny Munkegade 120, 
             Aarhus 8000 C, 
             Denmark}} \vspace{6pt}\received{v1.0 submitted September 2016}}

\email{lasse.kraghsorensen@kemi.uu.se}
\bigskip

\begin{abstract}
 We present the formulation and implementation of 
the Integral-Screened Configuration-Interaction method (ISCI).
The ISCI is a minimal-operational count integral-driven direct 
Configuration-Interaction (CI) method
with a simple and rigorous integral screening (IS).
With a novel derivation of the CI equations we show that
the time consuming $\sigma$-vector calculation is
separable up to an overall sign and that this
separability can lead to a rigorous IS.
The rigorous IS leads to linear scaling
in the $\sigma$-vector step
for large systems but can also lead 
to near linear scaling for smaller systems for the standard
CISD, CISDT and CISDTQ methods, where 
the exponent for the scaling is 1.27, 1.48 and 1.98, respectively, even
while retaining an accuracy of $10^{-14}$ or less in the energy.
In the ISCI the non-relativistic CI problem can be broken into
42 generalized-matrix-vector products in the $\sigma$-vector
calculation, which can be separately optimized.
Due to the IS the ISCI can use dramatically larger
orbital spaces combined with large CI expansions, on a single cpu,
as compared with traditional CI methods.
Further, it offers an intrinsic possibility for parallelization on modern
computing architectures.
We show examples of how IS leads to linear or near linear
scaling with respect to the size of the simulation box in calculations on the 
Beryllium atom.

\end{abstract}

\maketitle
\clearpage

\section{Introduction}
\label{SEC:intro}

Configuration interaction (CI) was from the early days of Kellner \cite{kellner_ci1,kellner_ci2} and Hylleraas \cite{hyll_ci}
and another 50 to 60 years the {\it{post}}-self-consistent field (SCF) method of choice \cite{shavitt_his,cremer_ci}.
With the development of M{\o}ller-Plesset (MP) perturbation theory \cite{mppt}, coupled-cluster (CC) theory \cite{cc1,cc2}
and lately density functional theory (DFT) \cite{dft_1,dft_2} CI has to a large degree been replaced
with these newer methods \cite{cremer_ci}. The reason for the replacement is understandable since
the MP and CC methods can be size-consistent and 
size-extensive \cite{bartlett_rev07} and represent a more
compact parametrization of the wavefunction compared to CI. For larger molcules
DFT has for a long time been the obvious choice due to the steep scaling
in the wavefunction based methods. 

CI in its most common hierarchical form as show in Eq. \ref{ci_hierarchy}
where the expansion has been truncated at the single doubles level, giving the CISD
method, is rarely used nowadays. While the single-reference CI
finds little use
the multi-reference configuration-interaction method (MRCI) is, however, used
extensively for multi-configurational problems due to the precision, accuracy
and reliability of the method. A reliability that is still not present
in the multi-reference versions of perturbation theory (MRPT) and the
multi-reference coupled-cluster methods (MRCC) \cite{bartlett_2011}. 
More general time-dependent versions of MRCI
like the restricted active space configuration interaction (RASCI) \cite{rasscf,sato_rasscf,rasscf2} and
the generalized active space configuration interaction (GASCI) \cite{Bauch,Hochstuhl,Hochstuhl2} along with
the CIS \cite{klamroth_cis,klamroth_cis2,rohringer_cis,greenman_cis} 
has recently become popular for 
dynamics simulations in atomic and small molecular systems.
The combination of DFT and MRCI \cite{dft_mrci,SPOCK} greatly extended 
the size and the complexity of systems where MRCI was applicable.
It was recently shown that long range screening in a local basis can
give linear scaling in the MRCI method \cite{lin_sca_mrci,lin_sca_mrci2} thereby extending the
range in the size of systems for which MRCI can be applied even more.
Furthermore there is still a great interest in the full configuration interaction (FCI)
for benchmarking methods \cite{cremer_ci}. The interest in tractable FCI methods for
slightly larger systems has lead to stochastic CI 
methods, like diffusion Quantum Monte Carlo (QMC) \cite{anderson_qmc,luchow_qmc},
configuration selecting CI methods \cite{selcting_ci,bunge_ci,bunge_ci2,diaz_ci}
where higher excitations are selected based on the size of
lower excitations or the multifacet graphically contracted function method \cite{mfgcf1,mfgcf2}. 
Another interesting new CI development 
is the configuration interaction generalized 
singles and doubles (CIGSD) by Nakatsuji \cite{nakatsuji_cigsd,nakatsuji_cigsd2} which
in principle is exact but unfortunately suffers from
divergences of the integrals \cite{nakatsuji_int}. So even if CI is no longer
the method of choice for most applications the reliability along
with new developments does show that CI is still competetive
in many new areas of research.

During the last almost 90 years a number of great advances
in the calculation of the CI method has been achieved which
has been summarized in several recent reviews \cite{shavitt_his,cremer_ci,sherrill_ci}.
We will therefore here only mention some of the 
developments which are directly related
to the ISCI method. A major breakthrough was achieved in 
1972 by Roos \cite{roos_ci}  with the direct CI method where the eigenvalues and
eigenvectors were calculated directly from the molecular integrals
and not from a set of stored matrix elements. The direct CI method, however,
presented a new problem in determining the coupling coefficients
which was elegantly solved with the unitary group approach (UGA) or
the graphical UGA (GUGA) as pioneered in CI by 
Paldus \cite{paldus_uga} and Shavitt \cite{shavitt_uga}. 
With a direct CI method, where the coupling coefficients are
easily found, the minimal-operational count method (MOC)
can be formulated \cite{olsen2}. In the MOC method the
operational count is identical to the theoretical minimum
when considering the Slater-Condon rules.

The ISCI method is a minimal-operational count integral-driven direct CI method
where the operational count of the MOC CI will be
an upper bound for the ISCI. Once integral screening (IS) is
used, the scaling of the ISCI method can be significantly reduced. The ISCI approach
has already proven to be a valuable tool for time-dependent
generalized active space configuration-interaction calculations (TD-GASCI)
for small systems in very large and sparse basis 
sets \cite{Bauch,Larsson,Bauch2,Sid}. The focus 
here will therefore be on presenting a derivation of the method
along with small time-independent calculations to show how the IS
gives reduced scaling and show why the ISCI method would also be interesting
for large molecular systems. In the derivation we will make the connection
to the direct CI \cite{roos_ci}, the linear scaling MRCI \cite{lin_sca_mrci,lin_sca_mrci2},
the GUGA approach \cite{paldus_uga,shavitt_uga}, the MOC \cite{olsen2} 
and the problem of having so many integrals 
that they can no longer be stored in memory but have to
be calculated on the fly. Since many of the standard
techniques developed over the years for CI are not
directly applicable to the ISCI there will also 
be a great emphasis on showing the algorithmic developments.

\section{Theory}
\label{SEC:theo}
\begin{figure}[h]
\bc
\includegraphics[angle=270,width=10.0cm]{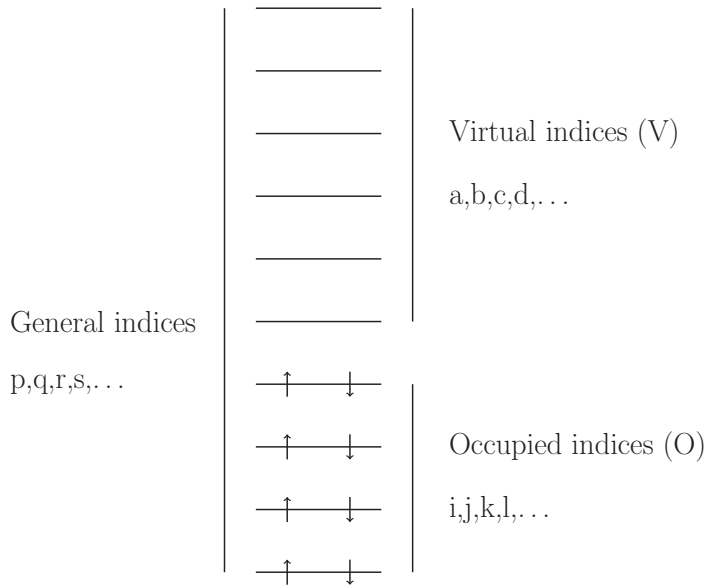}
\ec
\caption{\label{indices} A graphical view of indices used. The general
indices p,q,r,s\ldots  are running over both the occupied and the virtual orbital indices
while the indices i,j,k,l\ldots and a,b,c,d\ldots are restricted to the
occupied and virtual orbital indices, respectively.}
\end{figure}

Before commencing, a few definitions with regards to notation
need to be made. The indices p,q,r,s\ldots are general indices
running over both the occupied orbitals (O) and the virtual
orbitals (V) while a,b,c,d\ldots will be used for the virtual
orbitals and i,j,k,l\ldots for the occupied orbitals as illustrated
in Figure \ref{indices}. Since our CI formulation is based
on a single reference determinant, 
we can define the excitation and de-excitation
terms with respect to the reference determinant 
so any creation operator $\hat a^{\dagger}$
with indices a,b,c\ldots is an excitation operator while the
indices i,j,k\ldots will give a de-excitation operator while
the opposite is true for the annihilation operator $\hat a$.
We note that the de-excitation terms are only found in the 
Hamiltonian and that all de-excitation indices will have
to be matched by excitation terms from the CI excitation operator
to give non-zero contributions to the CI vector.

In this section we will present some of the differences between
the ISCI and the regular CI method in the course of presenting the
well known CI method. Here the concept of the generalized active space (GAS) will also be
introduced along with the general idea of IS
behind the ISCI.

\subsection{Hamiltonian}

While the implementation in this paper is based on the time-dependent
Schr\"odinger equation
the discussion will be kept very general and could therefore also be
used for the time-independent Schr\"odinger equation and
many other Hamiltonians like, e.g., the Dirac-Coulomb Hamiltonian \cite{krcc}. We
will, however, limit the discussion to a particle number conserving
Hamiltonians which can be expressed
in a restricted basis like a spin-restricted or
Kramers-restricted basis.
The only difference between different 
Hamiltonians will be the classes appearing in the excitation class
formalism \cite{soerensen_commcc,krcc} along with the appropriate one- and two-electron integrals.
The time-dependent non-relativistic Hamiltonian,
 in second quantization is a sum
of one- and two-body operators,
\beq
\label{ham}
 \hat{H}(t) = \sum_{pq} h_{pq}(t) \hat{a}_p^\dagger \hat{a}_q 
         + \frac{1}{2} \sum_{pqrs} g_{pqrs} \hat{a}_p^\dagger \hat{a}_r^\dagger \hat{a}_s \hat{a}_q ,
\eeq
where $h_{pq}(t) = \langle \phi_p | h(t) | \phi_q \rangle$ includes a
time-dependent coupling with, e.g., an external field, and where
$g_{pqrs} = \langle \phi_p \phi_r | g | \phi_s \phi_q \rangle $
is then constant in time provided we use time-independent orbitals.
The Hamiltonian in Eq. \ref{ham} can in a spin-restricted basis be expressed as
\beqa
\label{index_un}
\hat H &=& \sum_{pq} h_{pq}(t) (\hat{a}_{p{\alpha}}^\dagger \hat{a}_{q{\alpha}} + \hat{a}_{p{\beta}}^\dagger \hat{a}_{q{\beta}}) \\
       &+& \frac{1}{2} \sum_{pqrs} g_{pqrs} (\hat{a}_{p{\alpha}}^\dagger \hat{a}_{r{\alpha}}^\dagger \hat{a}_{s{\alpha}} \hat{a}_{q{\alpha}}
        +                                   \hat{a}_{p{\beta}}^\dagger \hat{a}_{r{\beta}}^\dagger \hat{a}_{s{\beta}} \hat{a}_{q{\beta}}) \nonumber \\
       &+& \frac{1}{2} \sum_{pqrs} g_{pqrs} (\hat{a}_{p{\alpha}}^\dagger \hat{a}_{r{\beta}}^\dagger \hat{a}_{s{\beta}} \hat{a}_{q{\alpha}}
        +                                   \hat{a}_{p{\beta}}^\dagger \hat{a}_{r{\alpha}}^\dagger \hat{a}_{s{\alpha}} \hat{a}_{q{\beta}}). \nonumber
\eeqa
Typically is the normal-ordered and index unrestricted form 
of Eq. \ref{index_un} rewritten in the UGA form \cite{paldus_uga,shavitt_uga},
where the Hamiltonian is written in terms of the generators of the 
unitary group,
\beq
\hat H = \sum_{pq} h_{pq} E_{pq} + \sum_{prqs} g_{psrq} (E_{ps} E_{rq} - E_{pq} \delta_{sr} ), \quad E_{pq} = \sum_m \hat{a}_{pm}^\dagger \hat{a}_{qm},
\eeq
or in terms of second quantized operators
\beqa
\label{uga}
\hat H &=& \sum_{pq} h_{pq} (\hat{a}_{p{\alpha}}^\dagger \hat{a}_{q{\alpha}} + \hat{a}_{p{\beta}}^\dagger \hat{a}_{q{\beta}}) \nonumber \\
       &+& \frac{1}{2} \sum_{prqs} g_{psrq} ( \hat{a}_{p{\alpha}}^\dagger \hat{a}_{s{\alpha}} \hat{a}_{r{\alpha}}^\dagger \hat{a}_{q{\alpha}}
        -                         \hat{a}_{p{\alpha}}^\dagger \hat{a}_{q{\alpha}} \delta_{sr} ) \nonumber \\
       &+& \frac{1}{2} \sum_{prqs} g_{psrq} ( \hat{a}_{p{\beta}}^\dagger \hat{a}_{s{\beta}} \hat{a}_{r{\beta}}^\dagger \hat{a}_{q{\beta}}
        -                         \hat{a}_{p{\beta}}^\dagger \hat{a}_{q{\beta}} \delta_{sr} ) \nonumber \\
       &+& \sum_{pqrs} g_{pqrs} \hat{a}_{p{\alpha}}^\dagger \hat{a}_{q{\alpha}} \hat{a}_{r{\beta}}^\dagger \hat{a}_{s{\beta}},
\eeqa
which can be derived by using the anti-commutation relations
of second quantized operators. 
We will, however, not use the UGA Hamiltonian
but rewrite the normal-ordered Hamiltonian in Eq. \ref{index_un}
to an index-restricted and spin-ordered form
\beqa
\label{index_re}
\hat H &=& \sum_{pq} h_{pq}(t) (\hat{a}_{p{\alpha}}^\dagger \hat{a}_{q{\alpha}} + \hat{a}_{p{\beta}}^\dagger \hat{a}_{q{\beta}}) \\
       &+& \sum_{p>r,q>s} (g_{psrq} - g_{pqrs})( \hat{a}_{p{\alpha}}^\dagger \hat{a}_{r{\alpha}}^\dagger \hat{a}_{q{\alpha}} \hat{a}_{s{\alpha}}
        +                                   \hat{a}_{p{\beta}}^\dagger \hat{a}_{r{\beta}}^\dagger \hat{a}_{q{\beta}} \hat{a}_{s{\beta}}) \nonumber \\
       &-& \sum_{pqrs} g_{pqrs} \hat{a}_{p{\alpha}}^\dagger \hat{a}_{r{\beta}}^\dagger \hat{a}_{q{\alpha}} \hat{a}_{s{\beta}}. \nonumber
\eeqa
In the index-restricted form of the Hamiltonian in Eq. \ref{index_re}, it
is noticed that all terms are normal ordered with
$\alpha$- before $\beta$ spins for the creation- and annihilation operators
unlike the Hamitonian used in the UGA Hamiltonian shown in Eq. \ref{uga}.
The order chosen for the spin and index restriction are
arbitrary and any other order would not change the
ISCI method since 
any sign change in the integrals in the Hamiltonian
would be compensated by an overall sign change
in Eq. \ref{final}. 
In the current implementation the index restricted time-dependent Hamiltonian
in Eq. \ref{index_re} has been used in a number of time-dependent 
simulations \cite{Bauch,Larsson,Bauch2,Sid} in very large basis sets.
We will therefore, here, focus on time-independent simulations
and the effect of IS.

\subsection{Configuration Interaction (CI)}

The CI wavefunction $|{\mathbf C} \rangle$ is parametrized by the action of 
an excitation operator $\hat X$ working on a reference determinant $|0 \rangle$,
\beq
\label{ci_ansatz}
 | {\mathbf C} \rangle = \hat C | 0 \rangle = \sum_i c_i \hat X_i | 0 \rangle ,
\eeq
which generates all possible determinants.
The expansion coefficients $c_i$ are found
by a variational optimization of the expectation value
of the electronic energy which is equivalent
to an eigenvalue problem for the coefficients and energy
\beq
\label{hcec}
{\mathbf H} | {\mathbf C} \rangle = E | {\mathbf C} \rangle.
\eeq 
While the FCI
ansatz in Eq. \ref{ci_ansatz} is exact in a complete basis and the best
approximation in an incomplete basis, it is, however, only tractable
for systems with few electrons in modest basis sets and is therefore
mostly used to benchmark other approximate theories \cite{cremer_ci}. In the development
of an approximate CI theory the excitation
operator in Eq. \ref{ci_ansatz} can be divided into a hierarchy
\beq
\label{ci_hierarchy}
\hat C = \sum_{i=0}^N \hat C_i 
                     = c_0 + \sum_{a,i}^{V,O} c_{i}^{a} \hat{a}_a^\dagger \hat{a}_i 
                     + \sum_{a>b,i>j}^{V,O} c_{ij}^{ab} \hat{a}_a^\dagger \hat{a}_b^\dagger \hat{a}_i \hat{a}_j
                     + \ldots  ,
\eeq
where the excitation operator is divided into excitation operators
with particle rank (see Eq. \ref{particle_rank}) 
spanning from zero to $N$, where $N$ is the total number
of particles and zero is the identity operator giving the reference
determinant, and where $V$ and $O$ are the numbers of virtual and occupied
orbitals, respectively. 
The sum in Eq. \ref{ci_hierarchy} has often been
truncated at $N=2$ giving the familiar CISD model.

Once a truncation in the CI hierarchy in Eq. \ref{ci_hierarchy}
has been set, the eigenvalue problem in Eq. \ref{hcec} is solved
iteratively by repeatedly applying the Hamiltonian to an
approximate eigenvector ${\mathbf v}$ to give the linearly transformed
approximate eigenvector ${\mathbf \sigma}$
\beq
\label{sigma}
{\mathbf H} {\mathbf v} = {\mathbf \sigma}.
\eeq
The application of the Hamiltonian to ${\mathbf v}$ shown in Eq. \ref{sigma}
is known as the $\sigma$-vector step and an efficient solution to
this problem is central in CI. Following
the $\sigma$-vector step is an optimization step where
typically a Davidson \cite{davidson_ci} or Lanczos \cite{lanczos} algorithm is used to
find the new approximate eigenvector ${\mathbf v}$ to be inserted
into Eq. \ref{sigma} until a desired convergence is obtained. 

While the introduction of a CI hierarchy in Eq. \ref{ci_hierarchy} has
made approximate CI calculations tractable the hierarchy, unfortunately,
converges very slowly towards the FCI and truncating the sum in Eq. \ref{ci_hierarchy}
at $N=2$ will often not suffice (for examples illustrating this point
see Ref. \cite{bible}). The problem associated with including excitation 
operators with higher particle rank shows 
when performing the $\sigma$-vector step in Eq. \ref{sigma}
since the scaling of including
an excitation operator of particle rank N is $V^{N+2} O^N$.
The scaling is seen by considering
the tensor contraction of the Hamiltonian with the
approximate eigenvector ${\mathbf v}$, with particle rank N,
\beq
\label{tensor_con}
\sum_{\substack{b_{1},b_{2} \\ a_{1},a_{2},\ldots,a_{N} \\ i_{1},i_{2},\ldots,i_{N}}}
       H^{b_{1},b_{2}}_{a_{1},a_{2}} \otimes v_{i_{1},i_{2},\ldots,i_{N}}^{a_{1},a_{2},\ldots,a_{N}} - 
                                \sigma_{i_{1},i_{2},\ldots,i_{N}}^{b_{1},b_{2},\ldots,a_{N}} = 0
\eeq
where the contraction is over the repeated indices $a_{1},a_{2}$.
We here use the convention with annihilation and creation indices
as sub- and superscript, respectively.
Counting the number of indices in the summation in Eq. \ref{tensor_con}
gives $N+2$ virtual indices and $N$ occupied indices and hence
the scaling of the tensor contraction is $V^{N+2}O^N$. 

\subsection{Generalized Active Space (GAS)}
\label{gas}

The Generalized Active Space (GAS) \cite{olsen2} concept is a generalization of the more familiar
Complete Active Space (CAS) and Restricted Active Space (RAS) approaches
where the orbitals can be divided into any number of orbital subspaces and where any
physically allowed excitation between the orbital subspaces can be performed. This
allows for complex and physically motivated truncation schemes of the
CI excitation operator in Eq. \ref{ci_hierarchy}.

The GAS is a very useful concept which can help to reduce the computational cost
and still make very accurate calculations by selecting the orbital subspaces
so that the most important CI coefficients for a 
particular problem are included. By only having
higher excitations in a subspace significantly smaller than the total
orbital space, the dimensions in the tensor contractions is dramatically
reduced for the higher excitations \ie $V$ will be much smaller for the
higher excitations.
Due to the large 
flexibility of the GAS it can, however, be difficult to make a direct comparison 
to the regular CI hierarchy without larger comparative studies
or comparison with experimental data. With
a systematic approach or good physical intuition a much more compact
and accurate wavefunction can be obtained with the GAS than with
the regular CI hierarchy in Eq. \ref{ci_hierarchy}.

The main usage of the GAS in the calculations presented in Sec. \ref{SEC:appl}
is, however, to exploit the integral sparsity of the two-electron integrals
since the GAS can be used to identify blocks 
in the $\sigma$-vector step, in Eq. \ref{sigma}, where all
integrals are trivially zero.
These blocks of trivially zero integrals then never enters the integral loop in the algorithm in
Sec. \ref{SEC:imp} and in this way the scaling of the integral loop can be
improved.

%
%
%
%



\subsection{The ISCI and Integral screening}
\label{screen}

Besides the trivially zero matrix elements from the Slater-Condon rules,
the CI Hamiltonian will also contain many very small elements
which can be considered numerically zero. The position of
the numerical zeros in the CI Hamiltonian are, however, significantly harder to predict,
especially if no prior knowledge about the structure of
the CI Hamiltonian is assumed or constructed. Furthermore larger
matrix elements, which is a sum of one or many integrals, can
contain many numerically zero integrals. It would therefore be
desirable to screen away all small matrix elements
and all small integrals, even those inside large matrix
elements. We will in the following sketch the main idea behind
the ISCI and the connection to integral and matrix element screening.

Any matrix element $H_{pq}$ can be written as a sum of integrals $I_r$
\beq
H_{pq} = \sum_r I_r .
\eeq
Since matrix multiplication is distributive the $\sigma$-vector
step in Eq. \ref{sigma} can be written as
\beq
\label{sigma_int}
\sum_t^{I_{all}} {\mathbf H}_t {\mathbf v} = {\mathbf \sigma}, \qquad 
({\mathbf H}_t){_{pq}} = \left\{\begin{array}{l} 0 \\ I_t \end{array}, \right.
\eeq
where the sum is over all integrals $I_{all}$. The matrix elements in
${\mathbf H}_t$ can now only take the values $0$ or $I_t$ and ${\mathbf H}_t$ is
therefore extremely sparse.
If some predefined threshold parameter $\epsilon$ for the 
IS is defined the
summation in Eq. \ref{sigma_int} can be split into two sums
\beq
\label{sigma_split}
\sum_t^{I_{all}} {\mathbf H}_t = \sum_l^{I_{large}} {\mathbf H}_l + \sum_s^{I_{small}} {\mathbf H}_s, 
\qquad |I_s | < \epsilon \le | I_l |.
\eeq
By having a division of the Hamiltonian in the $\sigma$-vector 
as shown in Eq. \ref{sigma_split} IS, of all
matrix elements containing a given integral $({\mathbf H}_t)$ is very simple
since such IS only requires knowing the value of $I_t$.
In this way the integral $I_t$ in principle only needs to be
calculated once in order to screen away $I_t$ in all of
${\mathbf H}$. Calculating the integral a minimal number of times
is ideal for very large basis sets where the integrals 
cannot be stored in memory but have to be calculated on the 
fly. 

The division of the Hamiltonian in Eq. \ref{sigma_split}
is of particular interest if $I_{small} >> I_{large}$
and if there is a fast way of finding and multiplying the
non-zero elements in ${\mathbf H}_t$ with the elements
in ${\mathbf v}$. If local orbitals are used then $I_{small}$
will grow as $N^4$, in a Gaussian basis set, with system size while $I_{large}$
only will grow as $N$ if the system is sufficiently spatially
extended since all integrals between orbitals
sufficiently far apart will be below $\epsilon$. The spatial 
distance between the local orbitals from where the interaction
can be neglected of course depends on the
size of $\epsilon$. 
It is therefore not surprising that the major 
gain from IS is 
seen in spatially extended systems, where much of the
long range interaction can be screened away \cite{lin_sca_mrci,pulay_lin_sca,pulay_lin_sca2,poul_lin_sca,werner_lin_sca}.
Here linear scaling in local orbitals \cite{ida_local_orb,ida_local_orb2} can be achieved.
For higher excitations 
the gain should be even larger as the same integral
will be used many more times.

The IS in Eq. \ref{sigma_split}
does not follow the usual strategy for CI algorithms 
since these will construct matrix elements, something
that will not be done using Eq. \ref{sigma_split}.
Following Eq. \ref{sigma_split} it is obvious that
the loop structure in the $\sigma$-vector step
must first consist of loops over the indices of
the integrals and the second over loops where
the integral is multiplied with the elements in ${\mathbf v}$.
While no matrix elements are contructed in the ISCI
the IS will, however, also lead to screening
of matrix elements 
\beq
| H_{pq} | < \epsilon,
\eeq
provided that all integrals for the matrix element
are numerically zero and not that the matrix element
is numerically zero due to fortuitus cancellation
of larger integrals with opposite signs. Fortuitus
cancellation will, for example, appear if two neutral systems are pulled
apart since van der Waals forces decay faster with distance than
the Coulomb interaction. By introducing modified
one-electron integrals, which are a sum of 
a one-electron integral and two-electron integrals,
for the IS the fortuitus cancellation would be
built into the one-electron integrals since these
would now also have the correct decay with distance
also for van der Waals forces.
An IS is therefore more desirable to have
than a matrix element screening since an IS can include
the matrix element screening and reduce the
cost of calculating many of the large matrix elements
simply by screening all the small integrals from these away.
In Section \ref{screen}, we will show the IS 
can be used wihtout any practical cost or effort
using the algorithm of Section \ref{second_part}.

\section{Excitation-class formalism}
\label{SEC:exclass}

In this section we recapitulate the excitation class formalism.
It was presented in Refs. \cite{soerensen_commcc,krcc} for the 
generalized-active space coupled-cluster (GASCC) method and will
here be used for the ISCI method. The
excitation class formalism is 
a way of mapping an operator, consisting of a string of second
quantized operators, onto a set of classes which can be helpful
in characterizing different parts of an operator. 
The classes we have chosen for the excitation class formalism
have a simple algebra and will be
used to determine trivially zero blocks of the Hamiltonian in Eq. \ref{hcec}.
Since we here wish to solve the time-independent Schr{\"o}dinger
equation we will use the language from the non-relativistic
framework \cite{krcc} even though there is no reference
to any specific Hamiltonian.
We assume that the orbitals
have been optimized in some restricted way so that the
orbitals can be related by the spin-flip operator \cite{krcc}.
The derivations will be done in strings of 
second quantized operators which will be connected to the GAS.
A possible parallelization strategy will be presented in the end of this section.

\subsection{Excitation-class Formalism}
\label{ex_class_form}

Starting out from basic relations depending on the number of spin alpha
and beta creation and annihilation operators it was 
shown \cite{soerensen_commcc} that one can map any second quantized normal-ordered operator 
into an operator-class formalism by introducing a set of auxiliary quantum numbers.
An extension from the $N\Delta M$-classes \cite{soerensen_commcc} to the $PHN\Delta M$-classes, 
was presented in Ref. \cite{krcc} since in the $PHN\Delta M$-classes
a way of addressing the number and types of de-excitation terms,
with respect to the chosen Fermi vacuum,
in an operator was needed in order to address the different 
contractions between the classes of ${\mathbf H}$ and ${\mathbf v}$
to ${\mathbf \sigma}$ in Eq. \ref{sigma}. These operator classes,
although they do have meaning, do {\it{not}} correspond to observables and are
devised to aid the formalism and to introduce sensible approximations. An
illustrative example will be given after the introduction of the 
$PHN\Delta M$-classes.

To obtain an efficient implementation of a relativistic or non-relativistic
CI or CC code it is important to group the operators into classes 
where each class shares one or several auxiliary quantum numbers. 
In the definitions of classes we only include information on 
the number of alpha and beta creation and annihilation operators,
$N^c_{\alpha}, N^c_{\beta},  N^a_{\alpha}$ and $N^a_{\beta}$, respectively,
thereby having four indices.
To uniquely identify a general number-conserving class of normal-ordered  
operators, only three indices are required since one linear combination
of $N^c_{\alpha}, N^c_{\beta},  N^a_{\alpha}, N^a_{\beta}$ is used
for number conservation (see Eq. \ref{ncons} below).
These three indices are chosen as particle rank $N$,
spin flip $\Delta M_s$ and the difference in the number of alpha and beta 
operators $M_{\alpha \beta}$. 

Different Hamiltonians and excitation operators
will differ in these quantities. 
The non-relativistic Hamiltonian will, for instance, have a spin flip of zero while
the Breit-Pauli Hamiltonian will have non-zero spin flip, due to spin-orbit interaction.
In cases where the spin-orbit interaction is small a reduction in the spin-flip
of the excitation operator can be introduced as an approximation
in a systematic way.
By choosing the classes along with the appropriate integrals
many different Hamiltonians can be considered within a common framework.

The classes are contructed such that an addition of an index represents a 
further division of the classes defined by the preceding indices.
These sets of indices are, however, not independent for 
a number-conserving operator since number conservation demands
\beq
\label{ncons}
N^{c \alpha} + N^{c \beta} = N^{a \alpha} + N^{a \beta},
\eeq
such that the number of creation operators equals the number
of annihilation operators.
The operator classes from a general operator like the Hamiltonian $\hat H$,
the excitation operator $\hat X$ or any other number-conserving 
normal-ordered second quantized operators
can all be divided in the $PHN\Delta M$-classes in the same way as shown below
\beq
\label{mixop2}
\hat O = \sum_{P,H} \sum_{N,\Delta ,M}  \hat O_{N,\Delta , M}^{P,H}.
\eeq
Here $N$ is the particle rank 
\beqa
\label{particle_rank}
N &=& \frac{1}{2}( N^{c \alpha} + N^{c \beta} + N^{a \alpha} + N^{a \beta} ) \nonumber \\
  &=& N^{c \alpha} + N^{c \beta} \quad \Rightarrow N^{c \alpha} + N^{c \beta} = N^{a \alpha} + N^{a \beta},
\eeqa
$\Delta$ is the spin flip of the spin orbitals \cite{jensen_saue,krcc} 
\beqa
\label{krflip}
\Delta M_s &=& \frac{1}{2}( N^{c \alpha} - N^{c \beta} + N^{a \alpha} - N^{a \beta} ) \nonumber \\                                  
           &=& N^{c \alpha} - N^{a \alpha} \quad \Rightarrow N^{c \alpha} + N^{c \beta} = N^{a \alpha} + N^{a \beta},
\eeqa
which is a pseudo-quantum number introduced as an auxiliary 
quantity to classify the different spin orbitals of a spatial orbital and
$M_{\alpha \beta}$ is the difference in the number of operators with alpha and beta spins
\beqa
M_{\alpha \beta} &=& \frac{1}{2}( N^{c \alpha} - N^{c \beta} + N^{a \alpha} - N^{a \beta} ) \nonumber \\                            
                 &=& N^{c \alpha} - N^{a \beta} \quad \Rightarrow N^{c \alpha} + N^{c \beta} = N^{a \alpha} + N^{a \beta}.
\eeqa
$P$ and $H$ denote the de-excitation part
that need to be contracted in the $\sigma$-vector step in Eq. \ref{sigma}.
$P$ gives the number of annihilation de-excitation terms while
$H$ the number of creation de-excitation terms.
$P$ and $H$ will be the subscript and superscript indices,
respectively, in Eq. \ref{tensor_con} due to the chosen
convention of annihilation and creation indices
as sub- and superscript.
While the Hamiltonian will have classes with non-zero $P$ and $H$
the excitation operator $\hat X$ will not, since $\hat X$ does not contain any
de-excitation terms according to the definition of
excitation and de-excitation operators in Section \ref{SEC:theo}.

By arranging the excitation operator $\hat X$ according to 
the excitation-class formalism, as seen in Figure \ref{cisdt}, we immediately gain the
characteristic tri-diagonal block structure in the Hamiltonian 
in non-relativistic CI theory. This block structure
is easily seen when applying the Slater-Condon rules
to excitation classes because
classes with a difference greater than two in any of the $N\Delta M$
classes are zero since the Hamiltonian contains at most two particle operators.
Additional rules may apply depending on the Hamiltonian and
the symmetry of the system.

In Figure \ref{cisdt} an example of the tri-diagonal block
structure is shown for a non-relativistic Hamiltonian
where excitations up to triples are included. 
Since every block in the Hamiltonian is addressed individually the 
introduction of the $PHN\Delta M$ classes reduces
the $\sigma$-vector step in Eq. \ref{sigma} to 
\beq
\label{sigma_block}
0 = \sum_j^c \sum_i^c {\mathbf H}_{ji} {\mathbf v}_i - {\mathbf \sigma}_j \quad \forall \quad {\mathbf H}_{ji} \ne 0,
\eeq
where $c$ is the number of $PHN\Delta M$ classes of the excitation operator defined in Eq. \ref{mixop2}.
In Eq. \ref{sigma_block} every non-zero block of the Hamiltonian ${\mathbf H}_{ji}$ is multiplied by the
approximate eigenvector ${\mathbf v}_i$ to the 
$\sigma$-vector ${\mathbf \sigma}_j$ thereby easily avoiding the calculation of
all trivially zero blocks shown in Figure \ref{cisdt}.

\begin{figure}[h!]
\bc
\includegraphics[angle=270,width=10.0cm]{cisdt.eps}
\ec
\caption{\label{cisdt} The CI matrix for a CISDT calculation where the
trivially zero blocks are white. The CI vector
has here been reordered according to the excitation classes in Eq. \ref{mixop2} whereby the
familiar tridiagonal structure of the diagonal blocks appear. The block structure
arises from the application of the Slater-Condon rules to the excitation classes
where all blocks with a difference of more than two in any of the $N \Delta M$ indices
is trivially zero. In our implementation each block in the CI matrix can
be addressed individually. We notice that the blocks
$| \hat X_{1,0,-1} \rangle \langle \hat X_{1,0,-1} | \hat H | \hat O \rangle \langle \hat O |$,
$| \hat X_{1,0,1} \rangle \langle \hat X_{1,0,1} | \hat H | \hat O \rangle \langle \hat O |$,
$| \hat O \rangle \langle \hat O | \hat H | \hat X_{1,0,-1} \rangle \langle \hat X_{1,0,-1} |$ and
$| \hat O \rangle \langle \hat O | \hat H | \hat X_{1,0,1} \rangle \langle \hat X_{1,0,1} |$ are not white
since the Brillouin theorem is not always fulfilled in our
calculations \cite{Bauch} since an outer spatial region is added after 
an SCF procedure in the inner region close to the core, 
as shown in Section \ref{basisandsetup}.}
\end{figure}

Every block of the Hamiltonian will only contain
a limited number of Hamiltonian classes. For the
reference $| 0 \rangle \langle 0 | \hat H | 0 \rangle \langle 0 |$ only the
following Hamiltonian classes will be present
(see Eq. \ref{mixop2} for notation):
\beqa
&&H_{1,0,-1}^{0,1},H_{1,0,1}^{0,1}, \\
&&H_{2,0,-2}^{0,2},H_{2,0,0}^{0,2},H_{2,0,2}^{0,2},
\eeqa
where only the diagonal integrals of the classes will be used.
In the $| X_{2,0,2} \rangle \langle X_{2,0,2} | H | X_{2,0,2} \rangle \langle X_{2,0,2} |$ block,
more Hamiltonian classes will be present:
\beqa
&&H_{1,0,-1}^{0,1},H_{1,0,1}^{0,1}, \\
\label{h01}
&&H_{2,0,-2}^{0,2},H_{2,0,0}^{0,2},H_{2,0,2}^{0,2}, \\
\label{h02}
&&H_{1,0,1}^{1,0},H_{2,0,2}^{2,0}, \\
\label{h10h20}
&&H_{2,0,0}^{1,1},H_{2,0,2}^{1,1},
\label{h11}
\eeqa
however, for $H_{1,0,-1}^{0,1},H_{1,0,1}^{0,1},H_{2,0,-2}^{0,2},H_{2,0,0}^{0,2},H_{2,0,2}^{0,2}$
a very limited number of integrals will be used,
for $H_{1,0,1}^{1,0},H_{2,0,0}^{1,1},H_{2,0,2}^{1,1}$
significantly more integrals is used while $H_{2,0,2}^{2,0}$
will be the by far most expensive term for large basis sets. Since only in the diagonal
of the $| X_{2,0,2} \rangle \langle X_{2,0,2} | H | X_{2,0,2} \rangle \langle X_{2,0,2} |$ block
all classes have non-zero contribution 
it is not worth making an exception for the diagonal elements
but instead all classes are calculated separately.
 
In the off-diagonal blocks of the Hamiltonian like $| X_{1,0,1} \rangle \langle X_{1,0,1} | H | X_{3,0,3} \rangle \langle X_{3,0,3} |$
only $H_{2,0,2}^{2,2}$ can give non-zero contributions while
for the adjoint block $| X_{3,0,3} \rangle \langle X_{3,0,3} | H | X_{1,0,1} \rangle \langle X_{1,0,1} |$
only $H_{2,0,2}^{0,0}$ will give non-zero contributions.
From this we see that the matrix elements in the 
$| X_{3,0,3} \rangle \langle X_{3,0,3} | H | X_{1,0,1} \rangle \langle X_{1,0,1} |$ block
can only be connected by a pure excitation operator ($H_{2,0,2}^{0,0}$) while 
for the $| X_{1,0,1} \rangle \langle X_{1,0,1} | H | X_{3,0,3} \rangle \langle X_{3,0,3} |$ block 
a pure de-excitation operator is needed ($H_{2,0,2}^{2,2}$).
This is part of a general observation:
The lower part of the blocks in the Hamiltonian is connected
by Hamitonian classes with more excitation than de-excitation terms
while the opposite is true for the upper part. The diagonal blocks
then only have contributions from Hamiltonian classes with the
same number of excitation and de-excitation terms.

\begin{figure}[h!]                                                                                                                              
\bc                                                                                                                                            
\includegraphics[angle=270,width=10.0cm]{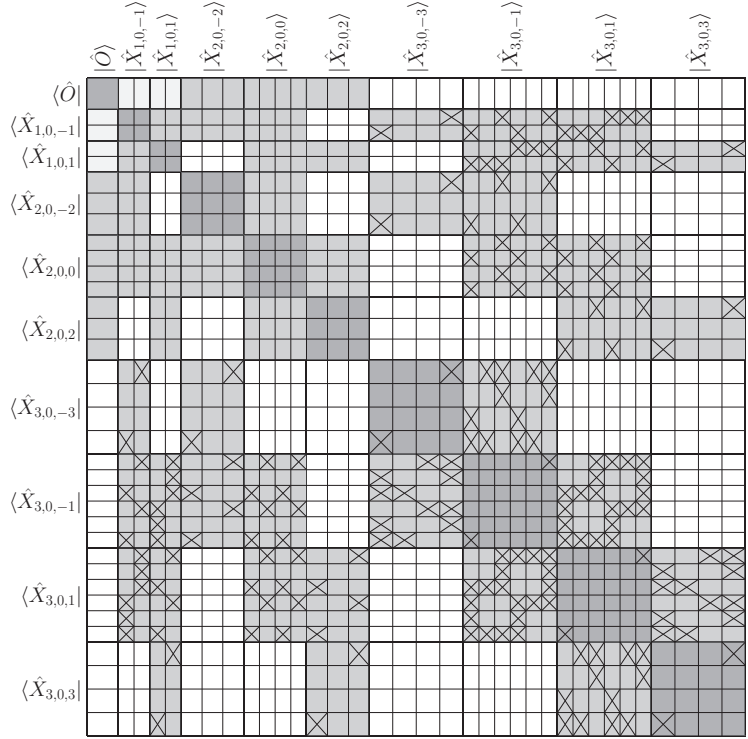}                                                                             
\ec                                                                                                                                            
\caption{\label{cisdt_gas} The CI matrix for a CISDT calculation where the virtual                                                             
space has been divided into two GASs. The coloring and ordering is
as in Figure \ref{cisdt}. The larger                                                                                      
blocks are all divided into many smaller blocks that                                                                                           
can be addressed individually. This division can help to create                                                                                
a better load balancing in a parallelization of the code. The many                                                                             
crosses indicate blocks which are trivially zero due to the                                                                            
ordering of the different types described in Appendix \ref{app_caab}.}                                                                         
\end{figure}

In the standard hierarchy in Eq. \ref{ci_hierarchy} the
size of the blocks increases as $O^{2N}V^{2N}$ which is 
illustrated in Figure \ref{cisdt} by the increase in the block sizes 
with the particle rank $N$.
As discussed in Section \ref{gas} the use of the GAS will often limit
higher excitations to a smaller part of the orbital space.
This will not change the structure
of the Hamiltonian shown in Figure \ref{cisdt} but only
the dimension of the blocks.
The GAS can, however, be used to further divide the blocks in the Hamiltonian
as shown in Figure \ref{cisdt_gas} where the virtual
space has been divided into two. 
Introducing an additional
active space need not change the calculation but can serve
to decrease the size of the individual blocks while increasing the number of blocks.
The advantage of having many smaller blocks, that can be addressed
individually, is to create a better load balancing for
a parallelization of the code. 
The load balancing is also helped by the fact that the larger the $N$, i.e., the larger
the block, the more smaller blocks will appear by
the introduction of additional GAS. 
For triple or higher excitations
the GAS will furthermore introduce additional trivially
zero blocks inside the otherwise non-zero blocks. Such sub-blocks are
shown by the elements with crosses in Figure \ref{cisdt_gas}.
These additional zero blocks are, however, only a by-product
of the sorting of the operators and will not give any speed up
of the $\sigma$-vector step for the ISCI method
presented in Section \ref{SEC:imp}. The additional zeros
will instead speed up the solution of Eqs. \ref{four}-\ref{last},
describing the different components of the $\sigma$-vector step,
due to fewer indices in the loop structure.

\section{The integral screened configuration interaction method}
\label{SEC:imp}
The aim of this section is to reformulate the CI method so that 
the IS from Section \ref{screen} will 
become possible. This will be accomplished by rewriting the Hamiltonian
in terms of excitation and de-excitation operators.
With the Hamitonian on such a form the operator strings in the $\sigma$-vector step will 
become separable up to a sign for any contraction. 
Due to the separability of the operator strings in the $\sigma$-vector step
any non-zero matrix element arising in CI
can be found by solving four short equations.
We will develop this
method from a well-defined reference determinant, 
the Fermi vacuum. 
Finally we will show that by combining the excitation-class formalism, the GAS concept 
and the rewritten Hamiltonian, a CI algorithm can
be constructed where the outer loops are over
the Hamiltonian indices and an easy and rigorous
IS can be accomplished without any cost.
The $\sigma$-vector step in
the ISCI method will be decomposed into
42 generalized matrix vector products which can be individually
optimized. 

We will in the derivation
omit all integrals and coefficients, like $h_{pq}(t)$ and $g_{pqrs}$ 
in Eq. \ref{index_re}, since these follow trivially from
the operators in question. In other words we are first
interested in finding all non-zero contributions and not
their actual value. The values, meaning integrals and
CI-coefficients, are uniquely defined from the spins and indices
of the second quantized operators and can therefore
easily be identified and evaluated.

\subsection{Hamiltonian}

Since the Hamiltonian in Eq. \ref{index_re} is normal ordered
with $\alpha$- before $\beta$ spins for the creation and annihilation operators,
every operator term in the Hamitonian can be written on the form
\beq
\label{any}
\hat H_{any} = \hat C_{\alpha} \hat C_{\beta} \hat A_{\alpha} \hat A_{\beta},
\eeq
where $\hat H_{any}$ is any term in the Hamiltonian in Eq. \ref{index_re}
and $\hat C_{\alpha}$, $\hat C_{\beta}$, $\hat A_{\alpha}$ and $\hat A_{\beta}$ 
are strings of second quantized operators with indices ordered according
to the order in the Hamiltonian in Eq. \ref{index_re}. $\hat C_{\alpha}$
covers the creation operators with $\alpha$ spin, $\hat A_{\alpha}$
covers the annihilation operators with $\alpha$ spin and likewise
for $\hat C_{\beta}$ and $\hat A_{\beta}$ with $\beta$ spin.
We note that if no $\hat C_{\alpha}$ string
is present in a given term in the Hamiltonian, e.g.,
for $h_{pq} \hat{a}_{p{\beta}}^\dagger \hat{a}_{q{\beta}}$, then
the $\hat C_{\alpha}$ string is the identity operator and likewise
for all other operator strings.

With the definition of excitation and de-excitation operators in
section \ref{ex_class_form} combined with the index restriction
on the Hamiltonian in Eq. \ref{index_re} and a canonical orbital
ordering, i.e., occupied before virtual orbitals it is seen that all
excitation terms for $\hat C_{\alpha}$ come before any 
de-excitation terms while for $\hat A_{\alpha}$ the opposite
is true. For $\hat C_{\beta}$ and $\hat A_{\beta}$ the same
is observed just with $\beta$ spin.
The relationship between the index-restricted Hamiltonian in 
Eq. \ref{index_re}, the way any Hamiltonian term can be written in
Eq. \ref{any} and how these operators are represented in the
computer is detailed in Appendix \ref{app_caab}. 
With the knowledge of the order of excitation and de-excitation
terms in the Hamiltonian, Eq. \ref{any} can be rewritten as
\beq
\label{exdx}
\hat H_{any} = \hat C_{\alpha}^{ex} \hat C_{\alpha}^{dx} \hat C_{\beta}^{ex} \hat C_{\beta}^{dx} 
               \hat A_{\alpha}^{dx} \hat A_{\alpha}^{ex} \hat A_{\beta}^{dx} \hat A_{\beta}^{ex},
\eeq
where the $ex$ and $dx$ superscripts denote if the operator is an
excitation or de-excitation operator, respectively. 
While the derivation of Eq. \ref{exdx} is
for a non-relativistic Hamiltonian it 
is general for any operator that can be written
in an index restricted 
form like the Hamiltonian in Eq. \ref{index_re}.
All 42 diferent terms in the non-relativistic Hamiltonian are written
in Eqs. \ref{class_to_ex_dx}-\ref{the_end} in terms of excitation
and de-excitation operators.

\subsection{The $\sigma$-vector step}

The $\sigma$-vector step in 
Eq. \ref{sigma_block} can be rewritten in a way
that any non-zero term in the Hamiltonian can be found by
solving four operator equations. First
the ${\mathbf \sigma}$ and ${\mathbf v}$ 
from Eq. \ref{sigma} is split into strings
of creation and annihilation operators
\beqa
 \hat {\sigma} &=& \hat \sigma_{c \alpha} \hat \sigma_{c \beta} \hat \sigma_{a \alpha} \hat \sigma_{a \beta}, \\
\label{sigmav}
 \hat v &=& \hat v_{c \alpha} \hat v_{c \beta} \hat v_{a \alpha} \hat v_{a \beta},
\eeqa
where the superscript has been omitted since $\hat {\sigma}$ and $\hat v$ 
are excitation operators like $ \hat X$.
Inserting Eqs. \ref{exdx}-\ref{sigmav} into the $\sigma$-vector
step in Eq. \ref{sigma_block} for any part of the Hamiltonian operator in Eq. \ref{index_re},
we obtain
\beq
\label{sig}
\hat \sigma_{c \alpha} \hat \sigma_{c \beta} \hat \sigma_{a \alpha} \hat \sigma_{a \beta} =
\hat C_{\alpha}^{ex} \hat C_{\alpha}^{dx} \hat C_{\beta}^{ex} \hat C_{\beta}^{dx}
\hat A_{\alpha}^{dx} \hat A_{\alpha}^{ex} \hat A_{\beta}^{dx} \hat A_{\beta}^{ex}
\hat v_{c \alpha} \hat v_{c \beta} \hat v_{a \alpha} \hat v_{a \beta}.
\eeq
Rearranging Eq. \ref{sig} using the elementary anti-commutation rules
for second quantized operators we find
\beq
\label{final}
\hat \sigma_{c \alpha} \hat \sigma_{c \beta} \hat \sigma_{a \alpha} \hat \sigma_{a \beta} =
\hat C_{\alpha}^{ex} \hat A_{\alpha}^{dx} \hat v_{c \alpha}
\hat C_{\beta}^{ex} \hat A_{\beta}^{dx} \hat v_{c \beta}
\hat C_{\alpha}^{dx} \hat A_{\alpha}^{ex} \hat v_{a \alpha}
\hat C_{\beta}^{dx} \hat A_{\beta}^{ex} \hat v_{a \beta} (-1)^M ,
\eeq
where $M$ is the number of transpositions needed to rearrange
the operators from Eq. \ref{sig} to Eq. \ref{final}.
Here no transpositions 
which would give anything but a sign change have been performed.
In finding non-zero elements in the $\sigma$-vector step 
it is seen that Eq. \ref{final} can be split into four parts: 
\beqa
\label{four}
\hat \sigma_{c \alpha} = \hat C_{\alpha}^{ex} \hat A_{\alpha}^{dx} \hat v_{c \alpha}, \\
\label{fourb}
\hat \sigma_{c \beta} = \hat C_{\beta}^{ex} \hat A_{\beta}^{dx} \hat v_{c \beta}, \\
\label{lasta}
\hat \sigma_{a \alpha} = \hat C_{\alpha}^{dx} \hat A_{\alpha}^{ex} \hat v_{a \alpha}, \\
\label{last}
\hat \sigma_{a \beta} = \hat C_{\beta}^{dx} \hat A_{\beta}^{ex} \hat v_{a \beta},
\eeqa
which each have to be fulfilled for a non-zero
contribution in the $\sigma$-vector calculation. 

Equations \ref{four}-\ref{last} are, in our implementation, solved by
applying a Hamiltonian term from Eqs. \ref{class_to_ex_dx}-\ref{the_end}
to $\hat v$, which in the operator form
is identical to the excitation operator $\hat X$. By using
elementary operations for second quantized operators, the 
different parts of the $\sigma$-vector can be found.
For every non-zero operation the indices of the Hamiltonian, 
$\hat v$ and $ \hat \sigma$ is tabulated and used
in the $\sigma$-vector step since each of these indices 
will give a part of the integral
multiplied with the coefficients in $\hat v$ to $\hat \sigma$.

The sign from the transposition of the operators will then be 
multiplied after the solutions of Eqs. \ref{four}-\ref{last}
is assembled in the $\sigma$-vector step
since this is an overall sign for the reorder. Equations \ref{four}-\ref{last}
represent the operator form of the $\sigma$-vector step in Eq. \ref{sigma_block}
split into four operator equations for the creation and annihilation operators with
$\alpha$ and $\beta$ spin. The operator part of the $\sigma$-vector step
is therefore separable up to an overall sign which only depends
on the number of transpositions performed.
With the tabulated solutions to Eqs. \ref{four}-\ref{last}
the $\sigma$-vector step can immediately be cast into a matrix
vector product form without any setup needed as shown in Section \ref{second_part}.

If $\hat H_{2,0,0}^{2,2}$ is applied to $\hat v$,
part of the solutions to Eqs. \ref{four}-\ref{last} for the $\hat H_{1,0,2}^{1,1}$
and $\hat H_{1,0,-2}^{1,1}$ Hamiltonian classes, 
plus partial solutions to 
many more of the Hamiltonian classes in Eqs. \ref{class_to_ex_dx}-\ref{the_end},
are also found. This happens because
$\hat C_{\beta}^{dx}$ and $\hat A_{\beta}^{dx}$ in
$\hat H_{2,0,0}^{2,2}$ and $\hat H_{1,0,-2}^{1,1}$ are identical
and $\hat C_{\alpha}^{dx}$ and $\hat A_{\alpha}^{dx}$ in
$\hat H_{2,0,0}^{2,2}$ and  $\hat H_{1,0,-2}^{1,1}$ also are
identical.
If all unique combinations of the excitation and de-excitation
creation strings for alpha and beta spins ($\hat C_{\sigma}^{xx}$),
for the annihilation strings ($\hat A_{\sigma}^{xx}$), and $\hat v$
are tabulated in the beginning, all
Hamiltonian classes can be constructed by suitable combinations of these.
The storage and addressing
of non-zero elements is therefore radically different compared
to the GUGA lexicographical addressing and storage
scheme \cite{duch_uga,duch_uga2,karwowski_uga,duch}.

In the current algorithm Eqs. \ref{four}-\ref{last} can either be calculated
in the beginning, stored in memory and then
used when needed or recalculated when needed. The storage
needed for the solution to Eqs. \ref{four}-\ref{last} in memory 
can become extremely large 
but can be dramatically reduced by noting
that many of the operator strings are
repeated in Eqs. \ref{class_to_ex_dx}-\ref{the_end}. Further
reductions can be achieved by utilizing identical sorting
in multiple GAS \cite{lasse_unpub}.

Solving Eqs. \ref{four}-\ref{last} immediately allows for
the calculation of 
the $\sigma$-vector step in Eq. \ref{sigma_block} by
only looping over the non-zero elements since
Eqs. \ref{four}-\ref{last} gives the indices for the integral along
with relative offsets and phase factors for ${\mathbf v}$ and ${\mathbf \sigma}$.
The ISCI method is therefore a 
minimal-operational count \cite{olsen2}
integral-driven direct CI approach where the operational count of the regular MOC CI method
will be an upper bound.
Once IS is used the scaling of the ISCI method
can be reduced as seen in Section \ref{calc_be}.

We show an example of the calculation of the $\sigma$-vector
$| \hat X_{2,0,0} \rangle \langle \hat X_{2,0,0} | \hat H_{2,0,0}^{2,0} | \hat X_{2,0,0} \rangle \langle \hat X_{2,0,0} | 
{\mathbf v}_{2,0,0} = {\mathbf \sigma}_{2,0,0}$,
which is the most expensive block in a CISD calculation.
A naive calculation, where the Hamiltonian 
$| \hat X_{2,0,0} \rangle \langle \hat X_{2,0,0} | \hat H_{2,0,0}^{2,0} | \hat X_{2,0,0} \rangle \langle \hat X_{2,0,0} |$ is setup
and multiplied with ${\mathbf v}_{2,0,0}$, will give
a scaling of $O^4 V^4$ which, compared with the
optimal scaling of $O^2 V^4$, is far too large.
The non-zero elements in this large
Hamiltonian block with a dimension of $O^4 V^4$
can be found by solving equations with
a scaling of $V^2$. Inserting the operators for the expression 
$| \hat X_{2,0,0} \rangle \langle \hat X_{2,0,0} | \hat H_{2,0,0}^{2,0} | \hat X_{2,0,0} \rangle \langle \hat X_{2,0,0} |$ into
Eqs. \ref{four}-\ref{last},
\beqa
\label{h200ca}
\sum_d \hat a_{d \alpha}^{\dagger} &=& \sum_a \hat a_{a \alpha}^{\dagger} \sum_b \hat a_{b \alpha} \sum_c \hat a_{c \alpha}^{\dagger}, \\
\label{h200cb}
\sum_d \hat a_{d \beta}^{\dagger} &=& \sum_a \hat a_{a \beta}^{\dagger} \sum_b \hat a_{b \beta} \sum_c \hat a_{c \beta}^{\dagger}, \\
\label{h200aa}
\sum_i \hat a_{i \alpha} &=& \sum_i \hat a_{i \alpha}, \\
\label{h200ab}
\sum_i \hat a_{i \beta} &=& \sum_i \hat a_{i \beta}, 
\eeqa
we see that while Eqs. \ref{h200ca} and \ref{h200cb} have a scaling of
$V^3$ this can easily be reduced to a $V^2$ scaling by applying
the last creation operator $(\hat a_{d \sigma}^{\dagger})$ only when indices $b$ and $c$
match or generating an intermediate where the indices $b$ and $c$
match; in our implementation the latter is done. Equations \ref{h200aa} and \ref{h200ab}
are trivially solved with a copy since no contraction or addition of operators
has to be performed. The information stored from every equation
is the relative offset from ${\mathbf v}$ and ${\mathbf \sigma}$, 
the Hamitonian indices and relative phases. In Eq. \ref{h200ca}
this corresponds to the indices $c$ and $d$ for a relative offset
for ${\mathbf v}$ and ${\mathbf \sigma}$, respectively, and $a$ and $b$
as part of the Hamiltonian indices. The relative phases are calculated
by the number of transpositions needed to bring a creation and annihilation
operator next to each other for a contraction and a second phase
to bring the left hand sides of Eqs. \ref{h200ca}-\ref{h200ab} in
an ordered manner shown for the excitation operator in Eq. \ref{ci_ansatz}.
In the regular lexicographical scheme the scaling is proportional
to the number of non-zero elements for operators like 
$\hat H_{2,0,0}^{2,0}$ while the scaling here is lower
because the non-zero elements are found as a product of
two or more equations.

\subsection{The ISCI Algorithm}

The ISCI algorithm can be divided into two parts where the first
part consists of setting up the calculation, by deriving
the terms needed for the calculation, and the second
part performs the actual calculation of the $\sigma$-vector step in Eq. \ref{sigma}.
These two parts are very distinctly separated and will 
therefore also be described here separately.

\subsubsection{First part}
\label{first_part}

In the first part we use the Creator Annihilator Alpha Beta (CAAB) representation
of a second quantized operator 
which is decribed in more detail in 
Appendix \ref{app_caab} and Ref. \cite{lasse_thesis}.
Parts of the code used in our program
stem from the Kramers Restricted Coupled-Cluster (KRCC) module \cite{krcc}
which is a part of the \dir program package \cite{DIRAC14}.
Every operator in the CAAB representation is
called a type. The type shows the number of creation and annihilation operators
with $\alpha$ or $\beta$ spin indices of an operator in
every GAS. 

The algorithm proceeds by finding all types of
the excitation operator and the Hamiltonian. The types
are sorted according to excitation classes in Eq. \ref{mixop2}.
Every non-zero block is found by letting
the Hamiltonian types operate on the excitation operator types
and the non-zero blocks are tabulated where
the type of the Hamiltonian is stored along with
the excitation types for ${\mathbf v}$ and ${\mathbf \sigma}$.
In this way all possible CI
calculations with any choices of GAS can be set up
in a very fast manner. With the current implemented setup
it is possible to use many hundreds of GAS.

We can now loop over the non-zero blocks in three
distinct ways as determined by the first loop.
The first way would be over the ${\mathbf \sigma}$ types
which would mean that we would build the $\sigma$-vector one
type after another. The second option would be over the
${\mathbf v}$ types where we would work with
all ${\mathbf H}$ types on one ${\mathbf v}$ type
after another. The last way would be over the ${\mathbf H}$ types, which is
the way that is currently implemented, where one ${\mathbf H}$
type after another is used. 

For very large basis sets, where the integrals no longer
can be stored in memory or even in disc, the advantage of having the
${\mathbf H}$ types in the outer loop is that 
all the integrals for the ${\mathbf H}$ type
can be calculated first and then
used in all non-zero blocks for the ${\mathbf H}$ type
instead of calculating every integral on the fly
in each non-zero block, which is currently done. 
Such an approach would give a minimum number of
times an integral would have to be calculated for
very large basis sets since 
any integral would only have to be calculated once in a $\sigma$-vector
step. Such a method we will call 
minimum integral calculation (MIC). 
The size of the ${\mathbf H}$ types in the 
MIC can be controlled by the GAS
and can therefore be of any desired size. The difference can be
illustrated with the following algorithm:
\begin{algorithmic}
\LOOP[H operator types]
 \STATE {IF MIC $\rightarrow$ fetch integrals for H operator type once}
 \LOOP[Non-zero $| \hat X_{type_a} \rangle \langle \hat X_{type_a} | \hat H_{type} | \hat X_{type_b} \rangle \langle \hat X_{type_b} |$ ]
  \STATE{Calculate ${\mathbf H}_{type} {\mathbf v}_{type_b} = {\mathbf \sigma}_{type_a}$ }
 \ENDLOOP[Non-zero $| \hat X_{type_a} \rangle \langle \hat X_{type_a} | \hat H_{type} | \hat X_{type_b} \rangle \langle \hat X_{type_b} |$]
\ENDLOOP[H operator types]
\end{algorithmic}
where the outer loop is over all ${\mathbf H}$ types followed
by a calculation of all integrals for the given ${\mathbf H}$ type
for the MIC. The second loop is over the non-zero
blocks in the $\sigma$-vector calculation 
which is in the second half
of the code. The two first loops are precomputed
in the setup and the only information passed between the
first and second part of the code is which part
of Eq. \ref{sigma_block} to calculate and which
way to fetch the integrals. 
For very large basis sets the integrals
for a given ${\mathbf H}$ type are calculated once 
for every non-zero block where in the MIC
the integrals are calculated once in the beginning and then fetched
in the non-zero blocks. If the calculation
of the integrals is cheap and the number of times
an integral needs to be recalculated is small then
the two methods will be practically identical. If, however,
the opposite is true, which it will be in
the case for higher excitations, then the MIC
method will be favored since fething is cheaper than
recalculating integrals. While the MIC would
not change the scaling of integral calculation, it
would, however, reduce the prefactor which would
be beneficial for the calculations performed in Section \ref{calc_be}
since the loop over the integrals quickly becomes
the dominant step. 

\subsubsection{Second part}
\label{second_part}

The second part of the code consists of finding
the solution to Eqs. \ref{four}-\ref{last} 
which will be used in the calculation of the $\sigma$-vector
step for each $| \hat X_{type_a} \rangle \langle \hat X_{type_a} | \hat H_{type} | \hat X_{type_b} \rangle \langle \hat X_{type_b} |$.
Eqs. \ref{four}-\ref{last} are solved
by splitting the CAAB operator into an excitation and
de-excitation part as shown in Eq. \ref{type} and
the solution can either
be tabulated or calculated when needed as 
discussed in Section \ref{ex_class_form}.

The first step is to represent the operators as strings, as shown in Eq. \ref{final},
where a string is a set of indices ordered according
to the Hamiltonian order in Eq. \ref{index_re}. The length
of the string will depend on the number of creation or
annihilation operators contained in a given operator in Eq. \ref{final}. 
This means that the length of a string is the shortest possible and can never
be longer than the highest excitation level included
in the CI expansion. The loop structure for Eq. \ref{four} is:
\begin{algorithmic}
\LOOP[ Strings $\hat A_{\alpha}^{dx}$ ]
 \LOOP[ Strings $\hat v_{c \alpha}$ ]
  \STATE{ Contract $\hat A_{\alpha}^{dx}$ with $\hat v_{c \alpha}$ to intermediate strings $\hat I$ with intermediate phase }
 \ENDLOOP[ Strings $\hat v_{c \alpha}$ ]
  \IF{ Any contraction between $\hat A_{\alpha}^{dx}$ and $\hat v_{c \alpha}$ is possible }
   \LOOP[ Strings $\hat C_{\alpha}^{ex}$ ]
    \LOOP[ Strings $\hat I$ ]
     \STATE{ Add $\hat C_{\alpha}^{ex}$ to intermediate string $\hat I$ for final string $\hat \sigma_{c \alpha}$ and phase }
     \IF{ Addition of $\hat C_{\alpha}^{ex}$ and $\hat I$ to $\hat \sigma_{c \alpha}$ is possible }
      \STATE{ Calculate a relative offset for $\hat \sigma_{c \alpha}$ string }
      \STATE{ Store relative offset from strings $\hat v_{c \alpha}$ and $\hat \sigma_{c \alpha}$ }
      \STATE{ Store Hamitonian indices from strings $\hat A_{\alpha}^{dx}$ and $\hat C_{\alpha}^{ex}$ }
      \STATE{ Store total phase for contraction and addition }
     \ENDIF{ Addition of $\hat C_{\alpha}^{ex}$ and $ \hat I$ to $\hat \sigma_{c \alpha}$ is possible }
    \ENDLOOP[ Strings $\hat I$ ]
   \ENDLOOP[ Strings $\hat C_{\alpha}^{ex}$ ]
  \ENDIF{ Any contraction between $\hat A_{\alpha}^{dx}$ and $\hat v_{c \alpha}$ is possible }
\ENDLOOP[ Strings $\hat A_{\alpha}^{dx}$ ]
\end{algorithmic}
First the indices in a given $\hat A_{\alpha}^{dx}$
annihilation string are contracted with the ${\mathbf v}$ creation string
$\hat v_{c \alpha}$ to a set of intermediate creation strings $\hat I$.
The creation strings $\hat C_{\alpha}^{ex}$ are then added to the intermediate
creation strings $\hat I$. The relative offsets from the
creation strings of $\hat v_{c \alpha}$
and $\hat \sigma_{c \alpha}$, the Hamiltonian
indices in $\hat A_{\alpha}^{dx}$ and $\hat C_{\alpha}^{ex}$
along with a total phase for the contraction and addition of the strings are stored.
Here the first loop is over the Hamiltonian indices
is crucial for a rigorous IS. For Eq. \ref{fourb} the
$\alpha$ spins in Eq. \ref{four} are substituted with $\beta$ spins.
The loop structure for Eq. \ref{lasta} is:
\begin{algorithmic}
\LOOP[ Strings $\hat A_{\alpha}^{ex}$ ]
 \LOOP[ Strings $\hat v_{a \alpha}$ ]
  \STATE{ Add $\hat A_{\alpha}^{ex}$ to $\hat v_{a \alpha}$ for intermediate strings $\hat I$ with intermediate phase }
 \ENDLOOP[ Strings $\hat v_{a \alpha}$ ]
 \IF{ Addition of $\hat A_{\alpha}^{ex}$ and $\hat v_{a \alpha}$ is possible }
  \LOOP[ Strings $\hat C_{\alpha}^{dx}$ ]
   \LOOP[ Strings $\hat I$ ]
   \STATE{ Contract $\hat C_{\alpha}^{dx}$ with $\hat I$ to final string $\hat \sigma_{a \alpha}$ and phase }
   \IF{ Contraction of $\hat C_{\alpha}^{dx}$ and $\hat I$ to $\hat \sigma_{a \alpha}$ is possible }
    \STATE{ Calculate a relative offset for $\hat \sigma_{a \alpha}$ string }
    \STATE{ Store relative offset from strings $\hat v_{a \alpha}$ and $\hat \sigma_{a \alpha}$ }
    \STATE{ Store Hamitonian indices from string $\hat A_{\alpha}^{ex}$ and $\hat C_{\alpha}^{dx}$ }
    \STATE{ Store total phase for contraction and addition }
   \ENDIF{ Contraction of $\hat C_{\alpha}^{dx}$ and $\hat I$ to $\hat \sigma_{a \alpha}$ is possible }
   \ENDLOOP[ Strings $\hat I$ ]
  \ENDLOOP[ Strings $\hat C_{\alpha}^{dx}$ ]
 \ENDIF{ Addition of $\hat A_{\alpha}^{ex}$ and $\hat v_{a \alpha}$ to intermediate string is possible }
\ENDLOOP[ Strings $\hat A_{\alpha}^{ex}$ ]
\end{algorithmic}
where the only difference to the loop structure for Eq. \ref{four} 
is the order in which the addition and contraction is performed.
For Eq. \ref{last} we again can substitute $\alpha$ for $\beta$.
The strings in the contraction step are symbolically manipulated so the 
creation and annihilation operator that should be contracted stand
next to each other, a sign for the number of tranpositions is
calculated and the contracted indices are removed for an intermediate $\hat I$ or resulting $\hat \sigma_{ax}$
string. 

As an example the matrix element 
$| \hat X_{3,0,3} \rangle \langle \hat X_{3,0,3} | \hat H_{2,0,2}^{2,0} | \hat X_{3,0,3} \rangle \langle \hat X_{3,0,3} |$
gives a string length of 3 and 2 for $\hat v_{c \alpha}$ and $\hat A_{\alpha}^{dx}$,
respectively. A contration between $\hat a_{c \alpha}^{\dagger} \hat a_{b \alpha}^{\dagger} \hat a_{a \alpha}^{\dagger}$
from $\hat v_{c \alpha}$ and $\hat a_{b \alpha} \hat a_{a \alpha}$ from $\hat A_{\alpha}^{dx}$ is
\beq
\label{contract}
\hat A_{\alpha}^{dx} \hat v_{c \alpha} =
\hat a_{b \alpha} \hat a_{a \alpha} \hat a_{c \alpha}^{\dagger} \hat a_{b \alpha}^{\dagger} \hat a_{a \alpha}^{\dagger} =
\hat a_{c \alpha}^{\dagger} \hat a_{b \alpha} \hat a_{b \alpha}^{\dagger} \hat a_{a \alpha} \hat a_{a \alpha}^{\dagger} (-1)^M =
\hat a_{c \alpha}^{\dagger} (-1)^M = - \hat a_{c \alpha}^{\dagger},
\eeq
where $M$ is the number of transpositions, which in this example is three. The code 
will perform the symbolic manipulation shown in Eq. \ref{contract}
and return the resulting string $\hat a_{c \alpha}^{\dagger}$ along
with the sign resulting from the contractions. 
Had for example $\hat a_{d \alpha}^{\dagger} \hat a_{c \alpha}^{\dagger} \hat a_{a \alpha}^{\dagger}$ from $\hat v_{c \alpha}$
been used instead of $\hat a_{c \alpha}^{\dagger} \hat a_{b \alpha}^{\dagger} \hat a_{a \alpha}^{\dagger}$
all indices would not have been contracted causing
no string to be returned but a new string in $\hat v_{c \alpha}$
tried. In the addition of operators a symbolic manipulation
of two strings is performed where the indices of the two strings
are sorted in a new string and where a sign depending
on the number of transpositions is returned. To continue the loop 
with the intermediate string $- \hat a_{c \alpha}^{\dagger}$ from Eq. \ref{contract}, this can be added
to a string from $\hat C_{\alpha}^{ex}$ which could be 
$\hat a_{d \alpha}^{\dagger} \hat a_{a \alpha}^{\dagger}$
\beq
\label{add}
- \hat C_{\alpha}^{ex} \hat a_{c \alpha}^{\dagger} = 
- \hat a_{d \alpha}^{\dagger} \hat a_{a \alpha}^{\dagger} \hat a_{c \alpha}^{\dagger} =
- \hat a_{d \alpha}^{\dagger} \hat a_{c \alpha}^{\dagger} \hat a_{a \alpha}^{\dagger} (-1)^M =
\hat a_{d \alpha}^{\dagger} \hat a_{c \alpha}^{\dagger} \hat a_{a \alpha}^{\dagger} = \hat \sigma_{c \alpha},
\eeq
where again $M$ is the number of transpositions for bringing the
final string into the desired order.
The indices from the Hamiltonian, $b$ and $a$ from 
$\hat A_{\alpha}^{dx}$ and $d$ and $a$ from 
$\hat C_{\alpha}^{ex}$, the relative offset from
$\hat v_{c \alpha}$ and $\hat \sigma_{c \alpha}$ along with the
final sign from Eq. \ref{add} are stored.

From the $| \hat X_{3,0,3} \rangle \langle \hat X_{3,0,3} | \hat H_{2,0,2}^{2,0} | \hat X_{3,0,3} \rangle \langle \hat X_{3,0,3} |$ example,
all the indices for fetching an integral have been found
and since a large number of matrix elements, in this example, use the same integral
\beq
\label{int_share}
\sum_{kjic}
\langle \hat a_{d \alpha}^{\dagger} \hat a_{c \alpha}^{\dagger} \hat a_{a \alpha}^{\dagger} 
\hat a_{k \alpha} \hat a_{j \alpha} \hat a_{i \alpha}|
(g_{daab}-g_{dbaa}) 
\hat a_{d \alpha}^{\dagger} \hat a_{a \alpha}^{\dagger} \hat a_{b \alpha} \hat a_{a \alpha} |
\hat a_{c \alpha}^{\dagger} \hat a_{b \alpha}^{\dagger} \hat a_{a \alpha}^{\dagger} 
\hat a_{k \alpha} \hat a_{j \alpha} \hat a_{i \alpha} \rangle ,
\eeq
it is desirable to have a loop structure
in which the same integral immediately can be used in all the
matrix elements like:
\begin{algorithmic}
\LOOP[ Integral indices ]
 \STATE{Fetch or calculate integral $I$}
 \LOOP[ Matrix elements ]
  \STATE{Multiply integral with element in $\hat v$ to element in $\hat \sigma$}
 \ENDLOOP[ Matrix elements ]
\ENDLOOP[ Integral indices ]
\end{algorithmic}
since this will minimize the number of times an integral will need to be calculated
or fetched but even more crucial will enable a rigorous IS
simply by introducing an if-statement as shown below
\begin{algorithmic}
\LOOP[ Integral indices ]
 \STATE{Fetch or calculate integral $I$}
 \IF{ $|I| > \epsilon $ }
 \LOOP[ Matrix elements ]
  \STATE{Multiply integral with element in $\hat v$ to element in $\hat \sigma$}
 \ENDLOOP[ Matrix elements ]
 \ENDIF
\ENDLOOP[ Integral indices ]
\end{algorithmic}
where only integrals above a given parameter $\epsilon$ are used. 

Having solved Eqs. \ref{four}-\ref{last} a general loop
structure for the $\sigma$-vector step can be written where
an integral is only fetched or calculated once in the outer loops and is
immediately multiplied with the approximate coefficient in $ \hat v$
and added to the $\sigma$-vector $ \hat \sigma$ in the inner loops.
\begin{algorithmic}
\LOOP[ $\hat C_{\beta}^{dx} \hat A_{\beta}^{ex}$ ]
 \STATE{Get indices from $\hat C_{\beta}^{dx}$ and $\hat A_{\beta}^{ex}$ if needed}
 \STATE{Get number of $\hat v_{a \beta}$ strings and offset}
 \LOOP[ $\hat C_{\alpha}^{dx} \hat A_{\alpha}^{ex} $ ]
  \STATE{Get indices from $\hat C_{\alpha}^{dx} $ and $\hat A_{\alpha}^{ex} $ if needed}
  \STATE{Get number of $\hat v_{a \alpha}$ strings and offset}
  \LOOP[ $\hat C_{\beta}^{ex} \hat A_{\beta}^{dx} $ ]
   \STATE{Get indices from $\hat C_{\beta}^{ex} $ and $\hat A_{\beta}^{dx} $ if needed}
   \STATE{Get number of $\hat v_{c \beta}$ strings and offset}
   \LOOP[ $\hat C_{\alpha}^{ex} \hat A_{\alpha}^{dx} $ ]
    \STATE{Get indices from $\hat C_{\alpha}^{ex} $ and $\hat A_{\alpha}^{dx} $ if needed}
    \STATE{Get number of $\hat v_{c \alpha}$ strings and offset}
    \STATE{Fetch or calculate integral $I$}
    \IF{ $|I| > \epsilon $ }
    \LOOP[ $\hat v_{a \beta}$ ]
     \STATE{Get relative offsets and phase for $\hat \sigma_{a \beta}$ and $\hat v_{a \beta}$}
     \LOOP[ $\hat v_{a \alpha}$ ]
      \STATE{Get relative offsets and phase for $\hat \sigma_{a \alpha} $ and $\hat v_{a \alpha} $}
      \LOOP[ $\hat v_{c \beta}$ ]
       \STATE{Get relative offsets and phase for $\hat \sigma_{c \beta} $ and $\hat v_{c \beta} $}
       \LOOP[ $\hat v_{c \alpha}$ ]
        \STATE{Get relative offsets and phase for $\hat \sigma_{c \alpha} $ and $\hat v_{c \alpha} $}
        \STATE{Calculate total offset from relative offsets for $\hat \sigma$ and $\hat v$}
        \STATE{Calculate total phase from relative phases and the overall phase}
        \STATE{Multiply integral with element in $\hat v$ and overall phase to element in $\hat \sigma$}
       \ENDLOOP[ $\hat v_{c \alpha}$ ]
      \ENDLOOP[ $\hat v_{c \beta}$ ]
     \ENDLOOP[ $\hat v_{a \alpha}$ ]
    \ENDLOOP[ $\hat v_{a \beta}$ ]
    \ENDIF
   \ENDLOOP[ $\hat C_{\alpha}^{ex} \hat A_{\alpha}^{dx} $ ]
  \ENDLOOP[ $\hat C_{\beta}^{ex} \hat A_{\beta}^{dx} $ ]
 \ENDLOOP[ $\hat C_{\alpha}^{dx} \hat A_{\alpha}^{ex} $ ]
\ENDLOOP[ $\hat C_{\beta}^{dx} \hat A_{\beta}^{ex}$ ]
\end{algorithmic}
In the general loop structure above, it is seen that by separating
the integral index loop, like $\hat C_{\beta}^{dx} \hat A_{\beta}^{ex}$,
and the $ \hat v$ loop, like $\hat v_{a \beta}$, we can
choose to have the integral loops as the outer loops 
and $ \hat v$ in the inner loops. In this way
the desired loop structure is achieved where every integral is only
fetched or calculated once, screened and then, if larger than $\epsilon$, immediately multiplied
with $\hat v$ and added to $\hat \sigma$.  
The indices for the integral fetched or
calculated is comprised of the indices from the outer loops where all
loops need not give any index. 
In the inner loops the precomputed relative offsets and phases,
form Eqs. \ref{four}-\ref{last}, are used 
to determine the total offset and phase. The total offsets and phase is easily calculated as products
of the relative offsets and phases, respectively. 
The multiplication in the end is just the
integral multiplied with the element with the total offset in $\hat v$ and the total phase
to the element with the total offset in $\hat \sigma$.
The general loop structure is, however, just an example
and the loops can be ordered in any desired way with the
exception that the integral indices loop should be before the
$ \hat v$ loop for each of the
Eqs. \ref{four}-\ref{last}. Not all combinations will
of course have the integral loops before the $ \hat v$ loops
and in those cases the same integral is calculated more than once.

The general loop structure is not computational
efficient due to the many redundant loops and if-statements
inside the nested loops.
The general loop structure can be optimized by returning
to the excitation classes of Section \ref{ex_class_form}
since the general
loop structure can be split into separate loops for all excitation classes
of the Hamiltonian where no redundant loops or if-statements are present. 
The elimination is easily accomplished since the if-statements
in the general loop structure
separates the different Hamiltonian classes and the
redundant loops ensure that all Hamiltonian
classes are included. Therefore by creating a separate
$\sigma$-vector call for each class
\beqa
\label{class_to_ex_dx}
H_{2,0,-2}^{2,2} &=& \hat C_{\beta}^{dx} \hat A_{\beta}^{dx} \\
H_{2,0,0}^{2,2} &=& \hat C_{\alpha}^{dx} \hat C_{\beta}^{dx} \hat A_{\alpha}^{dx} \hat A_{\beta}^{dx} \\
H_{2,0,2}^{2,2} &=& \hat C_{\alpha}^{dx} \hat A_{\alpha}^{dx} \\
H_{2,0,-2}^{2,1} &=& \hat C_{\beta}^{ex} \hat C_{\beta}^{dx} \hat A_{\beta}^{dx} \\
H_{2,0,0}^{2,1} &=& \hat C_{\alpha}^{ex} \hat C_{\beta}^{dx} \hat A_{\alpha}^{dx} \hat A_{\beta}^{dx} 
                + \hat C_{\alpha}^{dx} \hat C_{\beta}^{ex} \hat A_{\alpha}^{dx} \hat A_{\beta}^{dx} \\ 
H_{2,0,2}^{2,1} &=& \hat C_{\alpha}^{ex} \hat C_{\alpha}^{dx} \hat A_{\alpha}^{dx} \\
H_{2,0,-2}^{1,2} &=& \hat C_{\beta}^{dx} \hat A_{\beta}^{dx} \hat A_{\beta}^{ex} \\
H_{2,0,0}^{1,2} &=& \hat C_{\alpha}^{dx} \hat C_{\beta}^{dx} \hat A_{\alpha}^{ex} \hat A_{\beta}^{dx} 
                + \hat C_{\alpha}^{dx} \hat C_{\beta}^{dx} \hat A_{\alpha}^{dx} \hat A_{\beta}^{ex} \\ 
H_{2,0,2}^{1,2} &=& \hat C_{\alpha}^{dx} \hat A_{\alpha}^{dx} \hat A_{\alpha}^{ex} \\
H_{2,0,-2}^{2,0} &=& \hat C_{\beta}^{ex} \hat A_{\beta}^{dx} \\
H_{2,0,0}^{2,0} &=& \hat C_{\alpha}^{ex} \hat C_{\beta}^{ex} \hat A_{\alpha}^{dx} \hat A_{\beta}^{dx} \\
H_{2,0,2}^{2,0} &=& \hat C_{\alpha}^{ex} \hat A_{\alpha}^{dx} \\
H_{2,0,-2}^{1,1} &=& \hat C_{\beta}^{ex} \hat C_{\beta}^{dx} \hat A_{\beta}^{dx} \hat A_{\beta}^{ex} \\ 
H_{2,0,0}^{1,1} &=& \hat C_{\alpha}^{dx} \hat C_{\beta}^{ex} \hat A_{\alpha}^{dx} \hat A_{\beta}^{ex} 
                 +  \hat C_{\alpha}^{dx} \hat C_{\beta}^{ex} \hat A_{\alpha}^{ex} \hat A_{\beta}^{dx} 
                 +  \hat C_{\alpha}^{ex} \hat C_{\beta}^{dx} \hat A_{\alpha}^{dx} \hat A_{\beta}^{ex} 
                 +  \hat C_{\alpha}^{ex} \hat C_{\beta}^{dx} \hat A_{\alpha}^{ex} \hat A_{\beta}^{dx} \\
H_{2,0,2}^{1,1} &=& \hat C_{\alpha}^{ex} \hat C_{\alpha}^{dx} \hat A_{\alpha}^{dx} \hat A_{\alpha}^{ex} 
\eeqa
\beqa
H_{2,0,-2}^{0,2} &=& \hat C_{\beta}^{dx} \hat A_{\beta}^{ex} \\
H_{2,0,0}^{0,2} &=& \hat C_{\alpha}^{dx} \hat C_{\beta}^{dx} \hat A_{\alpha}^{ex} \hat A_{\beta}^{ex} \\
H_{2,0,2}^{0,2} &=& \hat C_{\alpha}^{dx} \hat A_{\alpha}^{ex} \\
H_{2,0,-2}^{1,0} &=& \hat C_{\beta}^{ex} \hat A_{\beta}^{dx} \hat A_{\beta}^{ex} \\
H_{2,0,0}^{1,0} &=& \hat C_{\alpha}^{ex} \hat C_{\beta}^{ex} \hat A_{\alpha}^{ex} \hat A_{\beta}^{dx}
                + \hat C_{\alpha}^{ex} \hat C_{\beta}^{ex} \hat A_{\alpha}^{dx} \hat A_{\beta}^{ex}  \\
H_{2,0,2}^{1,0} &=& \hat C_{\alpha}^{ex} \hat A_{\alpha}^{dx} \hat A_{\alpha}^{ex} \\
H_{2,0,-2}^{0,1} &=& \hat C_{\beta}^{ex} \hat C_{\beta}^{dx} \hat A_{\beta}^{ex} \\
H_{2,0,0}^{0,1} &=& \hat C_{\alpha}^{ex} \hat C_{\beta}^{dx} \hat A_{\alpha}^{ex} \hat A_{\beta}^{ex}
                + \hat C_{\alpha}^{dx} \hat C_{\beta}^{ex} \hat A_{\alpha}^{ex} \hat A_{\beta}^{ex} \\
H_{2,0,2}^{0,1} &=& \hat C_{\alpha}^{ex} \hat C_{\alpha}^{dx} \hat A_{\alpha}^{ex} \\
H_{2,0,-2}^{0,0} &=& \hat C_{\beta}^{ex} \hat A_{\beta}^{ex} \\
H_{2,0,0}^{0,0} &=& \hat C_{\alpha}^{ex} \hat C_{\beta}^{ex} \hat A_{\alpha}^{ex} \hat A_{\beta}^{ex} \\
H_{2,0,2}^{0,0} &=& \hat C_{\alpha}^{ex} \hat A_{\alpha}^{ex} \\
H_{1,0,-1}^{1,1} &=& \hat C_{\beta}^{dx} \hat A_{\beta}^{dx} \\
H_{1,0,1}^{1,1} &=& \hat C_{\alpha}^{dx} \hat A_{\alpha}^{dx} \\
H_{1,0,-1}^{1,0} &=& \hat C_{\beta}^{ex} \hat A_{\beta}^{dx} \\ 
H_{1,0,1}^{1,0} &=& \hat C_{\alpha}^{ex} \hat A_{\alpha}^{dx} \\
H_{1,0,-1}^{0,1} &=& \hat C_{\beta}^{dx} \hat A_{\beta}^{ex} \\ 
H_{1,0,1}^{0,1} &=& \hat C_{\alpha}^{dx} \hat A_{\alpha}^{ex} \\
H_{1,0,-1}^{0,0} &=& \hat C_{\beta}^{dx} \hat A_{\beta}^{ex} \\ 
\label{the_end}
H_{1,0,1}^{0,0} &=& \hat C_{\alpha}^{ex} \hat A_{\alpha}^{ex}
\eeqa
the general CI problem can be reduced to 
programming 42 resonably similar
matrix-vector products. By programming every term in Eqs. \ref{class_to_ex_dx}-\ref{the_end},
every loop can be performed
without any if-statements or redundant loops
and still have a simple and rigorous IS.

For $\hat H_{2,0,2}^{2,0}$, shown in Eq. \ref{int_share},
the loop structure can be
\begin{algorithmic}
\LOOP[ $\hat C_{\alpha}^{ex} \hat A_{\alpha}^{dx} $ ]
 \STATE{Get indices from $\hat C_{\alpha}^{ex} $ and $\hat A_{\alpha}^{dx} $}
 \STATE{Get number of $\hat v_{c \alpha}$ strings and offset}
 \STATE{Fetch or calculate integral $I$}
 \IF{ $|I| > \epsilon $ }
 \LOOP[ $\hat v_{c \alpha}$ ]
  \STATE{Get relative phase and offsets for $\hat \sigma_{c \alpha} $ and $\hat v_{c \alpha} $}
  \STATE{Calculate total phase from relative phase and the overall phase}
  \LOOP[ $\hat v_{a \beta}$ ]
   \STATE{Get relative offsets for $\hat \sigma_{a \beta}$ and $\hat v_{a \beta}$}
   \LOOP[ $\hat v_{a \alpha}$ ]
    \STATE{Get relative offsets for $\hat \sigma_{a \alpha} $ and $\hat v_{a \alpha} $}
    \LOOP[ $\hat v_{c \beta}$ ]
     \STATE{Get relative offsets for $\hat \sigma_{c \beta} $ and $\hat v_{c \beta} $}
     \STATE{Calculate total offset from relative offsets for $\hat \sigma$ and $\hat v$}
     \STATE{Multiply integral with element in $\hat v$ to element in $\hat \sigma$}
    \ENDLOOP[ $\hat v_{c \beta}$ ]
   \ENDLOOP[ $\hat v_{a \alpha}$ ]
  \ENDLOOP[ $\hat v_{a \beta}$ ]
 \ENDLOOP[ $\hat v_{c \alpha}$ ]
 \ENDIF
\ENDLOOP[ $\hat C_{\alpha}^{ex} \hat A_{\alpha}^{dx} $ ]
\end{algorithmic}
where we have moved the loop for $\hat v_{c \alpha}$ further out
in comparison to the general loop.
In the $\hat v_{c \alpha}$ loop, all integral indices and phases are
known and the integral, from the $\hat C_{\alpha}^{ex} \hat A_{\alpha}^{dx} $, can 
then be recycled in the inner loops where only precomputed offsets are found. 
With this loop structure,
every integral for any matrix element $\langle \hat X_p | \hat H_{2,0,2}^{2,0} | \hat X_q \rangle$ 
only needs to be fetched or calculated once for every block
regardless of how the CI expansion is truncated. The
loop structure of all terms in Eqs. \ref{class_to_ex_dx}-\ref{the_end}
can in this way be ordered such that a minimum number of integrals
is fetched or calculated, all redundant loops and
if-statements are removed and a separate optimization 
of all terms in Eqs. \ref{class_to_ex_dx}-\ref{the_end} 
can be performed.

With the present algorithm, any one- or two-electron operator
can be calculated in the same way as the Hamiltonian where the
only change needed is to fetch the correct integral for the
operator in question. This makes the algorithm presented
very general and easy to extend to many different properties.

\subsubsection{Density matrices}

The calculation of one- or two body density matrices, 
\beq
\label{d1}
{\mathbf D}_1 = \langle {\mathbf C} | \hat{a}_p^\dagger \hat{a}_q | {\mathbf C} \rangle
\eeq
and
\beq
\label{d2}
{\mathbf D}_2 = \langle {\mathbf C} | \hat{a}_p^\dagger \hat{a}_r^\dagger \hat{a}_s \hat{a}_q | {\mathbf C} \rangle
\eeq
is similar to calculating the Hamiltonian matrix except
instead of an integral being fetched or calculated a
product between two CI-coefficients is calculated. Unlike
the Hamiltonian a single general loop can be constructed
for the density matrices since no indices for an integral
needs to be found and only the total offsets and phase for $\hat \sigma$ and $\hat v$ 
is of interest. The loop structure for calculating the density
matrix can therefore be
\begin{algorithmic}
\LOOP[$\hat \sigma_{a \beta}$]
 \STATE{Get relative offsets for $\hat \sigma_{a \beta}$ and $\hat v_{a \beta}$}
 \STATE{Get relative phase from contraction and addition}
 \LOOP[$\hat \sigma_{a \alpha}$]
  \STATE{Get relative offsets for $\hat \sigma_{a \alpha} $ and $\hat v_{a \alpha} $}
  \STATE{Get relative phase from contraction and addition}
  \LOOP[$\hat \sigma_{c \beta}$]
   \STATE{Get relative offsets for $\hat \sigma_{c \beta} $ and $\hat v_{c \beta} $}
   \STATE{Get relative phase from contraction and addition}
   \LOOP[$\hat \sigma_{c \alpha}$]
    \STATE{Get relative offsets for $\hat \sigma_{c \alpha} $ and $\hat v_{c \alpha} $}
    \STATE{Get relative phase from contraction and addition}
    \STATE{Calculate total offset from relative offsets for $\hat \sigma$ and $\hat v$}
    \STATE{Calculate total phase from relative phases and the overall phase}
    \STATE{Multiply CI-coefficients from offsets in vector}
   \ENDLOOP[$\hat \sigma_{c \alpha}$]
  \ENDLOOP[$\hat \sigma_{c \beta}$]
 \ENDLOOP[$\hat \sigma_{a \alpha}$]
\ENDLOOP[$\hat \sigma_{a \beta}$]
\end{algorithmic}
where the $PHN\Delta M$-classes of the density matrix operators
can be constructed from the similar $PHN\Delta M$-classes of the
Hamiltonian. Higher order density matrices can be constructed
in a similar way since the division of the Hamiltonian in Eq. \ref{exdx}
is not limited to a two-body operator but can be used for any
N-body operators.

\section{Application}
\label{SEC:appl}
In this section we will present application of the ISCI on Be
which demonstrate how local orbitals and rigorous IS
leads to reduced scaling and finally linear scaling
in the $\sigma$-vector step with respect
to the simulation box size for CISD and all
the way to FCI. It is here important 
to keep in mind that what we here call the
$\sigma$-vector step is the multiplication of
integrals and coefficients and this does not include
the integral loop. 
While the ISCI method was
developed for atoms and small molecules in
very strong laser fields, the conclusion,
with regards to reduced scaling when
using local orbitals, will also carry
over to very large systems in local orbitals
since all conclusions are only based on having local orbitals
for spatially extended system. The exact scaling
factor may, however, differ but the trend will be the same. We will
here demonstrate that reduced scaling can be 
obtained even with very strict convergence
thresholds for systems where the system is not yet
large enough to obtain linear scaling
by throwing away distant interactions. This ability
to reduce the scaling for smaller systems will
in the end give a smaller prefactor once the system
is large enough that distant interactions
can be safely eliminated and linear scaling will set in.
The scaling in the $\sigma$-vector step in ISCI therefore 
gradually decreases with system size.

\subsection{Basis and calculation set up}
\label{basisandsetup}

Throughout we use a
spherical Finite-Element Discrete-Variable-Representation 
(FE-DVR) basis \cite{rescigno2000} which
is the only basis we currently have available. 
The calculations will follow
the strategy laid out in Refs. \cite{Bauch,Hochstuhl}
where we set up the FE-DVR basis in an inner region in which a
Hartree-Fock (HF) calculation is performed to
provide a set of starting orbitals for a
CI expansion. Secondly we extend the FE-DVR
basis with an outer region. Lastly we perform
the correlation calculation in the complete
space spanned by the inner and outer region.
We will in this way have a local set of
orbitals since the HF orbitals will be confined to the
inner region and the orbitals in the outer
region will be confined to their respective
finite element. Since the outer region is
added after the HF calculation the Brillouin theorem
will not be fulfilled as illustrated in Figure \ref{cisdt}.

We will show that if
the size of the inner region is fixed then the correlation
calculation will approach linear scaling with respect to the
size of the outer space where the scaling in the
$\sigma$-vector step gradually will diminish as the 
system size increases.
From an electronic structure viewpoint the choice of basis
and way of extending may seem odd since only the
long range part of the wavefunction is improved
when the size of the box is increased.
The aim of this study is, however, not high precision nor to study
the long range of the wavefunction but to simply to demonstrate
the IS for a simple spatially extended system
using local orbitals. While the vast majority of the
orbitals will have no real influence on the
electronic structure they will be important once
an external field is applied.

In a spherical FE-DVR basis the representation of the two-electron integrals 
is very sparse
\beq
\label{gsparse}
g_{k_{1}l_{1}m_{1}k_{2}l_{2}m_{2}k_{3}l_{3}m_{3}k_{4}l_{4}m_{4}} =  
g_{k_{1}l_{1}m_{1}k_{2}l_{2}m_{2}k_{3}l_{3}m_{3}k_{4}l_{4}m_{4}} 
\delta_{k_{1}k_{2}} \delta_{k_{3}k_{4}} \delta_{m_{1}-m_{2},m_{3}-m_{4}}
\eeq
where $klm$ is the radial, angular-
and magnetic moment index, respectively. In our calculations
the representation of the two-electron integrals
scale as $\mathcal{O}(N_{b,inner}^4)$ in the inner region,
$\mathcal{O}(N_{b,outer}^2)$ in the outer region and 
$\mathcal{O}(N_{b,inner}^2N_{b,outer})$ for the mixed region.
Since the cost of calculating
the integrals is very different in the three cases,
this has been omitted in comparison between the ISCI and CI.
To show that the sparsity of the representation
of the two-electron integrals can be
recovered with a simple IS,
we perform both calculations without
any prior knowledge of the radial sparsity
of the two-electron integrals and calculations
where the radial sparsity is taken into
account. 

We have thoughout made no assumption of
symmetry even if this would be 
advantageous for atomic systems. The IS
will here automatically screen away
all integrals that are trivially zero due to symmetry in 
the CI-matrix if the intial orbitals show
any symmetry.

All calculations have been converged to $10^{-10}$ and
the IS threshold set to $10^{-14}$ 
unless otherwise stated. We have chosen to not converge
all calculations but just run a few iterations
to collect statistics since currently the CI coefficients
are optimized using the short-iterative Arnoldi-Lanczos (SIAL)
algorithm \cite{park1986,beck2000} by propagation in imaginary time
which will require $3000$ to $4000+$ sigma vector
steps to converge. Since the converged results
is not the central point we decided not to use
excessive computational resources to converge all 
the calculations. Therefore the comparison of
converged results will be at the CISD level of theory.

\subsection{Be}
\label{calc_be}

For Be the size of the inner region was
chosen to be 6 Bohr with 1 FE per Bohr and 10 DVR functions
per $lm$ combination in each FE. Here we included s-functions
and p-functions with $m=0$ giving a total of 106
basis functions in the inner region once the bridge
functions is included. The outer region is then stepwise
increased from 0 to 144 Bohr with the total size of the
box ranging from 6 to 150 Bohr. 
The number of FE per Bohr
and DVR functions per FE in the outer region is kept
the same as in the inner region. The  
number of orbitals and amplitudes for CISD, CISDT and FCI
for differently sized simulation boxes are summarized 
in Table \ref{norb}. By only choosing
s- and p-functions with $m=0$ the magnetic sparsity
in the representation of the two electron integrals
will always give one and hence only the radial sparsity
will matter. Since the use of radial sparsity is implemented
we will also compare against this much smaller
set of integrals to show that there still is an additional and significant numerical
sparsity that can only be exploited by a rigorous IS.
The choice of setting $m=0$ was therefore motivated by
an easier comparison to the non-trivially zero integrals
and that for spatially extended systems in other basis sets
it will be the radial sparsity that can be compared.

\begin{table}[h]
\begin{center}
\begin{tabular}{r|rrrr}
 R [Bohr] & \# Orbitals & \# CISD & \# CISDT & \# CISDTQ\\ \hline
  6 &  106 &      54.393 &   2.282.489 &   30.969.225  \\
  8 &  142 &      98.421 &   5.547.221 &  100.220.121  \\
 10 &  178 &     155.409 &  10.997.009 &  248.157.009  \\
 14 &  250 &     308.265 &  30.691.241 &  968.765.625  \\
 20 &  358 &     634.749 &  90.617.309 & 4.083.593.409 \\
 30 &  538 &   1.438.089 & 30.8844.809 & \\
 40 &  718 &   2.565.429 & 735.663.509 & \\
 50 &  898 &   4.016.769 & & \\
 60 & 1078 &   5.792.109 & & \\
 70 & 1258 &   7.891.449 & & \\
 80 & 1438 &  10.314.789 & & \\
 90 & 1618 &  13.062.129 & & \\
100 & 1798 &  16.133.469 & & \\
120 & 2158 &  23.248.149 & & \\
150 & 2698 &  36.350.169 & & 
\end{tabular}
\end{center}
\caption{\label{norb} Number of orbitals and coefficients for a given box size.}
\end{table}

With this choice of basis the IS will predominantly screen away
the long range interaction since within each FE there
will be large integrals in order to describe the local interaction.
The setup will therefore be reminiscent of how IS works in larger
molecular systems since including more FE will always
increase the number of large integrals in the local region.

\subsubsection{Energy as a function of the simulation box size}
\label{sec:energy}

\begin{figure}[h!]                                                                                                                              
\bc                                                                                                                                            
\includegraphics[angle=270,width=10.0cm]{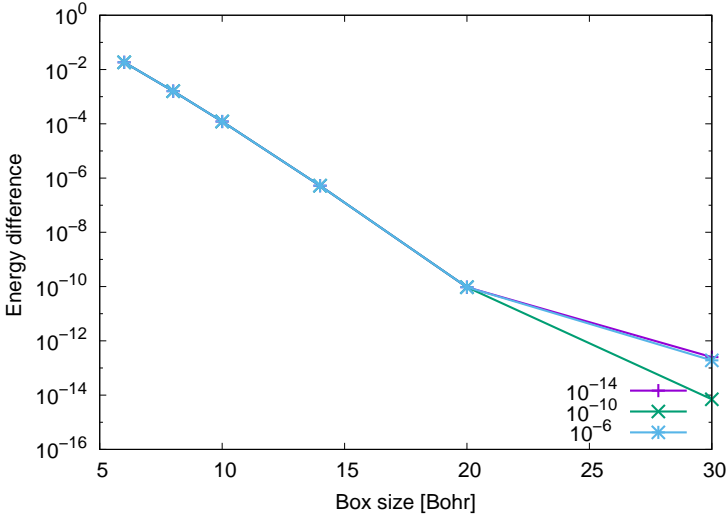}   
\ec                                                                                                                                            
\caption{\label{ediff} Energy difference in a CISD calculation from a simulation box                                                       
of 40 Bohr with different IS thresholds of $10^{-14}$, $10^{-10}$ and $10^{-6}$. 
Due to the small energy                                                                          
differences at 20 and 30 Bohr compared to 40 Bohr these have been                                                                              
converged to $10^{-14}$ for all IS unlike the rest to $10^{-10}$.                                                                              
Like the exponential decay of the wave function an                                                                                             
exponential decay of the energy contribution is seen                                                                                           
irrespectively of the IS.}
\end{figure}

By extending the size of the simulation box the tail
of the wave function can be probed in a systematical way.
In Figure \ref{ediff} 
the difference between the energy at 40 Bohr for different IS 
and the energy for
different sizes of the simulation box is plottet.
The contribution from the tail of the wave function
to the total energy 
for different IS thresholds can in this way be analyzed. 
Due to the very small energy
differences at 20 and 30 Bohr compared to 40 Bohr the
calculations have been converged to $10^{-14}$.
Like the exponential decay of the wave function an
exponential decay of the energy contribution is seen
irrespectively of the IS. 
Depending on the
desired accuracy of the calculation the correct size
of the simulation box can then be found or in order to obtain
linear scaling the range from where all interaction
can be neglected.

The energy difference between an IS of $10^{-14}$ and
$10^{-10}$ is too close to the convergence threshold of $10^{-10}$
and $10^{-14}$ in order to distinguish them. For an IS of
$10^{-6}$ there is, however, a minor energy difference
compared with an IS of $10^{-14}$ of around $10^{-8}$ which
can be seen by comparing the total energy at 
40 Bohr which are -14.594490354839278, -14.594490354839515                                                                         
and -14.594490344288744 for an IS of $10^{-14}$, $10^{-10}$ and                                                                                
$10^{-6}$, respectively. The error in the energy difference
with the IS of $10^{-6}$ increases when enlarging the box from
6 to 8 Bohrs.
By comparing the energy with
an IS of $10^{-14}$ and $10^{-10}$ it is seen that
decreasing the IS does not automatically                                                                      
give a lower energy which is understandable since the IS will eliminate both                                                                             
positive and negative contributions.

The difference
at 30 Bohr does not lie on a straight line from the other
points which is most likely due to the fact that we here
are close to the convergence threshold of $10^{-14}$ and
because the iterations in the SIAL are stopped based on the energy
change from the previous iteration which does not decrease
monotonically when the convergence threshold is very small.

\subsubsection{Integral Screening}
\label{sec:int_scr}

The number of two-electron integrals grows as $N_{orb}^4$
in Gaussian basis sets and these can therefore be
problematic to store in memory or on
disc if $N_{orb}$ is large. For
large $N_{orb}$ the integrals will therefore have to be calculated
on the fly when constructing the matrix elements 
in a traditional CI and
the repeated calculation of integrals will become
a bottleneck for very large basis sets.

\begin{figure}[h!]                                                                                                                              
\bc                                                                                                                                            
\includegraphics[angle=270,width=10.0cm]{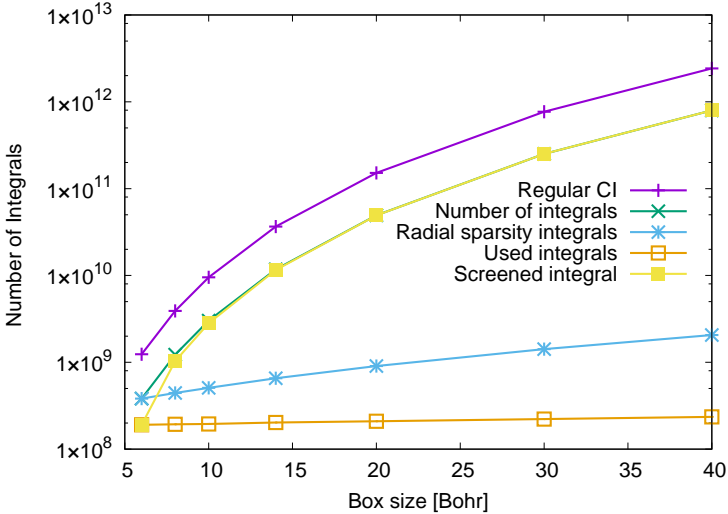}
\ec                                                                                                                                            
\caption{\label{int_scr} Number of integrals as a
function of the simulation box size. Here the number of
integrals are shown for a regular CISD calculation, an ISCISD and
an ISCISD, where the radial sparsity of the FE-DVR
basis set is used, are compared to the number of integrals used in the
ISCI method and the number of screened integrals. }
\end{figure}

The number of times an
integral is calculated on the fly 
as a function of the simulation box
for a regular CISD, an ISCISD and an ISCISD 
using the radial sparsity of the two-electron
integrals are compared in Figure \ref{int_scr}.
The regular CI will
recalculate both the trivially zero, in the FE-DVR basis,
and numerically zero integrals for every matrix element
while the ISCI will only recalculate these for every block
of elements. If the number of electrons would scale
with system size the difference between the regular CI
and the ISCI would be significantly more pronounced
due to the difference in scaling.
Using the two-electron integral sparsity the scaling in the
integral loop will be $N_{orb}^2$
and not $N_{orb}^4$ but even with the reduced scaling 
there is still a
substantial numerical sparsity that can be
exploited by rigorous IS.

\begin{figure}[h!]                                                                                                                              
\bc                                                                                                                                            
\includegraphics[angle=270,width=10.0cm]{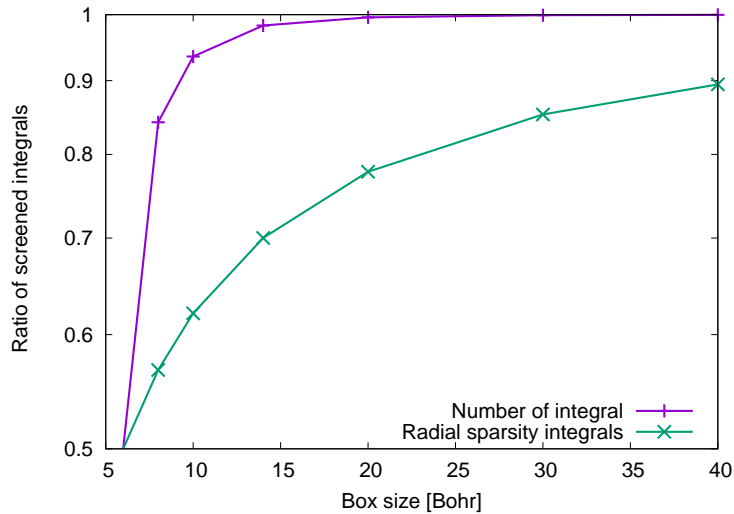}                                                                                            
\ec                                                                                                                                            
\caption{\label{int_scr2} IS efficiency as a function
of the size of the simulation box. The efficiency is here
taken as the ratio between the number of screened integrals,
and hence discharged integrals, and
the total number of integrals both for an ISCISD and for
an ISCISD using the two-electron integral sparsity of the FE-DVR basis.}
\end{figure}

From Figure \ref{int_scr2} it is seen that in the
inner region half the integrals are screened away
even though the construction of the orbitals in the SCF
procedure destroys the sparsity of the FE-DVR basis in the
inner region. Once the size of the box is increased
the ratio of the screened integrals grows
significantly and at 40 Bohrs less than $0.03$ percent of the integrals are
used in the ISCI when the IS is $10^{-14}$. Even
if the radial sparsity of the two-electron integrals is used only $11.5$
percent are used with a simulation box of 40 Bohrs.

For the vast majority of electronic structure calculations 
of atoms a simulation box of 40 Bohrs is not needed, however,
once an external laser field is applied in order to study
higher harmonic generation, photoionization and other processes
where electrons move in the continuum a simulation box
of 40 Bohrs is very small \cite{Larsson}.
For even larger simulation boxes the number of orbitals
will typically be around $10^{3}$ to $10^{6}$ and here
the benefits obtained by the IS procedure will be even more pronounced
as can be seen from Figure \ref{int_scr2}. For such large
basis sets it will no longer be possible to store all
integrals and hence they will have to be calculated on the
fly. Due to the numerous times an integral will have
to be recalculated in a regular CI this becomes
an insurmountable bottleneck and the regular
CI is therefore not even able to solve 
the time-independent problem for small systems
in the very large basis sets typically used in time-dependent 
simulations involving the continuum.

For large molecular systems we expect to see similar
trends in the IS as a function of distance 
if local orbitals are used \cite{ida_local_orb,ida_local_orb2}.
Here the screening efficiency will of course not be as rapid 
as seen in Figure \ref{int_scr2} due
to interactions within the molecule.

\subsubsection{Floating point operations}

\begin{figure}[h!]                                                                                                                              
\bc                                                                                                                                            
\includegraphics[angle=270,width=10.0cm]{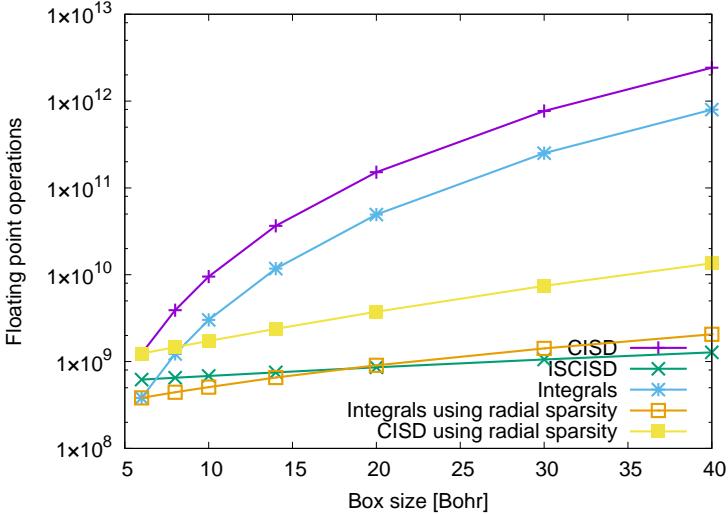}
\ec                                                                                                                                            
\caption{\label{floating} Floating point operations
as a function of the size of the simulation box for 
a CISD, a CISD using radial sparsity and an ISCISD
calculation along with
the integral loops both with and without using the
radial sparsity of the integrals.}
\end{figure}

As seen in Section \ref{sec:int_scr} the vast majority
of the integrals can be screened away and since the
ISCI only uses the integrals above the screening
threshold in the $\sigma$-vector step this will
result in a reduced scaling of the $\sigma$-vector step.
In Figure \ref{floating} the number of floating
point operations in the $\sigma$-vector step 
as a function of the size of the simulation box for 
a CISD, a CISD using the radial sparsity and an ISCISD 
calculation are compared.
It is readily seen that the scaling of the ISCISD
is lower than the CISD and also significantly
lower than a CISD using the radial sparsity. Once
the simulation box is extended to 40 Bohrs the
regular CISD and
the CISD using radial sparsity is using $10^{3}$ and $10$
times as many floating point operations as the ISCI, respectively.
It is therefore obvious that for even larger simulation
boxes just using the radial sparsity of the two-electron
integrals in a CI calculation will not suffice.

The outer loop of the ISCI method is over all integrals
and as can be seen in Figure \ref{floating} this
loop will be the dominating step for the ISCI method.
If the radial sparsity of the two-electron integrals is not used then the
integral loop will only reduce the
savings compared to the regular CI by around
two thirds in this case. However, if the radial sparsity
of the two-electron integrals is used, the integral
loop will have a scaling of $N_{orb}^2$ and a significant
reduction is observed. Despite this significant
reduction in the scaling of the integral loop
it is still more expensive than 
the $\sigma$-vector step in the ISCI approach when the
size of the simulation box is increased.

\begin{figure}[h!]                                                                                                                              
\bc                                                                                                                                            
\includegraphics[angle=270,width=10.0cm]{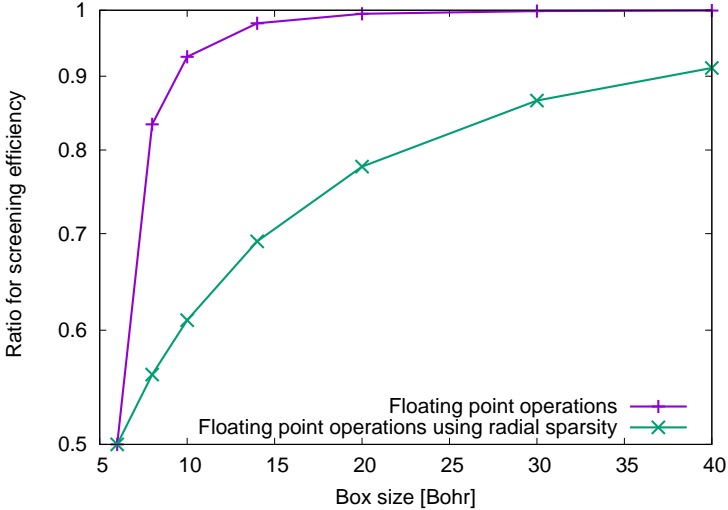}
\ec                                                                                                                                            
\caption{\label{floating_scr} Floating point operations screening efficiency 
in the $\sigma$-vector step
as a function of the size of the simulation box. The efficiency is here
taken as the ratio between the floating point operations screened away in the ISCISD
and the total floating point operations for the CISD and for 
a CISD using the two-electron integral sparsity of the FE-DVR basis.}
\end{figure}

The IS in the ISCI is directly
translated into a screening of the floating
point operation in the $\sigma$-vector step 
as can be seen by comparing Figures \ref{int_scr2} 
and \ref{floating_scr} since
the floating point operations screening follows
the IS closely. Here only $0.05$ percent of the
floating points operations in the $\sigma$-vector step are carried out
in the ISCISD compared to the CISD
with a simulation box of 40 Bohrs and an IS threshold of $10^{-14}$.
When the radial sparsity of the two-electron integrals is used,
only $9.8$ percent of the $\sigma$-vector step
have to be performed in the ISCI.

\begin{figure}[h!]                                                                                                                              
\bc                                                                                                                                            
\includegraphics[angle=270,width=10.0cm]{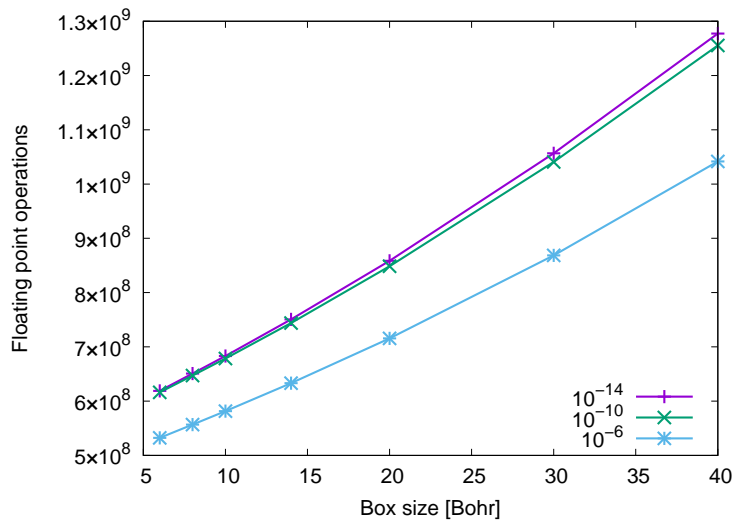}
\ec                                                                                                                                            
\caption{\label{effort_scr_thr} Comparison of the floating
point operations for different IS thresholds
as a function of the simulation box size.}
\end{figure}

The screening threshold has an impact on both the accuracy
and efficiency of the floating point screening. In Figure \ref{effort_scr_thr}
the floating point operations with different IS
threshold are compared. Not surprisingly,
the number of floating point operations in the $\sigma$-vector step
is reduced when the IS threshold is increased.
There is, however, no abrupt stop for integrals above a certain
size depending on distance since there will always be integrals
with large values within each FE and it is therefore only
the long range interaction that is screened
away when the size of the simulation box is increased.

These large integrals that reside locally in the FE-DVR basis set
are essential in order to describe electron-electron interaction
away from the atom or molecule once an external field is applied to the
system. For electronic structure calculations the large
integrals far away from the nuclei are only very weakly
coupled to the system and mainly help to describe the
tail of the wave function as seen from Figure \ref{ediff}. 
These integrals are also the reason why the tail of the wave function 
for the highest IS threshold of $10^{-6}$ also gives small energy contributions.
Without any prior knowledge of the distance within the system
these integrals will have to be included in the ISCI. If, however,
the distance between orbitals in different GAS are known
this can be used to perform additional IS
which depends on distance.
An additional distance dependent IS
will give linear scaling in both the integral loop and
the $\sigma$-vector step.

\begin{figure}[h!]                                                                                                                              
\bc                                                                                                                                            
\includegraphics[angle=270,width=10.0cm]{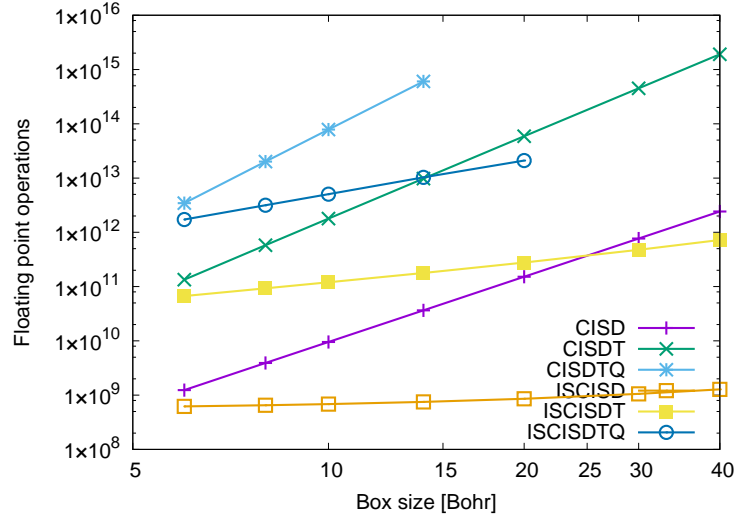}
\ec                                                                                                                                            
\caption{\label{effort_scr_all} Comparison of the floating
point operations in the $\sigma$-vector step between the CI and ISCI
for different orders in the hierarchy.}
\end{figure}

The reduced scaling in the $\sigma$-vector step in the ISCI is not 
restricted to CISD but goes for all order in the CI hierarchy
as seen in Figure \ref{effort_scr_all}. When increasing the 
box size the curve for the floating point operations of the higher order ISCI 
will cross with a lower order curve of the regular CI due 
to the reduced scaling of all orders in the ISCI.
At 14 Bohrs the floating point operations for the ISCISDTQ
is similar to the regular CISDT despite the ISCISDTQ
at this point has 968.765.625 coefficients and the CISDT
30.691.241 coefficients as seen from Table \ref{norb}. Furthermore 
at 40 Bohrs the floating point operations for the
regular CISD is more than three times that of the
ISCISDT despite the ISCISDT having 287 times more 
coefficients than the regular CISD. 

\subsubsection{Reduced scaling}

The ISCI method gives reduced scaling, for spatially extended systems
in local orbitals, for all orders of the CI hierarchy as can be
seen in Figure \ref{effort_scr_all}.
To obtain the scaling of the different orders
of the ISCI method
a simple fit of the floating point operations is performed.
The fit is from the inner 
region and until 20 or 40 Bohrs to show that there is a reduction
in the scaling. These fits are performed with an
IS threshold of $10^{-14}$ and the energy is therefore
not changed by this. Afterwards we analyse the gradual
reduction of the scaling in the  
$\sigma$-vector step and show that this will become linear for
sufficiently large systems. Finally a comparison of different
IS thresholds is performed in order to show the interplay
between the IS and the scaling.

\begin{figure}[h!]                                                                                                                              
\bc                                                                                                                                            
\includegraphics[angle=270,width=10.0cm]{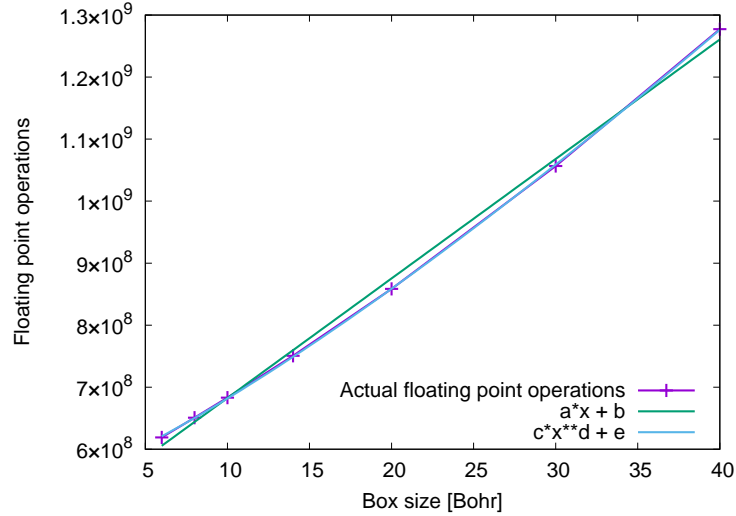}
\ec                                                                                                                                            
\caption{\label{effort_scr} 
The exponent for the scaling and a linear curve
have been fitted to the floating point operations as a function
of box size for the ISCISD. The coefficients from the linear 
fit are $a=1.93 \cdot 10^7$ and $b=4.90 \cdot 10^8$ and for the exponent fit
$c=7.48 \cdot 10^6$, $d=1.24$ and $e=5.52 \cdot 10^8$.}
\end{figure}

In Figure \ref{effort_scr} the exponent for the scaling
along with a linear curve
have been fitted to the floating point operations as
a function of box size for the ISCISD. We see that
the ISCISD, in this case, scales as $1.24 \pm 0.01$,
very close to linear scaling, when the curve
is fitted from 6 to 40 Bohrs.
With this low scaling for the ISCISD in the $\sigma$-vector step it is therefore
also evident why the integral loop, which has
a scaling of 2 when using the two-electron integral 
sparsity above a certain box size, becomes
the dominant step.

\begin{figure}[h!]                                                                                                                              
\bc                                                                                                                                            
\includegraphics[angle=270,width=10.0cm]{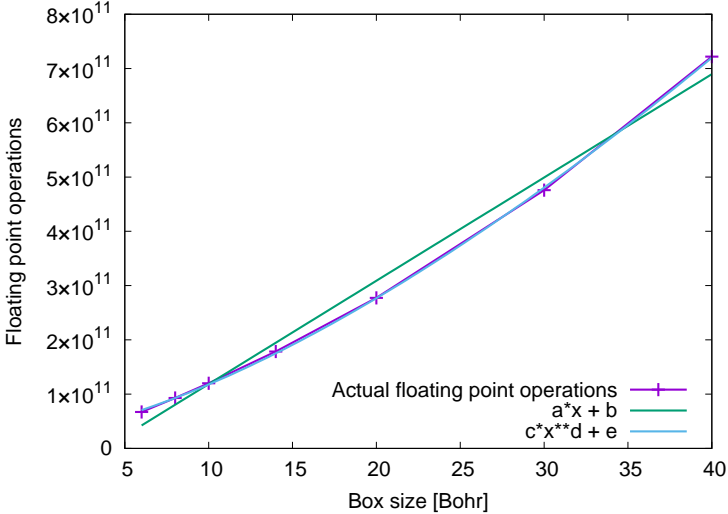}
\ec                                                                                                                                            
\caption{\label{effort_scr_t} 
The exponent for the scaling and a linear curve
have been fitted to the floating point operations as a function
of box size for the ISCISDT. The coefficients from the linear 
fit are $a=1.90 \cdot 10^{10}$ and $b=-7.18 \cdot 10^{10}$ and for the exponent fit
$c=2.97 \cdot 10^9$, $d=1.48$ and $e=2.87 \cdot 10^{10}$.}
\end{figure}

\begin{figure}[h!]                                                                                                                              
\bc                                                                                                                                            
\includegraphics[angle=270,width=10.0cm]{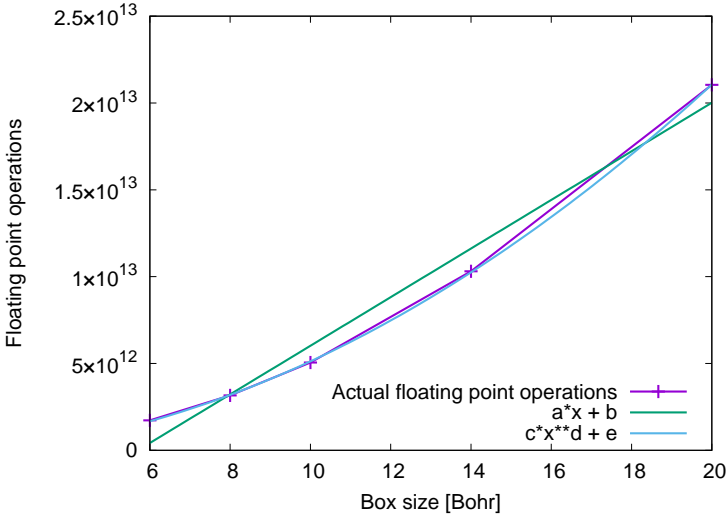}
\ec                                                                                                                                            
\caption{\label{effort_scr_q} 
The exponent for the scaling and a linear curve
have been fitted to the floating point operations as a function
of box size for the ISCISDTQ. The coefficients from the linear 
fit are $a=1.40 \cdot 10^{12}$ and $b=-7.98 \cdot 10^{12}$ and for the exponent fit
$c=5.68 \cdot 10^{10}$, $d=1.98$ and $e=-2.93 \cdot 10^{11}$.}
\end{figure}

The low scaling of the ISCI continues to all orders of the
CI hierarchy. The scaling going from the ISCISD to the 
ISCISDT, with the same fitting region, is only slightly increased 
to $1.48 \pm 0.03$ as can be seen in Figure \ref{effort_scr_t}.
Even though the scaling of the ISCISDT is lower than
the scaling of the integral loop, even when using the
two-electron integral sparsity, the prefactor will
be dominant in the inner region but these curves will at some
point cross when the size of the simulation box is increased
sufficiently and then loop over the integrals will be
the dominant one.
For the ISCISDTQ the scaling is approximately quadratic,
for a fit from 6 to 20 Bohrs,
since the exponent is $1.98 \pm 0.03$, see Figure \ref{effort_scr_q},
which is significantly lower than the usual scaling
of CISDTQ which is $O^4 V^6$. 

The increase of the exponent with the excitation level
in the ISCI is expectable since the scaling in 
CI grows as $O^N V^{N+2}$ and the decrease
in the scaling in the ISCI is gradually reduced when
the system size is increased. The reduction in scaling 
is accomplished without any prior
knowledge about the structure of the CI-matrix, i.e.,
without using knowledge about the distance between orbitals, which
can be used to screen large integrals which are centered
far from the system. Such a distance dependent
IS is often used to obtain linear
scaling \cite{pulay_lin_sca2,werner_lin_sca,poul_lin_sca,werner_lin_sca2,bartlett_lin,neese_lin,poul_lin_sca2}.

Usually when the system size is increased an
increase in the floating point operations per FE
in the $\sigma$-vector step, which depend on the
scaling of the method, is expected.
Due to the gradual reduction in scaling, until linear,
the ISCI will instead approach a constant value.

\begin{figure}[h!]                                                                                                                              
\bc                                                                                                                                            
\includegraphics[angle=270,width=10.0cm]{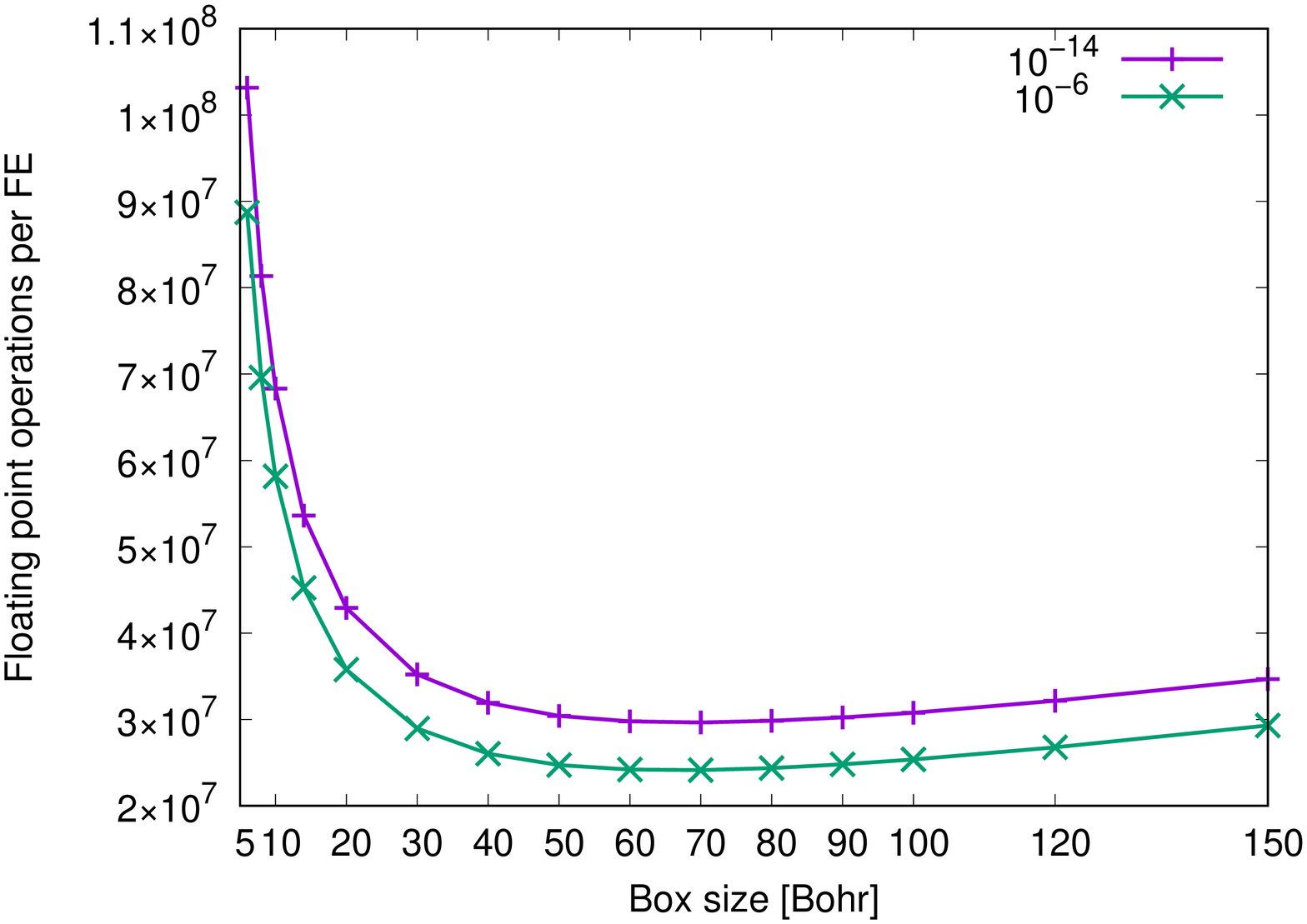}
\ec                                                                                                                                            
\caption{\label{gradual_sca} 
Floating point operations per FE for different box sizes with
an IS cut-off of $10^{-14}$ and $10^{-6}$.}
\end{figure}

In Figure \ref{gradual_sca}, a
gradual reduction in the floating point operations 
per FE with system size can be seen. The gradual
reduction is caused by the orbital rotation in
the inner region, which destroys the sparsity of
the FE-DVR basis, which makes the FE's in the
inner region significantly more expensive
and the approach to a constant value is therefore
not from below as normally seen.
The floating point operations per FE therefore
hits a minimum at around 70 Bohrs and then
slightly increases again. With the lower IS threshold
of $10^{-6}$ a reduction of 15-20 percent in the
floating point operations in the $\sigma$-vector step
is seen in this case.

\subsubsection{Distribution of coefficients}

\begin{figure}[h!]                                                                                                                              
\bc                                                                                                                                            
\includegraphics[angle=270,width=10.0cm]{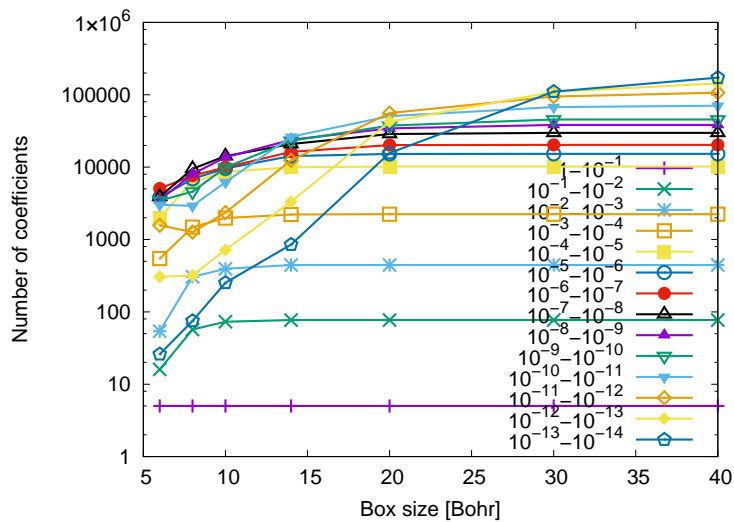}
\ec                                                                                                                                            
\caption{\label{coe_log} Number of coefficients in a given
interval as a function of the size of the simulation box.}
\end{figure}

The main problem in the ISCI is the number of coefficients.
These are not screened since no prior knowledge
about the CI-matrix is assumed. In Figure \ref{coe_log}
the number of CI coefficients within a given range is
plottet as a function of the size of the simulation box.
The number of very small coefficients in Figure \ref{coe_log}
is very dependent on the convergence threshold in the
CI calculation. The convergence threshold for the
energy in these calculations is $10^{-10}$.
The plot in Figure \ref{coe_log} clearly shows that the larger the coefficients
the less change in the number of these with distance.
From around 14 Bohrs there is barely a change in the
number of coefficients larger than $10^{-6}$ and at this
distance the energy is converged to $5.2 \times 10^{-7}$ which
would be sufficient for most electronic structure calculations.
For dynamics simulations the convergence in general 
have to be significantly stricter but for distances
above 30-40 Bohrs it does appear possible to introduce
sensible numerical approximations which do not effect
the dynamics.

Here, approximation in the CI-vector similar to those
done for the Hamiltonian in order to exploit the
integral sparsity,
can also be introduced which will significantly reduce
the number of CI coefficients for large molecules and simulation
boxes without compromising the accuracy of the
calculations. By considering the sparsity of the CI-vector
and a screening, the ISCI will be better able
to handle higher excitations and large molecules 
in good basis sets simultaneously.

\begin{figure}[h!]                                                                                                                              
\bc                                                                                                                                            
\includegraphics[angle=270,width=10.0cm]{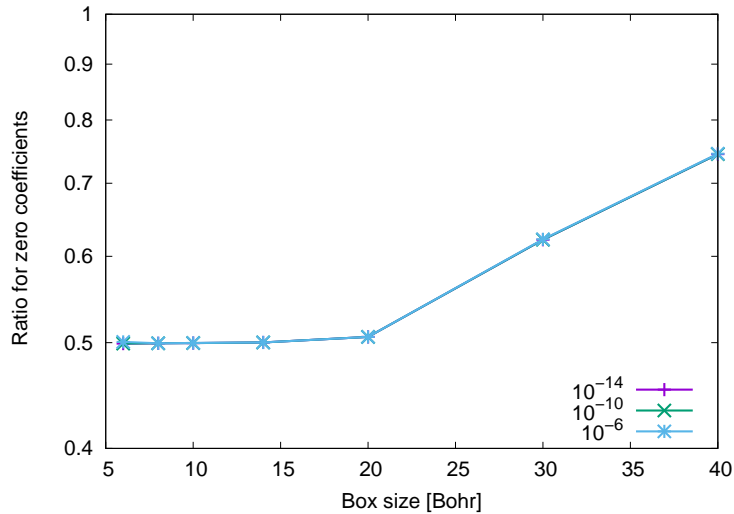}
\ec                                                                                                                                            
\caption{\label{coe_scr} Ratio between the number of
CI coefficients below $10^{-14}$ and the total number of 
CI coefficients for different IS thresholds of $10^{-6}$, $10^{-10}$ and $10^{-14}$.}
\end{figure}

The long tail of small coefficients is almost completely
independent of the IS threshold as can be seen from Figure \ref{coe_scr}
where the ratio between the number of 
CI coefficients below $10^{-14}$ and the total number of
CI coefficients for different IS thresholds have been plotted. 
Since the number of small coefficients is practically the
same for all IS thresholds we conclude that the long range contribution to the tail of
the wave function comes from large
integrals between local orbitals some distance 
away from the origin of the system. 

The IS in the ISCI method can therefore
still be numerically exact by letting the IS threshold be
distance dependent. This would still give the tail
for the wave function as seen in Figure \ref{ediff} and
at the same time allow for a reduction in the scaling to linear
for any order in the ISCI and reduce the prefactor. 
The distance dependent IS 
can be combined with a screening of the CI coefficients
this would make it possible to treat large systems
using higher excitations.

\section{Summary and prospects}
\label{SEC:summ}

We have presented a novel derivation of the 
well known CI equations which allows for
an \map integral screening (IS) of the integrals before
the $\sigma$-vector step. 
The new method is dubbed the integral screened
configuration interaction (ISCI) method.
The IS appears naturally when the Hamiltonian is
written in terms of 
normal- and spin-ordered and index restricted operators which is
unlike the UGA Hamiltonian where the operators are
written in terms of the
generators of the unitary group.

By rewriting the Hamiltonian it is shown that the
strings of second quantized operators
in the $\sigma$-vector step are separable up to 
a sign. This separability allows for a 
construction of the $\sigma$-vector step
where the outer loops are over the indices
of the integrals in the Hamiltonian. Having
the outer loops over the integrals allows
for a very simple, efficient and rigorous \map IS
where only integrals above a predefined threshold
are used in the $\sigma$-vector step.
Such a procedure will automatically give
a reduction in the scaling of the ISCI method for
spatially extended systems in local basis sets.
We here show that the scaling of the
$\sigma$-vector step in the ISCI method gradually
is reduced for all orders of the CI hierarchy
when the system size grows until eventually a linear scaling
is achieved.

Since the usual machinery from CI, such as the
lexicographical scheme, does not appear
to be directly applicable to the ISCI method we
give a detailed description of the current
way of solving these problems in the ISCI approach.
Here we also show some of the 42 unique ways
of contracting a non-relativistic Hamiltoian
with the CI-vector and how each of these can
be separately optimized. Furthermore
we also show how the same code can be used
to calculate density matrices.

A numerical example on Be is shown where the
size of the simulation box is extended from 6 to
150 Bohrs and the scaling of the ISCI method is examined.
We furthermore show that from a low IS threshold
of $10^{-14}$ we can obtain a scaling for the ISCISD
of $1.24 \pm 0.01$, for the ISCISDT of $1.48 \pm 0.03$
and finally for the ISCISDTQ of $1.98 \pm 0.03$.
We note that the scaling depends on the IS threshold
so by increasing the IS threshold the scaling will be even
further decreased. We also show that for
sufficiently large simulation boxes the loop
over the integrals, even when using the 
two-electron integral sparsity of the FE-DVR basis set,
will be the dominant step for all orders
of the ISCI method.

In the application part we discuss the effect
of the IS for both electronic structure theory
and for dynamics simulations in strong laser
fields since the requirements here sometimes are
different. We also show how the 
two-electron integral sparsity of the FE-DVR basis set
can be exploited using the generalized-active-space (GAS)
approach. Furthermore we lay out the path to
linear scaling for all orders of the ISCI and discuss how
to obtain distance dependent integral and coefficient
screening in an efficient way that does not compromise
the accuracy in neither electronic structure
calculation nor in time-dependent simulations.

Currently the main aim of the method is accurate simulations of
physical processes where one or more electrons move in the continuum for 
atoms and molecules in strong laser fields.
Including
both electron correlation and simulating electron processes 
involving the continuum is a major theoretical and numerical
challenge and usually involves severe approximations in
either the particle interaction, the
description of the continuum, the initial and subsequent correlation level
or combinations of these.
With the ISCI method we aim for being able to
treat up to two electrons in the continuum accurately.
We also believe that the present type of IS would not only be of interest
for time-dependent simulations of dynamics of electrons
in the continuum but also suitable for
electron structure theory since the electron-electron interaction
at large distances is dominated by the local field.
In electronic structure theory the rigorous IS will 
give a very good error control and in combination with the 
GAS will allow for low or linear scaling ISMRCI calculations
which could serve as an alternative to the
embedding schemes. The main problem with extending the method to very
large systems is the lack of
size-extensivity in truncated CI.
In the near future we
plan to interface the GASCI part with a quantum
chemistry code which can provide local orbitals
for large molecules to show the scaling properties of
the ISCI for these kind of systems. Here we also 
plan to introduce the distance dependent integral
and coefficient screening.

\section*{Acknowledgement(s)}

This work was supported by the ERC-Stg (Project No. 277767-TDMET),
the VKR Center of Excellence, QUSCOPE. The calculations
were performed at the Centre for Scientific Computing Aarhus (CSCAA).


\begin{appendix}
 \section{The CAAB representation}
\label{app_caab}

The Creator Annihilator Alpha Beta (CAAB) representation was 
first introduced in the LUCIA code \cite{olsen_lucia,olsen_cc} and
has since been used in several other codes \cite{Soerensen_spinfree,soerensen_mrcc_zpc,krcc,soerensen_commcc}.
The CAAB operator in our implementation is similar 
to the one in LUCIA with the exception
that we can vary the number of spin orbitals within the same GAS completely
independent of each other.
The CAAB operator is an abstract representation of a second quantized operator
which can easily be manipulated on a computer and presents an extremely
convenient way to setup the generalized active space (GAS). 
The representation of a CAAB operator with a GAS can be set in a matrix
form with four rows, the creator $\alpha$, creator $\beta$, annihilator $\alpha$ and annihilator $\beta$,
and NGAS column, where NGAS is the number of GAS.
The entries in the matrix in the CAAB representation 
is the number second quantized operators in a given GAS. 

As an example of the CAAB representation the $\hat X_{2,0,0}$ operator in 
two GAS is chosen. Here the first GAS will be the occupied space
while the second will be the virtual space since we in this way will
have a well-defined Fermi vacuum. The $\hat X_{2,0,0}$ operator in 
the CAAB representation, where the operator is called a type, is
\beqa
\label{2gas}
CA\qquad 0 \quad 1 \nonumber \\
CB\qquad 0 \quad 1 \nonumber \\
AA\qquad 1 \quad 0 \nonumber \\
AB\qquad 1 \quad 0
\eeqa
where $CA$ stands for creation $\alpha$, $CB$ for creation $\beta$,
$AA$ for annihilation $\alpha$ and $AB$ for annihilation $\beta$.
In Eq. \ref{2gas} it is seen that an electron with $\alpha$ and $\beta$ spin
is excited from the occupied space to the virtual space by
annihilation of an electron with $\alpha$ and $\beta$ spin in the occupied space
and creation of an electron with $\alpha$ and $\beta$ spin in the virtual space. Splitting
the virtual space into two, the $\hat X_{2,0,0}$ operator
from Eq. \ref{2gas} is split into four types
\beqa
\label{3gas1}
&CA&\qquad 0 \quad 1 \quad 0 \nonumber \\
&CB&\qquad 0 \quad 1 \quad 0 \nonumber \\
&AA&\qquad 1 \quad 0 \quad 0 \nonumber \\
&AB&\qquad 1 \quad 0 \quad 0 
\eeqa
\beqa
\label{3gas2}
&CA&\qquad 0 \quad 0 \quad 1 \nonumber \\
&CB&\qquad 0 \quad 1 \quad 0 \nonumber \\
&AA&\qquad 1 \quad 0 \quad 0 \nonumber \\
&AB&\qquad 1 \quad 0 \quad 0 
\eeqa
\beqa
\label{3gas3}
&CA&\qquad 0 \quad 1 \quad 0 \nonumber \\
&CB&\qquad 0 \quad 0 \quad 1 \nonumber \\
&AA&\qquad 1 \quad 0 \quad 0 \nonumber \\
&AB&\qquad 1 \quad 0 \quad 0 
\eeqa
\beqa
\label{3gas4}
&CA&\qquad 0 \quad 0 \quad 1 \nonumber \\
&CB&\qquad 0 \quad 0 \quad 1 \nonumber \\
&AA&\qquad 1 \quad 0 \quad 0 \nonumber \\
&AB&\qquad 1 \quad 0 \quad 0 
\eeqa
The splitting of the types in the CAAB representation is also 
visible in the $| \hat X_{2,0,0} \rangle \langle \hat X_{2,0,0} | \hat H | \hat X_{2,0,0} \rangle \langle \hat X_{2,0,0} |$
block when going from Figure \ref{cisdt} to Figure \ref{cisdt_gas}.
In Figure \ref{cisdt} the matrix element $| \hat X_{2,0,0} \rangle \langle \hat X_{2,0,0} | \hat H | \hat X_{2,0,0} \rangle \langle \hat X_{2,0,0} |$
is a single block represented by the CAAB type in Eq. \ref{2gas} while
in Figure \ref{cisdt_gas} $| \hat X_{2,0,0} \rangle \langle \hat X_{2,0,0} | \hat H | \hat X_{2,0,0} \rangle \langle \hat X_{2,0,0} |$
is split into four times four smaller blocks from the splitting
of the $\hat X_{2,0,0}$ CAAB type into the four $\hat X_{2,0,0}$ CAAB types
in Eqs. \ref{3gas1}-\ref{3gas4}.
Increasing the number of GAS will 
increase the number of blocks while at the same time reduce the size of the
individual blocks which will help to give a good load balancing for 
a parallelization of the code since every block in the CI matrix is addressed individually.
It is important to stress that using the CAAB representation of $\hat X_{2,0,0}$
in Eq. \ref{2gas} or those in Eqs. \ref{3gas1}-\ref{3gas4} will give
identical results and the only difference is how the
algorithm performs the calculation.

For approximations with only a single electron in the continuum \cite{Bauch}
the three GASs can be divided into occupied orbitals, unoccpied bound
orbitals $(E<0)$ and continuum orbital $(E>0)$. Since only one electron should be in the
continuum the $\hat X_{2,0,0}$ CAAB representation will then just
consists of the types in Eqs. \ref{3gas1}-\ref{3gas3} since the
type in Eq. \ref{3gas4} would place two electrons in the continuum.
Since the continuum typically contains the vast majority
of the orbitals removing the operator in Eq. \ref{3gas4} from
the calculation will often reduce the computational effort
by several magnitudes.

Connecting the CAAB representation of the Hamiltonian to the
excitation and de-excitation operators in Eq. \ref{exdx} can
be done by a two-GAS setup where the occupied orbitals
are in the first space and the unoccupied in the second.
In this way all 8 terms from Eq. \ref{exdx} can be
distributed in the entries of the types
\beqa
\label{type}
&CA&\qquad \hat C_{\alpha}^{dx} \quad \hat C_{\alpha}^{ex} \nonumber \\
&CB&\qquad \hat C_{\beta}^{dx} \quad \hat C_{\beta}^{ex} \nonumber \\
&AA&\qquad \hat A_{\alpha}^{ex} \quad \hat A_{\alpha}^{dx} \nonumber \\
&AB&\qquad \hat A_{\beta}^{ex} \quad \hat A_{\beta}^{dx} 
\eeqa
whereby any term in the Hamiltonian can be written. From the generic
type in Eq. \ref{type} any excitation or de-excitation term can
easily be found, addressed and manipulated. The generalization of
Eq. \ref{type} to many GASs is simple as long as the occupied and virtual
orbitals are in separate GASs and follow directly from the example
of $\hat X_{2,0,0}$ above. Combining the index-restricted Hamiltonian
in Eq. \ref{index_re} with the generic Hamiltonian type in Eq. \ref{type} any
term in the Hamiltonian can be written as shown in Eq. \ref{exdx}.

The manipulation of the types
is the first step performed in any calculation where 
the non-zero blocks in the CI matrix are identified. Here we have two
operations where the first is contraction and the second addition.
In the $\sigma$-vector step in Eq. \ref{sigma} all de-excitation terms in the Hamiltonian
are contracted with the approximate eigenvector ${\mathbf v}$
while all the excitation terms in the Hamiltonian will add to ${\mathbf v}$.
The result of the contraction and addition will give
the linearly transformed approximate eigenvector ${\mathbf \sigma}$.
In the CAAB representation of operators the application of 
\beq
\hat H_{2,0,-2}^{2,1} \hat v_{3,0,-1} = \hat \sigma_{2,0,0}
\eeq
in the $\sigma$-vector step is
\beqa
\label{h20m2}
&CA&\qquad 0 \quad 0 \nonumber \\
\hat H_{2,0,-2}^{2,1} = \quad &CB&\qquad 1 \quad 1 \nonumber \\
&AA&\qquad 0 \quad 0 \nonumber \\
&AB&\qquad 0 \quad 2 
\eeqa
\beqa
\label{v30m1}
&CA&\qquad 0 \quad 1 \nonumber \\
\hat v_{3,0,-1} = \quad &CB&\qquad 0 \quad 2 \nonumber \\
&AA&\qquad 1 \quad 0 \nonumber \\
&AB&\qquad 2 \quad 0 
\eeqa
\beqa
\label{s200}
&CA&\qquad 0 \quad 1 \nonumber \\
\hat \sigma_{2,0,0} = \quad &CB&\qquad 0 \quad 1 \nonumber \\
&AA&\qquad 1 \quad 0 \nonumber \\
&AB&\qquad 1 \quad 0 
\eeqa
where $\hat \sigma_{2,0,0}$ results from the application of
$\hat H_{2,0,-2}^{2,1}$ on $\hat v_{3,0,-1}$. The contraction
of the $\hat A_{\beta}^{dx}$ term in $\hat H_{2,0,-2}^{2,1}$
with $\hat v_{3,0,-1}$ happens by subtracting the two in the
$AB$ line in Eq. \ref{h20m2} with the two in the $CB$ line in 
Eq. \ref{v30m1} which are the two particle contractions.
The hole contraction, which is the $\hat C_{\beta}^{dx}$ part
of $\hat H_{2,0,-2}^{2,1}$, happens by subtraction of the 
first one in the $CB$ line in Eq. \ref{h20m2} with one of the
two in the $AB$ line in Eq. \ref{v30m1}. The resulting 
CAAB operator for the Hamiltonian after the contraction is
\beqa
\label{h20m2c}
&CA&\qquad 0 \quad 0 \nonumber \\
&CB&\qquad 0 \quad 1 \nonumber \\
&AA&\qquad 0 \quad 0 \nonumber \\
&AB&\qquad 0 \quad 0 
\eeqa
and for $\hat v_{3,0,-1}$ is
\beqa
\label{v30m1c}
&CA&\qquad 0 \quad 1 \nonumber \\
&CB&\qquad 0 \quad 0 \nonumber \\
&AA&\qquad 1 \quad 0 \nonumber \\
&AB&\qquad 1 \quad 0 
\eeqa
where both now only contain excitation terms which are added
to give $\hat \sigma_{2,0,0}$ in Eq. \ref{s200}. In this way all possible non-zero
blocks in the CI matrix can quickly be found and tabulated.
The algorithm can in this way be guided by which
part of the $\sigma$-vector step should be calculated.

From these examples it is seen how a second quantized operator
can be represented in the computer and how approximations easily 
can be introduced just by dividing the GAS appropriately and
then eliminating the unwanted types. Furthermore the direct connection
between the excitation and de-excitation operators in Eq. \ref{exdx}
and the types for the CAAB representation is shown. It is the types
in the CAAB representation that is setup and manipulated in the first
part of the algorithm in Section \ref{first_part} and the types
for a contraction is then passed down to the $\sigma$-vector step, shown
in Section \ref{second_part}, where this is translated into 
one of the 42 loop structures and calculated.

\end{appendix}

\bibliography{full}

\newcommand{\Aa}[0]{Aa}
\begin{thebibliography}{75}%
\makeatletter
\providecommand \@ifxundefined [1]{%
 \@ifx{#1\undefined}
}%
\providecommand \@ifnum [1]{%
 \ifnum #1\expandafter \@firstoftwo
 \else \expandafter \@secondoftwo
 \fi
}%
\providecommand \@ifx [1]{%
 \ifx #1\expandafter \@firstoftwo
 \else \expandafter \@secondoftwo
 \fi
}%
\providecommand \natexlab [1]{#1}%
\providecommand \enquote  [1]{``#1''}%
\providecommand \bibnamefont  [1]{#1}%
\providecommand \bibfnamefont [1]{#1}%
\providecommand \citenamefont [1]{#1}%
\providecommand \href@noop [0]{\@secondoftwo}%
\providecommand \href [0]{\begingroup \@sanitize@url \@href}%
\providecommand \@href[1]{\@@startlink{#1}\@@href}%
\providecommand \@@href[1]{\endgroup#1\@@endlink}%
\providecommand \@sanitize@url [0]{\catcode `\\12\catcode `\$12\catcode
  `\&12\catcode `\#12\catcode `\^12\catcode `\_12\catcode `\%12\relax}%
\providecommand \@@startlink[1]{}%
\providecommand \@@endlink[0]{}%
\providecommand \url  [0]{\begingroup\@sanitize@url \@url }%
\providecommand \@url [1]{\endgroup\@href {#1}{\urlprefix }}%
\providecommand \urlprefix  [0]{URL }%
\providecommand \Eprint [0]{\href }%
\providecommand \doibase [0]{http://dx.doi.org/}%
\providecommand \selectlanguage [0]{\@gobble}%
\providecommand \bibinfo  [0]{\@secondoftwo}%
\providecommand \bibfield  [0]{\@secondoftwo}%
\providecommand \translation [1]{[#1]}%
\providecommand \BibitemOpen [0]{}%
\providecommand \bibitemStop [0]{}%
\providecommand \bibitemNoStop [0]{.\EOS\space}%
\providecommand \EOS [0]{\spacefactor3000\relax}%
\providecommand \BibitemShut  [1]{\csname bibitem#1\endcsname}%
\let\auto@bib@innerbib\@empty
\bibitem [{\citenamefont {Kellner}(1927{\natexlab{a}})}]{kellner_ci1}%
  \BibitemOpen
  \bibfield  {author} {\bibinfo {author} {\bibfnamefont {G.~W.}\ \bibnamefont
  {Kellner}},\ }\href@noop {} {\bibfield  {journal} {\bibinfo  {journal} {Z.
  Phys.}\ }\textbf {\bibinfo {volume} {44}},\ \bibinfo {pages} {91} (\bibinfo
  {year} {1927}{\natexlab{a}})}\BibitemShut {NoStop}%
\bibitem [{\citenamefont {Kellner}(1927{\natexlab{b}})}]{kellner_ci2}%
  \BibitemOpen
  \bibfield  {author} {\bibinfo {author} {\bibfnamefont {G.~W.}\ \bibnamefont
  {Kellner}},\ }\href@noop {} {\bibfield  {journal} {\bibinfo  {journal} {Z.
  Phys.}\ }\textbf {\bibinfo {volume} {44}},\ \bibinfo {pages} {110} (\bibinfo
  {year} {1927}{\natexlab{b}})}\BibitemShut {NoStop}%
\bibitem [{\citenamefont {Hylleraas}(1928)}]{hyll_ci}%
  \BibitemOpen
  \bibfield  {author} {\bibinfo {author} {\bibfnamefont {E.~A.}\ \bibnamefont
  {Hylleraas}},\ }\href@noop {} {\bibfield  {journal} {\bibinfo  {journal} {Z.
  Phys.}\ }\textbf {\bibinfo {volume} {48}},\ \bibinfo {pages} {469} (\bibinfo
  {year} {1928})}\BibitemShut {NoStop}%
\bibitem [{\citenamefont {Shavitt}(1998)}]{shavitt_his}%
  \BibitemOpen
  \bibfield  {author} {\bibinfo {author} {\bibfnamefont {I.}~\bibnamefont
  {Shavitt}},\ }\href@noop {} {\bibfield  {journal} {\bibinfo  {journal} {Mol.
  Phys.}\ }\textbf {\bibinfo {volume} {94}},\ \bibinfo {pages} {3} (\bibinfo
  {year} {1998})}\BibitemShut {NoStop}%
\bibitem [{\citenamefont {Cremer}(2013)}]{cremer_ci}%
  \BibitemOpen
  \bibfield  {author} {\bibinfo {author} {\bibfnamefont {D.}~\bibnamefont
  {Cremer}},\ }\href@noop {} {\bibfield  {journal} {\bibinfo  {journal} {WIREs
  Comput. Mol. Sci.}\ }\textbf {\bibinfo {volume} {3}},\ \bibinfo {pages} {482}
  (\bibinfo {year} {2013})}\BibitemShut {NoStop}%
\bibitem [{\citenamefont {M{\o}ller}\ and\ \citenamefont
  {Plesset}(1934)}]{mppt}%
  \BibitemOpen
  \bibfield  {author} {\bibinfo {author} {\bibfnamefont {C.}~\bibnamefont
  {M{\o}ller}}\ and\ \bibinfo {author} {\bibfnamefont {M.~S.}\ \bibnamefont
  {Plesset}},\ }\href@noop {} {\bibfield  {journal} {\bibinfo  {journal} {Phys.
  Rev.}\ }\textbf {\bibinfo {volume} {46}},\ \bibinfo {pages} {618} (\bibinfo
  {year} {1934})}\BibitemShut {NoStop}%
\bibitem [{\citenamefont {{F. Coester}}(1958)}]{cc1}%
  \BibitemOpen
  \bibfield  {author} {\bibinfo {author} {\bibnamefont {{F. Coester}}},\
  }\href@noop {} {\bibfield  {journal} {\bibinfo  {journal} {Nuc. Phys.}\
  }\textbf {\bibinfo {volume} {7}},\ \bibinfo {pages} {421} (\bibinfo {year}
  {1958})}\BibitemShut {NoStop}%
\bibitem [{\citenamefont {{F. Coester and H. K{\"u}mmel}}(1960)}]{cc2}%
  \BibitemOpen
  \bibfield  {author} {\bibinfo {author} {\bibnamefont {{F. Coester and H.
  K{\"u}mmel}}},\ }\href@noop {} {\bibfield  {journal} {\bibinfo  {journal}
  {Nuc. Phys.}\ }\textbf {\bibinfo {volume} {17}},\ \bibinfo {pages} {477}
  (\bibinfo {year} {1960})}\BibitemShut {NoStop}%
\bibitem [{\citenamefont {Hohenberg}\ and\ \citenamefont {Kohn}(1964)}]{dft_1}%
  \BibitemOpen
  \bibfield  {author} {\bibinfo {author} {\bibfnamefont {P.}~\bibnamefont
  {Hohenberg}}\ and\ \bibinfo {author} {\bibfnamefont {W.}~\bibnamefont
  {Kohn}},\ }\href@noop {} {\bibfield  {journal} {\bibinfo  {journal} {Phys.
  Rev. B}\ }\textbf {\bibinfo {volume} {136}},\ \bibinfo {pages} {864}
  (\bibinfo {year} {1964})}\BibitemShut {NoStop}%
\bibitem [{\citenamefont {Kohn}\ and\ \citenamefont {Sham}(1965)}]{dft_2}%
  \BibitemOpen
  \bibfield  {author} {\bibinfo {author} {\bibfnamefont {W.}~\bibnamefont
  {Kohn}}\ and\ \bibinfo {author} {\bibfnamefont {L.~J.}\ \bibnamefont
  {Sham}},\ }\href@noop {} {\bibfield  {journal} {\bibinfo  {journal} {Phys.
  Rev.}\ }\textbf {\bibinfo {volume} {140}},\ \bibinfo {pages} {1133} (\bibinfo
  {year} {1965})}\BibitemShut {NoStop}%
\bibitem [{\citenamefont {Bartlett}\ and\ \citenamefont
  {Musial}(2007)}]{bartlett_rev07}%
  \BibitemOpen
  \bibfield  {author} {\bibinfo {author} {\bibfnamefont {R.~J.}\ \bibnamefont
  {Bartlett}}\ and\ \bibinfo {author} {\bibfnamefont {M.}~\bibnamefont
  {Musial}},\ }\href@noop {} {\bibfield  {journal} {\bibinfo  {journal} {Rev.
  Mod. Phys.}\ }\textbf {\bibinfo {volume} {79}},\ \bibinfo {pages} {291}
  (\bibinfo {year} {2007})}\BibitemShut {NoStop}%
\bibitem [{\citenamefont {Lyakh}\ \emph {et~al.}(2012)\citenamefont {Lyakh},
  \citenamefont {Musia{\l}}, \citenamefont {Lotrich},\ and\ \citenamefont
  {Bartlett}}]{bartlett_2011}%
  \BibitemOpen
  \bibfield  {author} {\bibinfo {author} {\bibfnamefont {D.~I.}\ \bibnamefont
  {Lyakh}}, \bibinfo {author} {\bibfnamefont {M.}~\bibnamefont {Musia{\l}}},
  \bibinfo {author} {\bibfnamefont {V.~F.}\ \bibnamefont {Lotrich}}, \ and\
  \bibinfo {author} {\bibfnamefont {R.~J.}\ \bibnamefont {Bartlett}},\
  }\href@noop {} {\bibfield  {journal} {\bibinfo  {journal} {Chem. Rev.}\
  }\textbf {\bibinfo {volume} {112}},\ \bibinfo {pages} {182} (\bibinfo {year}
  {2012})}\BibitemShut {NoStop}%
\bibitem [{\citenamefont {Miyagi}\ and\ \citenamefont {Madsen}(2013)}]{rasscf}%
  \BibitemOpen
  \bibfield  {author} {\bibinfo {author} {\bibfnamefont {H.}~\bibnamefont
  {Miyagi}}\ and\ \bibinfo {author} {\bibfnamefont {L.~B.}\ \bibnamefont
  {Madsen}},\ }\href@noop {} {\bibfield  {journal} {\bibinfo  {journal} {Phys.
  Rev. A}\ }\textbf {\bibinfo {volume} {87}},\ \bibinfo {pages} {062511}
  (\bibinfo {year} {2013})}\BibitemShut {NoStop}%
\bibitem [{\citenamefont {Sato}\ and\ \citenamefont
  {Ishikawa}(2013)}]{sato_rasscf}%
  \BibitemOpen
  \bibfield  {author} {\bibinfo {author} {\bibfnamefont {T.}~\bibnamefont
  {Sato}}\ and\ \bibinfo {author} {\bibfnamefont {K.~L.}\ \bibnamefont
  {Ishikawa}},\ }\href@noop {} {\bibfield  {journal} {\bibinfo  {journal}
  {Phys. Rev. A}\ }\textbf {\bibinfo {volume} {88}},\ \bibinfo {pages} {023402}
  (\bibinfo {year} {2013})}\BibitemShut {NoStop}%
\bibitem [{\citenamefont {Miyagi}\ and\ \citenamefont
  {Madsen}(2014)}]{rasscf2}%
  \BibitemOpen
  \bibfield  {author} {\bibinfo {author} {\bibfnamefont {H.}~\bibnamefont
  {Miyagi}}\ and\ \bibinfo {author} {\bibfnamefont {L.~B.}\ \bibnamefont
  {Madsen}},\ }\href@noop {} {\bibfield  {journal} {\bibinfo  {journal} {Phys.
  Rev. A}\ }\textbf {\bibinfo {volume} {89}},\ \bibinfo {pages} {063416}
  (\bibinfo {year} {2014})}\BibitemShut {NoStop}%
\bibitem [{\citenamefont {Bauch}\ \emph {et~al.}(2014)\citenamefont {Bauch},
  \citenamefont {S{\o}rensen},\ and\ \citenamefont {Madsen}}]{Bauch}%
  \BibitemOpen
  \bibfield  {author} {\bibinfo {author} {\bibfnamefont {S.}~\bibnamefont
  {Bauch}}, \bibinfo {author} {\bibfnamefont {L.~K.}\ \bibnamefont
  {S{\o}rensen}}, \ and\ \bibinfo {author} {\bibfnamefont {L.~B.}\ \bibnamefont
  {Madsen}},\ }\href@noop {} {\bibfield  {journal} {\bibinfo  {journal} {Phys.
  Rev. A}\ }\textbf {\bibinfo {volume} {90}},\ \bibinfo {pages} {062508}
  (\bibinfo {year} {2014})}\BibitemShut {NoStop}%
\bibitem [{\citenamefont {Hochstuhl}\ and\ \citenamefont
  {Bonitz}(2012)}]{Hochstuhl}%
  \BibitemOpen
  \bibfield  {author} {\bibinfo {author} {\bibfnamefont {D.}~\bibnamefont
  {Hochstuhl}}\ and\ \bibinfo {author} {\bibfnamefont {M.}~\bibnamefont
  {Bonitz}},\ }\href@noop {} {\bibfield  {journal} {\bibinfo  {journal} {Phys.
  Rev. A}\ }\textbf {\bibinfo {volume} {86}},\ \bibinfo {pages} {053424}
  (\bibinfo {year} {2012})}\BibitemShut {NoStop}%
\bibitem [{\citenamefont {Hochstuhl}\ \emph {et~al.}(2014)\citenamefont
  {Hochstuhl}, \citenamefont {Hinz},\ and\ \citenamefont
  {Bonitz}}]{Hochstuhl2}%
  \BibitemOpen
  \bibfield  {author} {\bibinfo {author} {\bibfnamefont {D.}~\bibnamefont
  {Hochstuhl}}, \bibinfo {author} {\bibfnamefont {C.~M.}\ \bibnamefont {Hinz}},
  \ and\ \bibinfo {author} {\bibfnamefont {M.}~\bibnamefont {Bonitz}},\
  }\href@noop {} {\bibfield  {journal} {\bibinfo  {journal} {Eur. Phys. J.
  Special Topics}\ }\textbf {\bibinfo {volume} {223}},\ \bibinfo {pages} {177}
  (\bibinfo {year} {2014})}\BibitemShut {NoStop}%
\bibitem [{\citenamefont {Klamroth}(2003)}]{klamroth_cis}%
  \BibitemOpen
  \bibfield  {author} {\bibinfo {author} {\bibfnamefont {T.}~\bibnamefont
  {Klamroth}},\ }\href@noop {} {\bibfield  {journal} {\bibinfo  {journal}
  {Phys. Rev. B}\ }\textbf {\bibinfo {volume} {68}},\ \bibinfo {pages} {245421}
  (\bibinfo {year} {2003})}\BibitemShut {NoStop}%
\bibitem [{\citenamefont {Krause}\ \emph {et~al.}(2005)\citenamefont {Krause},
  \citenamefont {Klamroth},\ and\ \citenamefont {Saalfrank}}]{klamroth_cis2}%
  \BibitemOpen
  \bibfield  {author} {\bibinfo {author} {\bibfnamefont {P.}~\bibnamefont
  {Krause}}, \bibinfo {author} {\bibfnamefont {T.}~\bibnamefont {Klamroth}}, \
  and\ \bibinfo {author} {\bibfnamefont {P.}~\bibnamefont {Saalfrank}},\
  }\href@noop {} {\bibfield  {journal} {\bibinfo  {journal} {J. Chem. Phys.}\
  }\textbf {\bibinfo {volume} {123}},\ \bibinfo {pages} {074105} (\bibinfo
  {year} {2005})}\BibitemShut {NoStop}%
\bibitem [{\citenamefont {Rohringer}\ \emph {et~al.}(2006)\citenamefont
  {Rohringer}, \citenamefont {Gordon},\ and\ \citenamefont
  {Santra}}]{rohringer_cis}%
  \BibitemOpen
  \bibfield  {author} {\bibinfo {author} {\bibfnamefont {N.}~\bibnamefont
  {Rohringer}}, \bibinfo {author} {\bibfnamefont {A.}~\bibnamefont {Gordon}}, \
  and\ \bibinfo {author} {\bibfnamefont {R.}~\bibnamefont {Santra}},\
  }\href@noop {} {\bibfield  {journal} {\bibinfo  {journal} {Phys. Rev. A}\
  }\textbf {\bibinfo {volume} {74}},\ \bibinfo {pages} {043420} (\bibinfo
  {year} {2006})}\BibitemShut {NoStop}%
\bibitem [{\citenamefont {Greenman}\ \emph {et~al.}(2010)\citenamefont
  {Greenman}, \citenamefont {Ho}, \citenamefont {Pabst}, \citenamefont
  {Kamarchik}, \citenamefont {Mazziotti},\ and\ \citenamefont
  {Santra}}]{greenman_cis}%
  \BibitemOpen
  \bibfield  {author} {\bibinfo {author} {\bibfnamefont {L.}~\bibnamefont
  {Greenman}}, \bibinfo {author} {\bibfnamefont {P.~J.}\ \bibnamefont {Ho}},
  \bibinfo {author} {\bibfnamefont {S.}~\bibnamefont {Pabst}}, \bibinfo
  {author} {\bibfnamefont {E.}~\bibnamefont {Kamarchik}}, \bibinfo {author}
  {\bibfnamefont {D.~A.}\ \bibnamefont {Mazziotti}}, \ and\ \bibinfo {author}
  {\bibfnamefont {R.}~\bibnamefont {Santra}},\ }\href@noop {} {\bibfield
  {journal} {\bibinfo  {journal} {Phys. Rev. A}\ }\textbf {\bibinfo {volume}
  {82}},\ \bibinfo {pages} {023406} (\bibinfo {year} {2010})}\BibitemShut
  {NoStop}%
\bibitem [{\citenamefont {Grimme}\ and\ \citenamefont
  {Waletzke}(1999)}]{dft_mrci}%
  \BibitemOpen
  \bibfield  {author} {\bibinfo {author} {\bibfnamefont {S.}~\bibnamefont
  {Grimme}}\ and\ \bibinfo {author} {\bibfnamefont {M.}~\bibnamefont
  {Waletzke}},\ }\href@noop {} {\bibfield  {journal} {\bibinfo  {journal} {J.
  Chem. Phys.}\ }\textbf {\bibinfo {volume} {111}},\ \bibinfo {pages} {5645}
  (\bibinfo {year} {1999})}\BibitemShut {NoStop}%
\bibitem [{\citenamefont {Kleinschmidt}\ \emph {et~al.}(2002)\citenamefont
  {Kleinschmidt}, \citenamefont {Tatchen},\ and\ \citenamefont
  {Marian}}]{SPOCK}%
  \BibitemOpen
  \bibfield  {author} {\bibinfo {author} {\bibfnamefont {M.}~\bibnamefont
  {Kleinschmidt}}, \bibinfo {author} {\bibfnamefont {J.}~\bibnamefont
  {Tatchen}}, \ and\ \bibinfo {author} {\bibfnamefont {C.~M.}\ \bibnamefont
  {Marian}},\ }\href@noop {} {\bibfield  {journal} {\bibinfo  {journal} {J.
  Comput. Chem.}\ }\textbf {\bibinfo {volume} {23}},\ \bibinfo {pages} {824}
  (\bibinfo {year} {2002})}\BibitemShut {NoStop}%
\bibitem [{\citenamefont {Chwee}\ \emph {et~al.}(2008)\citenamefont {Chwee},
  \citenamefont {Szilva}, \citenamefont {Lindh},\ and\ \citenamefont
  {Carter}}]{lin_sca_mrci}%
  \BibitemOpen
  \bibfield  {author} {\bibinfo {author} {\bibfnamefont {T.~S.}\ \bibnamefont
  {Chwee}}, \bibinfo {author} {\bibfnamefont {A.~B.}\ \bibnamefont {Szilva}},
  \bibinfo {author} {\bibfnamefont {R.}~\bibnamefont {Lindh}}, \ and\ \bibinfo
  {author} {\bibfnamefont {E.~A.}\ \bibnamefont {Carter}},\ }\href@noop {}
  {\bibfield  {journal} {\bibinfo  {journal} {J. Chem. Phys.}\ }\textbf
  {\bibinfo {volume} {128}},\ \bibinfo {pages} {224106} (\bibinfo {year}
  {2008})}\BibitemShut {NoStop}%
\bibitem [{\citenamefont {Chwee}\ and\ \citenamefont
  {Carter}(2010)}]{lin_sca_mrci2}%
  \BibitemOpen
  \bibfield  {author} {\bibinfo {author} {\bibfnamefont {T.~S.}\ \bibnamefont
  {Chwee}}\ and\ \bibinfo {author} {\bibfnamefont {E.~A.}\ \bibnamefont
  {Carter}},\ }\href@noop {} {\bibfield  {journal} {\bibinfo  {journal} {J.
  Chem. Phys.}\ }\textbf {\bibinfo {volume} {132}},\ \bibinfo {pages} {074104}
  (\bibinfo {year} {2010})}\BibitemShut {NoStop}%
\bibitem [{\citenamefont {Anderson}(1975)}]{anderson_qmc}%
  \BibitemOpen
  \bibfield  {author} {\bibinfo {author} {\bibfnamefont {J.~B.}\ \bibnamefont
  {Anderson}},\ }\href@noop {} {\bibfield  {journal} {\bibinfo  {journal} {J.
  Chem. Phys.}\ }\textbf {\bibinfo {volume} {63}},\ \bibinfo {pages} {1499}
  (\bibinfo {year} {1975})}\BibitemShut {NoStop}%
\bibitem [{\citenamefont {L{\"u}chow}\ and\ \citenamefont
  {Anderson}(2000)}]{luchow_qmc}%
  \BibitemOpen
  \bibfield  {author} {\bibinfo {author} {\bibfnamefont {A.}~\bibnamefont
  {L{\"u}chow}}\ and\ \bibinfo {author} {\bibfnamefont {J.~B.}\ \bibnamefont
  {Anderson}},\ }\href@noop {} {\bibfield  {journal} {\bibinfo  {journal} {Ann.
  Rev. Phys. Chem.}\ }\textbf {\bibinfo {volume} {51}},\ \bibinfo {pages} {501}
  (\bibinfo {year} {2000})}\BibitemShut {NoStop}%
\bibitem [{\citenamefont {Caballol}\ and\ \citenamefont
  {Malrieu}(1992)}]{selcting_ci}%
  \BibitemOpen
  \bibfield  {author} {\bibinfo {author} {\bibfnamefont {R.}~\bibnamefont
  {Caballol}}\ and\ \bibinfo {author} {\bibfnamefont {J.-P.}\ \bibnamefont
  {Malrieu}},\ }\href@noop {} {\bibfield  {journal} {\bibinfo  {journal} {Chem.
  Phys. Lett.}\ }\textbf {\bibinfo {volume} {180}},\ \bibinfo {pages} {543}
  (\bibinfo {year} {1992})}\BibitemShut {NoStop}%
\bibitem [{\citenamefont {Bunge}(2006)}]{bunge_ci}%
  \BibitemOpen
  \bibfield  {author} {\bibinfo {author} {\bibfnamefont {C.~F.}\ \bibnamefont
  {Bunge}},\ }\href@noop {} {\bibfield  {journal} {\bibinfo  {journal} {J.
  Chem. Phys.}\ }\textbf {\bibinfo {volume} {125}},\ \bibinfo {pages} {014107}
  (\bibinfo {year} {2006})}\BibitemShut {NoStop}%
\bibitem [{\citenamefont {Bunge}\ and\ \citenamefont
  {Carb{\'o}-Dorca}(2006)}]{bunge_ci2}%
  \BibitemOpen
  \bibfield  {author} {\bibinfo {author} {\bibfnamefont {C.~F.}\ \bibnamefont
  {Bunge}}\ and\ \bibinfo {author} {\bibfnamefont {R.}~\bibnamefont
  {Carb{\'o}-Dorca}},\ }\href@noop {} {\bibfield  {journal} {\bibinfo
  {journal} {J. Chem. Phys.}\ }\textbf {\bibinfo {volume} {125}},\ \bibinfo
  {pages} {014108} (\bibinfo {year} {2006})}\BibitemShut {NoStop}%
\bibitem [{\citenamefont {Almora-D{\'i}az}(2014)}]{diaz_ci}%
  \BibitemOpen
  \bibfield  {author} {\bibinfo {author} {\bibfnamefont {C.~X.}\ \bibnamefont
  {Almora-D{\'i}az}},\ }\href@noop {} {\bibfield  {journal} {\bibinfo
  {journal} {J. Chem. Phys.}\ }\textbf {\bibinfo {volume} {140}},\ \bibinfo
  {pages} {184302} (\bibinfo {year} {2014})}\BibitemShut {NoStop}%
\bibitem [{\citenamefont {Shephard}\ \emph
  {et~al.}(2014{\natexlab{a}})\citenamefont {Shephard}, \citenamefont
  {Gidofalvi},\ and\ \citenamefont {Brozell}}]{mfgcf1}%
  \BibitemOpen
  \bibfield  {author} {\bibinfo {author} {\bibfnamefont {R.}~\bibnamefont
  {Shephard}}, \bibinfo {author} {\bibfnamefont {G.}~\bibnamefont {Gidofalvi}},
  \ and\ \bibinfo {author} {\bibfnamefont {S.~R.}\ \bibnamefont {Brozell}},\
  }\href@noop {} {\bibfield  {journal} {\bibinfo  {journal} {J. Chem. Phys.}\
  }\textbf {\bibinfo {volume} {141}},\ \bibinfo {pages} {064105} (\bibinfo
  {year} {2014}{\natexlab{a}})}\BibitemShut {NoStop}%
\bibitem [{\citenamefont {Shephard}\ \emph
  {et~al.}(2014{\natexlab{b}})\citenamefont {Shephard}, \citenamefont
  {Gidofalvi},\ and\ \citenamefont {Brozell}}]{mfgcf2}%
  \BibitemOpen
  \bibfield  {author} {\bibinfo {author} {\bibfnamefont {R.}~\bibnamefont
  {Shephard}}, \bibinfo {author} {\bibfnamefont {G.}~\bibnamefont {Gidofalvi}},
  \ and\ \bibinfo {author} {\bibfnamefont {S.~R.}\ \bibnamefont {Brozell}},\
  }\href@noop {} {\bibfield  {journal} {\bibinfo  {journal} {J. Chem. Phys.}\
  }\textbf {\bibinfo {volume} {141}},\ \bibinfo {pages} {064106} (\bibinfo
  {year} {2014}{\natexlab{b}})}\BibitemShut {NoStop}%
\bibitem [{\citenamefont {Nakatsuji}(2000)}]{nakatsuji_cigsd}%
  \BibitemOpen
  \bibfield  {author} {\bibinfo {author} {\bibfnamefont {H.}~\bibnamefont
  {Nakatsuji}},\ }\href@noop {} {\bibfield  {journal} {\bibinfo  {journal} {J.
  Chem. Phys.}\ }\textbf {\bibinfo {volume} {113}},\ \bibinfo {pages} {2949}
  (\bibinfo {year} {2000})}\BibitemShut {NoStop}%
\bibitem [{\citenamefont {Nakatsuji}\ and\ \citenamefont
  {Davidson}(2001)}]{nakatsuji_cigsd2}%
  \BibitemOpen
  \bibfield  {author} {\bibinfo {author} {\bibfnamefont {H.}~\bibnamefont
  {Nakatsuji}}\ and\ \bibinfo {author} {\bibfnamefont {E.~R.}\ \bibnamefont
  {Davidson}},\ }\href@noop {} {\bibfield  {journal} {\bibinfo  {journal} {J.
  Chem. Phys.}\ }\textbf {\bibinfo {volume} {115}},\ \bibinfo {pages} {2000}
  (\bibinfo {year} {2001})}\BibitemShut {NoStop}%
\bibitem [{\citenamefont {Nakatsuji}(2005)}]{nakatsuji_int}%
  \BibitemOpen
  \bibfield  {author} {\bibinfo {author} {\bibfnamefont {H.}~\bibnamefont
  {Nakatsuji}},\ }\href@noop {} {\bibfield  {journal} {\bibinfo  {journal}
  {Phys. Rev. A}\ }\textbf {\bibinfo {volume} {72}},\ \bibinfo {pages} {062110}
  (\bibinfo {year} {2005})}\BibitemShut {NoStop}%
\bibitem [{\citenamefont {Sherrill}\ and\ \citenamefont {{Schaefer
  {III}}}(1999)}]{sherrill_ci}%
  \BibitemOpen
  \bibfield  {author} {\bibinfo {author} {\bibfnamefont {C.~D.}\ \bibnamefont
  {Sherrill}}\ and\ \bibinfo {author} {\bibfnamefont {H.~F.}\ \bibnamefont
  {{Schaefer {III}}}},\ }\href@noop {} {\bibfield  {journal} {\bibinfo
  {journal} {Adv. Quant. Chem.}\ }\textbf {\bibinfo {volume} {34}},\ \bibinfo
  {pages} {143} (\bibinfo {year} {1999})}\BibitemShut {NoStop}%
\bibitem [{\citenamefont {Roos}(1972)}]{roos_ci}%
  \BibitemOpen
  \bibfield  {author} {\bibinfo {author} {\bibfnamefont {B.~O.}\ \bibnamefont
  {Roos}},\ }\href@noop {} {\bibfield  {journal} {\bibinfo  {journal} {Chem.
  Phys. Lett.}\ }\textbf {\bibinfo {volume} {15}},\ \bibinfo {pages} {153}
  (\bibinfo {year} {1972})}\BibitemShut {NoStop}%
\bibitem [{\citenamefont {Paldus}(1974)}]{paldus_uga}%
  \BibitemOpen
  \bibfield  {author} {\bibinfo {author} {\bibfnamefont {J.}~\bibnamefont
  {Paldus}},\ }\href@noop {} {\bibfield  {journal} {\bibinfo  {journal} {J.
  Chem. Phys.}\ }\textbf {\bibinfo {volume} {61}},\ \bibinfo {pages} {5321}
  (\bibinfo {year} {1974})}\BibitemShut {NoStop}%
\bibitem [{\citenamefont {Shavitt}(1977)}]{shavitt_uga}%
  \BibitemOpen
  \bibfield  {author} {\bibinfo {author} {\bibfnamefont {I.}~\bibnamefont
  {Shavitt}},\ }\href@noop {} {\bibfield  {journal} {\bibinfo  {journal} {Int.
  J. Quantum Chem.: Quantum Chem. Symp.}\ }\textbf {\bibinfo {volume} {11}},\
  \bibinfo {pages} {131} (\bibinfo {year} {1977})}\BibitemShut {NoStop}%
\bibitem [{\citenamefont {Olsen}\ \emph {et~al.}(1988)\citenamefont {Olsen},
  \citenamefont {Roos}, \citenamefont {J{\o}rgensen},\ and\ \citenamefont
  {Jensen}}]{olsen2}%
  \BibitemOpen
  \bibfield  {author} {\bibinfo {author} {\bibfnamefont {J.}~\bibnamefont
  {Olsen}}, \bibinfo {author} {\bibfnamefont {B.~O.}\ \bibnamefont {Roos}},
  \bibinfo {author} {\bibfnamefont {P.}~\bibnamefont {J{\o}rgensen}}, \ and\
  \bibinfo {author} {\bibfnamefont {H.~J.~{\Aa}.}\ \bibnamefont {Jensen}},\
  }\href@noop {} {\bibfield  {journal} {\bibinfo  {journal} {J. Chem. Phys.}\
  }\textbf {\bibinfo {volume} {89}},\ \bibinfo {pages} {2185} (\bibinfo {year}
  {1988})}\BibitemShut {NoStop}%
\bibitem [{\citenamefont {Larsson}\ \emph {et~al.}(2016)\citenamefont
  {Larsson}, \citenamefont {Bauch}, \citenamefont {S{\o}rensen},\ and\
  \citenamefont {Bonitz}}]{Larsson}%
  \BibitemOpen
  \bibfield  {author} {\bibinfo {author} {\bibfnamefont {H.~R.}\ \bibnamefont
  {Larsson}}, \bibinfo {author} {\bibfnamefont {S.}~\bibnamefont {Bauch}},
  \bibinfo {author} {\bibfnamefont {L.~K.}\ \bibnamefont {S{\o}rensen}}, \ and\
  \bibinfo {author} {\bibfnamefont {M.}~\bibnamefont {Bonitz}},\ }\href@noop {}
  {\bibfield  {journal} {\bibinfo  {journal} {Phys. Rev. A}\ }\textbf {\bibinfo
  {volume} {93}},\ \bibinfo {pages} {013426} (\bibinfo {year}
  {2016})}\BibitemShut {NoStop}%
\bibitem [{\citenamefont {Bauch}\ \emph {et~al.}(2016)\citenamefont {Bauch},
  \citenamefont {Larsson}, \citenamefont {Hinz},\ and\ \citenamefont
  {Bonitz}}]{Bauch2}%
  \BibitemOpen
  \bibfield  {author} {\bibinfo {author} {\bibfnamefont {S.}~\bibnamefont
  {Bauch}}, \bibinfo {author} {\bibfnamefont {H.~R.}\ \bibnamefont {Larsson}},
  \bibinfo {author} {\bibfnamefont {C.}~\bibnamefont {Hinz}}, \ and\ \bibinfo
  {author} {\bibfnamefont {M.}~\bibnamefont {Bonitz}},\ }\href@noop {}
  {\bibfield  {journal} {\bibinfo  {journal} {Phys. Rev. A}\ }\textbf {\bibinfo
  {volume} {696}},\ \bibinfo {pages} {012008} (\bibinfo {year}
  {2016})}\BibitemShut {NoStop}%
\bibitem [{\citenamefont {Chattopadhyay}\ \emph {et~al.}(2015)\citenamefont
  {Chattopadhyay}, \citenamefont {Bauch},\ and\ \citenamefont {Madsen}}]{Sid}%
  \BibitemOpen
  \bibfield  {author} {\bibinfo {author} {\bibfnamefont {S.}~\bibnamefont
  {Chattopadhyay}}, \bibinfo {author} {\bibfnamefont {S.}~\bibnamefont
  {Bauch}}, \ and\ \bibinfo {author} {\bibfnamefont {L.~B.}\ \bibnamefont
  {Madsen}},\ }\href@noop {} {\bibfield  {journal} {\bibinfo  {journal} {Phys.
  Rev. A}\ }\textbf {\bibinfo {volume} {92}},\ \bibinfo {pages} {063423}
  (\bibinfo {year} {2015})}\BibitemShut {NoStop}%
\bibitem [{\citenamefont {S{\o}rensen}\ and\ \citenamefont
  {Olsen}(2016)}]{krcc}%
  \BibitemOpen
  \bibfield  {author} {\bibinfo {author} {\bibfnamefont {L.~K.}\ \bibnamefont
  {S{\o}rensen}}\ and\ \bibinfo {author} {\bibfnamefont {J.}~\bibnamefont
  {Olsen}},\ }\href@noop {} {\bibfield  {journal} {\bibinfo  {journal} {Mol.
  Phys.}\ } (\bibinfo {year} {2016})},\ \bibinfo {note} {{DOI}:
  10.1080/00268976.2016.1195926}\BibitemShut {NoStop}%
\bibitem [{\citenamefont {S{\o}rensen}\ \emph {et~al.}(2011)\citenamefont
  {S{\o}rensen}, \citenamefont {Olsen},\ and\ \citenamefont
  {Fleig}}]{soerensen_commcc}%
  \BibitemOpen
  \bibfield  {author} {\bibinfo {author} {\bibfnamefont {L.~K.}\ \bibnamefont
  {S{\o}rensen}}, \bibinfo {author} {\bibfnamefont {J.}~\bibnamefont {Olsen}},
  \ and\ \bibinfo {author} {\bibfnamefont {T.}~\bibnamefont {Fleig}},\
  }\href@noop {} {\bibfield  {journal} {\bibinfo  {journal} {J. Chem. Phys.}\
  }\textbf {\bibinfo {volume} {134}},\ \bibinfo {pages} {214102} (\bibinfo
  {year} {2011})}\BibitemShut {NoStop}%
\bibitem [{\citenamefont {Davidson}(1975)}]{davidson_ci}%
  \BibitemOpen
  \bibfield  {author} {\bibinfo {author} {\bibfnamefont {E.~R.}\ \bibnamefont
  {Davidson}},\ }\href@noop {} {\bibfield  {journal} {\bibinfo  {journal} {J.
  Comput. Phys.}\ }\textbf {\bibinfo {volume} {17}},\ \bibinfo {pages} {87}
  (\bibinfo {year} {1975})}\BibitemShut {NoStop}%
\bibitem [{\citenamefont {Lanczos}(1950)}]{lanczos}%
  \BibitemOpen
  \bibfield  {author} {\bibinfo {author} {\bibfnamefont {C.}~\bibnamefont
  {Lanczos}},\ }\href@noop {} {\bibfield  {journal} {\bibinfo  {journal} {J.
  Res. Natl. Bur. Stand.}\ }\textbf {\bibinfo {volume} {45}},\ \bibinfo {pages}
  {225} (\bibinfo {year} {1950})}\BibitemShut {NoStop}%
\bibitem [{\citenamefont {Helgaker}\ \emph {et~al.}(2000)\citenamefont
  {Helgaker}, \citenamefont {J{\o}rgensen},\ and\ \citenamefont
  {Olsen}}]{bible}%
  \BibitemOpen
  \bibfield  {author} {\bibinfo {author} {\bibfnamefont {T.}~\bibnamefont
  {Helgaker}}, \bibinfo {author} {\bibfnamefont {P.}~\bibnamefont
  {J{\o}rgensen}}, \ and\ \bibinfo {author} {\bibfnamefont {J.}~\bibnamefont
  {Olsen}},\ }\href@noop {} {\emph {\bibinfo {title} {Molecular
  Electronic-Structure Theory}}}\ (\bibinfo  {publisher} {John Wiley \& Sons,
  Ltd},\ \bibinfo {address} {Chichester},\ \bibinfo {year} {2000})\BibitemShut
  {NoStop}%
\bibitem [{\citenamefont {Pulay}(1983)}]{pulay_lin_sca}%
  \BibitemOpen
  \bibfield  {author} {\bibinfo {author} {\bibfnamefont {P.}~\bibnamefont
  {Pulay}},\ }\href@noop {} {\bibfield  {journal} {\bibinfo  {journal} {Chem.
  Phys. Lett.}\ }\textbf {\bibinfo {volume} {100}},\ \bibinfo {pages} {151}
  (\bibinfo {year} {1983})}\BibitemShut {NoStop}%
\bibitem [{\citenamefont {S{\ae}b{\o}}\ and\ \citenamefont
  {Pulay}(1985)}]{pulay_lin_sca2}%
  \BibitemOpen
  \bibfield  {author} {\bibinfo {author} {\bibfnamefont {S.}~\bibnamefont
  {S{\ae}b{\o}}}\ and\ \bibinfo {author} {\bibfnamefont {P.}~\bibnamefont
  {Pulay}},\ }\href@noop {} {\bibfield  {journal} {\bibinfo  {journal} {Chem.
  Phys. Lett.}\ }\textbf {\bibinfo {volume} {113}},\ \bibinfo {pages} {13}
  (\bibinfo {year} {1985})}\BibitemShut {NoStop}%
\bibitem [{\citenamefont {Zi{\'o}{\l}kowski}\ \emph {et~al.}(2010)\citenamefont
  {Zi{\'o}{\l}kowski}, \citenamefont {Jans{\'i}k}, \citenamefont
  {Kj{\ae}rgaard},\ and\ \citenamefont {J{\o}rgensen}}]{poul_lin_sca}%
  \BibitemOpen
  \bibfield  {author} {\bibinfo {author} {\bibfnamefont {M.}~\bibnamefont
  {Zi{\'o}{\l}kowski}}, \bibinfo {author} {\bibfnamefont {B.}~\bibnamefont
  {Jans{\'i}k}}, \bibinfo {author} {\bibfnamefont {T.}~\bibnamefont
  {Kj{\ae}rgaard}}, \ and\ \bibinfo {author} {\bibfnamefont {P.}~\bibnamefont
  {J{\o}rgensen}},\ }\href@noop {} {\bibfield  {journal} {\bibinfo  {journal}
  {J. Chem. Phys.}\ }\textbf {\bibinfo {volume} {133}},\ \bibinfo {pages}
  {014107} (\bibinfo {year} {2010})}\BibitemShut {NoStop}%
\bibitem [{\citenamefont {Hampel}\ and\ \citenamefont
  {Werner}(1996)}]{werner_lin_sca}%
  \BibitemOpen
  \bibfield  {author} {\bibinfo {author} {\bibfnamefont {C.}~\bibnamefont
  {Hampel}}\ and\ \bibinfo {author} {\bibfnamefont {H.~J.}\ \bibnamefont
  {Werner}},\ }\href@noop {} {\bibfield  {journal} {\bibinfo  {journal} {J.
  Chem. Phys.}\ }\textbf {\bibinfo {volume} {104}},\ \bibinfo {pages} {6286}
  (\bibinfo {year} {1996})}\BibitemShut {NoStop}%
\bibitem [{\citenamefont {H{\o}yvik}\ \emph {et~al.}(2014)\citenamefont
  {H{\o}yvik}, \citenamefont {Kristensen}, \citenamefont {Kj{\ae}rgaard},\ and\
  \citenamefont {J{\o}rgensen}}]{ida_local_orb}%
  \BibitemOpen
  \bibfield  {author} {\bibinfo {author} {\bibfnamefont {I.}~\bibnamefont
  {H{\o}yvik}}, \bibinfo {author} {\bibfnamefont {K.}~\bibnamefont
  {Kristensen}}, \bibinfo {author} {\bibfnamefont {T.}~\bibnamefont
  {Kj{\ae}rgaard}}, \ and\ \bibinfo {author} {\bibfnamefont {P.}~\bibnamefont
  {J{\o}rgensen}},\ }\href@noop {} {\bibfield  {journal} {\bibinfo  {journal}
  {Theoret. Chim. Acta}\ }\textbf {\bibinfo {volume} {133}},\ \bibinfo {pages}
  {1417} (\bibinfo {year} {2014})}\BibitemShut {NoStop}%
\bibitem [{\citenamefont {H{\o}yvik}\ \emph {et~al.}(2012)\citenamefont
  {H{\o}yvik}, \citenamefont {Jans{\'i}k},\ and\ \citenamefont
  {J{\o}rgensen}}]{ida_local_orb2}%
  \BibitemOpen
  \bibfield  {author} {\bibinfo {author} {\bibfnamefont {I.}~\bibnamefont
  {H{\o}yvik}}, \bibinfo {author} {\bibfnamefont {B.}~\bibnamefont
  {Jans{\'i}k}}, \ and\ \bibinfo {author} {\bibfnamefont {P.}~\bibnamefont
  {J{\o}rgensen}},\ }\href@noop {} {\bibfield  {journal} {\bibinfo  {journal}
  {J. Chem. Phys.}\ }\textbf {\bibinfo {volume} {137}},\ \bibinfo {pages}
  {224114} (\bibinfo {year} {2012})}\BibitemShut {NoStop}%
\bibitem [{\citenamefont {Jensen}\ \emph {et~al.}(1996)\citenamefont {Jensen},
  \citenamefont {Dyall}, \citenamefont {Saue},\ and\ \citenamefont
  {F{\ae}gri}}]{jensen_saue}%
  \BibitemOpen
  \bibfield  {author} {\bibinfo {author} {\bibfnamefont {H.~J.~{\Aa}.}\
  \bibnamefont {Jensen}}, \bibinfo {author} {\bibfnamefont {K.~G.}\
  \bibnamefont {Dyall}}, \bibinfo {author} {\bibfnamefont {T.}~\bibnamefont
  {Saue}}, \ and\ \bibinfo {author} {\bibfnamefont {K.}~\bibnamefont
  {F{\ae}gri}},\ }\href@noop {} {\bibfield  {journal} {\bibinfo  {journal} {J.
  Chem. Phys.}\ }\textbf {\bibinfo {volume} {104}},\ \bibinfo {pages} {4083}
  (\bibinfo {year} {1996})}\BibitemShut {NoStop}%
\bibitem [{\citenamefont {Duch}(1985)}]{duch_uga}%
  \BibitemOpen
  \bibfield  {author} {\bibinfo {author} {\bibfnamefont {W.}~\bibnamefont
  {Duch}},\ }\href@noop {} {\bibfield  {journal} {\bibinfo  {journal} {Int. J.
  Quantum Chem.: Quantum Chem. Symp.}\ }\textbf {\bibinfo {volume} {27}},\
  \bibinfo {pages} {59} (\bibinfo {year} {1985})}\BibitemShut {NoStop}%
\bibitem [{\citenamefont {Duch}\ and\ \citenamefont
  {Karwowski}(1982)}]{duch_uga2}%
  \BibitemOpen
  \bibfield  {author} {\bibinfo {author} {\bibfnamefont {W.}~\bibnamefont
  {Duch}}\ and\ \bibinfo {author} {\bibfnamefont {J.}~\bibnamefont
  {Karwowski}},\ }\href@noop {} {\bibfield  {journal} {\bibinfo  {journal}
  {Int. J. Quantum Chem.: Quantum Chem. Symp.}\ }\textbf {\bibinfo {volume}
  {22}},\ \bibinfo {pages} {783} (\bibinfo {year} {1982})}\BibitemShut
  {NoStop}%
\bibitem [{\citenamefont {Wasilewski}(1989)}]{karwowski_uga}%
  \BibitemOpen
  \bibfield  {author} {\bibinfo {author} {\bibfnamefont {J.}~\bibnamefont
  {Wasilewski}},\ }\href@noop {} {\bibfield  {journal} {\bibinfo  {journal}
  {Int. J. Quantum Chem.: Quantum Chem. Symp.}\ }\textbf {\bibinfo {volume}
  {36}},\ \bibinfo {pages} {503} (\bibinfo {year} {1989})}\BibitemShut
  {NoStop}%
\bibitem [{\citenamefont {Duch}(1986)}]{duch}%
  \BibitemOpen
  \bibfield  {author} {\bibinfo {author} {\bibfnamefont {W.}~\bibnamefont
  {Duch}},\ }\href@noop {} {\emph {\bibinfo {title} {Grms or Graphical
  Representation of Model Spaces: Vol 1 Basics}}}\ (\bibinfo  {publisher}
  {Lecture Notes in Chemistry Vol. 42, Springer-Verlag},\ \bibinfo {year}
  {1986})\BibitemShut {NoStop}%
\bibitem [{\citenamefont {S{\o}rensen}(2016)}]{lasse_unpub}%
  \BibitemOpen
  \bibfield  {author} {\bibinfo {author} {\bibfnamefont {L.~K.}\ \bibnamefont
  {S{\o}rensen}},\ }\href@noop {} {} (\bibinfo {year} {2016}),\ \bibinfo {note}
  {unpublished}\BibitemShut {NoStop}%
\bibitem [{\citenamefont {S{\o}rensen}(2010)}]{lasse_thesis}%
  \BibitemOpen
  \bibfield  {author} {\bibinfo {author} {\bibfnamefont {L.~K.}\ \bibnamefont
  {S{\o}rensen}},\ }\emph {\bibinfo {title} {General Order Coupled-Cluster in
  the 4-Component Framework}},\ \href@noop {} {\bibinfo {type}
  {Dissertation}},\ \bibinfo  {school} {Mathematisch-Naturwissenschaftliche
  Fakult{\"a}t}, \bibinfo {address} {Heinrich-Heine-Universit{\"a}t
  D{\"u}sseldorf} (\bibinfo {year} {2010}),\ \bibinfo {note}
  {\url{http://www.theochem.hhu.de/fileadmin/redaktion/Fakultaeten/Mathematisch-Naturwissenschaftliche_Fakultaet/Chemie/Theo_Chem/Abschlussarbeiten/sorensen.lasse.diss.pdf}}\BibitemShut
  {NoStop}%
\bibitem [{DIR()}]{DIRAC14}%
  \BibitemOpen
  \href@noop {} {}\bibinfo {note} {{DIRAC}, a relativistic ab initio electronic
  structure program, Release {DIRAC14} (2014), written by T.~Saue, L.~Visscher,
  H.~J.~{\relax Aa}.~Jensen, and R.~Bast. with contributions from V.~Bakken,
  K.~G.~Dyall, S.~Dubillard, U.~Ekstr{\"o}m, E.~Eliav, T.~Enevoldsen,
  E.~Fa{\ss}hauer, T.~Fleig, O.~Fossgaard, A.~S.~P.~Gomes, T.~Helgaker,
  J.~K.~L{\ae}rdahl, Y.~S.~Lee, J.~Henriksson, M.~Ilia{\v{s}}, Ch.~R.~Jacob,
  S.~Knecht, S.~Komorovsk{\'y}, O.~Kullie, C.~V.~Larsen, H.~S.~Nataraj,
  P.~Norman, G.~Olejniczak, J.~Olsen, Y.~C.~Park, J.~K.~Pedersen,
  M.~Pernpointner, R.~di~Remigio, K.~Ruud, P.~Sa{\l}ek, B.~Schimmelpfennig,
  J.~Sikkema, A.~J.~Thorvaldsen, J.~Thyssen, J.~van~Stralen, S.~Villaume,
  O.~Visser, T.~Winther, and S.~Yamamoto (see
  \url{http://www.diracprogram.org})}\BibitemShut {NoStop}%
\bibitem [{\citenamefont {Rescigno}\ and\ \citenamefont
  {McCurdy}(2000)}]{rescigno2000}%
  \BibitemOpen
  \bibfield  {author} {\bibinfo {author} {\bibfnamefont {T.}~\bibnamefont
  {Rescigno}}\ and\ \bibinfo {author} {\bibfnamefont {C.}~\bibnamefont
  {McCurdy}},\ }\href@noop {} {\bibfield  {journal} {\bibinfo  {journal} {Phys.
  Rev. A}\ }\textbf {\bibinfo {volume} {62}},\ \bibinfo {pages} {032706}
  (\bibinfo {year} {2000})}\BibitemShut {NoStop}%
\bibitem [{\citenamefont {Park}\ and\ \citenamefont {Light}(1986)}]{park1986}%
  \BibitemOpen
  \bibfield  {author} {\bibinfo {author} {\bibfnamefont {T.~J.}\ \bibnamefont
  {Park}}\ and\ \bibinfo {author} {\bibfnamefont {J.~C.}\ \bibnamefont
  {Light}},\ }\href@noop {} {\bibfield  {journal} {\bibinfo  {journal} {J.
  Chem. Phys.}\ }\textbf {\bibinfo {volume} {85}},\ \bibinfo {pages} {5870}
  (\bibinfo {year} {1986})}\BibitemShut {NoStop}%
\bibitem [{\citenamefont {Beck}\ \emph {et~al.}(2000)\citenamefont {Beck},
  \citenamefont {J\"ackle}, \citenamefont {Worth},\ and\ \citenamefont
  {Meyer}}]{beck2000}%
  \BibitemOpen
  \bibfield  {author} {\bibinfo {author} {\bibfnamefont {M.~H.}\ \bibnamefont
  {Beck}}, \bibinfo {author} {\bibfnamefont {A.}~\bibnamefont {J\"ackle}},
  \bibinfo {author} {\bibfnamefont {G.}~\bibnamefont {Worth}}, \ and\ \bibinfo
  {author} {\bibfnamefont {H.-D.}\ \bibnamefont {Meyer}},\ }\href@noop {}
  {\bibfield  {journal} {\bibinfo  {journal} {Phys. Rev.}\ }\textbf {\bibinfo
  {volume} {324}},\ \bibinfo {pages} {1} (\bibinfo {year} {2000})}\BibitemShut
  {NoStop}%
\bibitem [{\citenamefont {Sch{\"u}tz}\ and\ \citenamefont
  {Werner}(2001)}]{werner_lin_sca2}%
  \BibitemOpen
  \bibfield  {author} {\bibinfo {author} {\bibfnamefont {M.}~\bibnamefont
  {Sch{\"u}tz}}\ and\ \bibinfo {author} {\bibfnamefont {H.~J.}\ \bibnamefont
  {Werner}},\ }\href@noop {} {\bibfield  {journal} {\bibinfo  {journal} {J.
  Chem. Phys.}\ }\textbf {\bibinfo {volume} {114}},\ \bibinfo {pages} {661}
  (\bibinfo {year} {2001})}\BibitemShut {NoStop}%
\bibitem [{\citenamefont {Flocke}\ and\ \citenamefont
  {Bartlett}(2004)}]{bartlett_lin}%
  \BibitemOpen
  \bibfield  {author} {\bibinfo {author} {\bibfnamefont {N.}~\bibnamefont
  {Flocke}}\ and\ \bibinfo {author} {\bibfnamefont {R.~J.}\ \bibnamefont
  {Bartlett}},\ }\href@noop {} {\bibfield  {journal} {\bibinfo  {journal} {J.
  Chem. Phys.}\ }\textbf {\bibinfo {volume} {121}},\ \bibinfo {pages} {10935}
  (\bibinfo {year} {2004})}\BibitemShut {NoStop}%
\bibitem [{\citenamefont {Pinski}\ \emph {et~al.}(2015)\citenamefont {Pinski},
  \citenamefont {Riplinger}, \citenamefont {Valeev},\ and\ \citenamefont
  {Neese}}]{neese_lin}%
  \BibitemOpen
  \bibfield  {author} {\bibinfo {author} {\bibfnamefont {P.}~\bibnamefont
  {Pinski}}, \bibinfo {author} {\bibfnamefont {C.}~\bibnamefont {Riplinger}},
  \bibinfo {author} {\bibfnamefont {E.~F.}\ \bibnamefont {Valeev}}, \ and\
  \bibinfo {author} {\bibfnamefont {F.}~\bibnamefont {Neese}},\ }\href@noop {}
  {\bibfield  {journal} {\bibinfo  {journal} {J. Chem. Phys.}\ }\textbf
  {\bibinfo {volume} {143}},\ \bibinfo {pages} {034108} (\bibinfo {year}
  {2015})}\BibitemShut {NoStop}%
\bibitem [{\citenamefont {Ettenhuber}\ \emph {et~al.}(2010)\citenamefont
  {Ettenhuber}, \citenamefont {Baudin}, \citenamefont {Kj{\ae}rgaard},
  \citenamefont {J{\o}rgensen},\ and\ \citenamefont
  {Kristensen}}]{poul_lin_sca2}%
  \BibitemOpen
  \bibfield  {author} {\bibinfo {author} {\bibfnamefont {P.}~\bibnamefont
  {Ettenhuber}}, \bibinfo {author} {\bibfnamefont {P.}~\bibnamefont {Baudin}},
  \bibinfo {author} {\bibfnamefont {T.}~\bibnamefont {Kj{\ae}rgaard}}, \bibinfo
  {author} {\bibfnamefont {P.}~\bibnamefont {J{\o}rgensen}}, \ and\ \bibinfo
  {author} {\bibfnamefont {K.}~\bibnamefont {Kristensen}},\ }\href@noop {}
  {\bibfield  {journal} {\bibinfo  {journal} {J. Chem. Phys.}\ }\textbf
  {\bibinfo {volume} {133}},\ \bibinfo {pages} {014107} (\bibinfo {year}
  {2010})}\BibitemShut {NoStop}%
\bibitem [{ols()}]{olsen_lucia}%
  \BibitemOpen
  \href@noop {} {}\bibinfo {note} {Program {LUCIA}, a general CI code written
  by J Olsen, University of Aarhus, with contributions from H. Larsen and M. P.
  F\"ulscher}\BibitemShut {NoStop}%
\bibitem [{\citenamefont {Olsen}(2000)}]{olsen_cc}%
  \BibitemOpen
  \bibfield  {author} {\bibinfo {author} {\bibfnamefont {J.}~\bibnamefont
  {Olsen}},\ }\href@noop {} {\bibfield  {journal} {\bibinfo  {journal} {J.
  Chem. Phys.}\ }\textbf {\bibinfo {volume} {113}},\ \bibinfo {pages} {7140}
  (\bibinfo {year} {2000})}\BibitemShut {NoStop}%
\bibitem [{\citenamefont {S{\o}rensen}\ \emph {et~al.}(2009)\citenamefont
  {S{\o}rensen}, \citenamefont {Fleig},\ and\ \citenamefont
  {Olsen}}]{Soerensen_spinfree}%
  \BibitemOpen
  \bibfield  {author} {\bibinfo {author} {\bibfnamefont {L.~K.}\ \bibnamefont
  {S{\o}rensen}}, \bibinfo {author} {\bibfnamefont {T.}~\bibnamefont {Fleig}},
  \ and\ \bibinfo {author} {\bibfnamefont {J.}~\bibnamefont {Olsen}},\
  }\href@noop {} {\bibfield  {journal} {\bibinfo  {journal} {J. Phys. B}\
  }\textbf {\bibinfo {volume} {42}},\ \bibinfo {pages} {165102} (\bibinfo
  {year} {2009})}\BibitemShut {NoStop}%
\bibitem [{\citenamefont {S{\o}rensen}\ \emph {et~al.}(2010)\citenamefont
  {S{\o}rensen}, \citenamefont {Fleig},\ and\ \citenamefont
  {Olsen}}]{soerensen_mrcc_zpc}%
  \BibitemOpen
  \bibfield  {author} {\bibinfo {author} {\bibfnamefont {L.~K.}\ \bibnamefont
  {S{\o}rensen}}, \bibinfo {author} {\bibfnamefont {T.}~\bibnamefont {Fleig}},
  \ and\ \bibinfo {author} {\bibfnamefont {J.}~\bibnamefont {Olsen}},\
  }\href@noop {} {\bibfield  {journal} {\bibinfo  {journal} {Z. Phys. Chem.}\
  }\textbf {\bibinfo {volume} {224}},\ \bibinfo {pages} {671} (\bibinfo {year}
  {2010})}\BibitemShut {NoStop}%
\end{thebibliography}%

\end{document}